\documentclass[useAMS,usenatbib]{mn2e}

\usepackage[dvips]{graphicx}
\usepackage{amssymb}
\usepackage{amsmath}
\usepackage{amstext}

\newcommand\apj  {ApJ}

\newcommand\mnras {MNRAS}
\newcommand\aj {AJ}

\newcommand{\nodata}{...}

\DeclareRobustCommand{\ion}[2]{%
\relax\ifmmode
\ifx\testbx\f@series
{\mathbf{#1\,\mathsc{#2}}}\else
{\mathrm{#1\,\mathsc{#2}}}\fi
\else\textup{#1\,{\mdseries\textsc{#2}}}%
\fi}

\newcommand{\Mo}{$M_{\odot}$}
\newcommand{\Zo}{$Z_{\odot}$}
\newcommand{\Ha}{H$\alpha$}

\title[Properties of SF regions within WR galaxies]{Photometric and spectroscopic studies of\\ star-forming regions within Wolf-Rayet galaxies}       
\author[Chrisphin Karthick et al.]{Chrisphin Karthick, M,$^{1,2}$\thanks{E-mail: chrisphin@gmail.com, chrisphin@aries.res.in} \'Angel R. L\'opez-S\'anchez,$^{3,4}$
D. K. Sahu,$^5$ \newauthor {B. B. Sanwal,$^1 $ Shuchi Bisht,$^6$} \\
$^1$Aryabhatta Research Institute of Observational Sciences (ARIES), Nainital-263 002, India\\
$^2$Department  of Astrophysics, Pondicherry University, Puducherry - 605 014, India\\
$^3$Australian Astronomical Observatory, PO Box 915, North Ryde, NSW 1670, Australia\\
$^4$Department of Physics and Astronomy, Macquarie University, NSW 2109, Australia\\
$^5$Indian Institute of Astrophysics(IIA), Bangalore-560 034, India\\
$^6$Kumaun University, Nainital-263 001, India}

\date{Received date: 20 Mar 2012; Accepted date: 28 Nov 2013 }

\begin{document}


\maketitle

\begin{abstract}

We present a study of the properties of star-forming regions within a sample of 7 Wolf-Rayet (WR) galaxies. We analyze their morphologies, colours, star-formation rate (SFR), metallicities, and stellar populations combining broad-band and narrow-band photometry with low-resolution optical spectroscopy. The $UBVRI$ observations were made through the 2m HCT (Himalayan Chandra Telescope) and 1m ARIES telescope. The spectroscopic data were obtained using the {\it Hanle Faint Object Spectrograph Camera} (HFOSC) mounted on the 2m HCT. The observed galaxies are NGC~1140, IRAS~07164+5301, NGC~3738, UM~311, NGC~6764, NGC~4861 and NGC~3003. 
The optical spectra have been used to search for the faint WR features, to
confirm that the ionization of the gas is consequence of the massive stars, and to quantify the oxygen abundance of each galaxy using several and independent empirical calibrations. We  detected the broad features originated by WR stars in  NGC~1140 and NGC~4861 and used them to derive their population of massive stars. For these two galaxies we also
derive the oxygen abundance using a direct estimation of the electron temperature of the ionized gas.
The N/O ratio in NGC~4861 is $\sim$0.25-0.35~dex higher than expected, which may be a consequence
of the chemical pollution by N-rich material released by WR stars. 
Using our H$\alpha$ images we have identified tens of regions within these galaxies, for which we  derived the SFR. 
Our H$\alpha$-based SFR usually agrees with the SFR computed using the far-infrared and the radio-continuum flux.
For all regions we found that the most recent star-formation event is 3 -- 6~Myr old.
We used the optical broad-band colours in combination with Starburst99 models to estimate the internal reddening and the age of the dominant underlying stellar population within all these regions.  Knots in NGC~3738, NGC~6764 and NGC~3003 generally show the presence of an important old (400 -- 1000~Myr) stellar population. However, the optical colours are not able to detect stars older than 20 -- 50~Myr in the knots of the other four galaxies. This fact suggests both the intensity of the starbursts and that the star-formation activity has been ongoing for at least some few tens of million years in these objects.

\end{abstract}


\begin{table*}
 \caption{List of WR galaxies observed in our study$^{a}$. \label{Table1} }
\begin{tabular}{lccccccc}\hline
Object Name              &  NGC~1140   &  IRAS~07164+5301$^{d}$ & NGC~3738  & UM~311   & NGC~6764 & NGC~4861 & NGC~3003    \\    \hline     
RA  [{\it  J2000}]       &  02 54 33   &   07 20 25             & 11 35 49  & 01 15 34 & 19 08 16 & 12 59 02 & 09 48 36    \\
   
Dec [{\it  J2000}]       &--10 01 40   &  +52 55 32             & +54 31 26 &--00 51 46 &+50 56 00 &+34 51 34 & +33 25 17   \\
$m_V$  [ mag ]       & 12.8        &  14.13                  &12.13      & 17.9     &12.56     &12.9      & 12.33       \\
V$_{r}$  [ km\,s$^{-1}$ ]& 1501    &  12981                 &~~229      & 1675     & 2416     & 833      & 1478        \\     
Distance$^{e}$ [ Mpc ]   &17.90        &  177                   &~~5.56     &18.7      &31.3      &14.8      &24.00        \\
E(B-V)$_{Galactic}^{c}$[ mag ] & 0.038       &  0.075                  & 0.010     & 0.039    &0.067     &0.010     & 0.013       \\ 
  $M_V$  [ mag ]    & $-$18.43    & $-$22.03                  &$-$16.59   & $-$13.42 & $-$19.85 & $-$17.94 & $-$19.56    \\
 $Z^b$                   &  0.010      & 0.014                  & 0.0089     & 0.0074   &  0.0195  & 0.0033   & 0.016       \\     
1.2 arcmin$^f$ [ kpc ]        & 6.1   & 61.0              & 1.9     & 6.4  &  10.6  & 5.1  & 8.16  \\

 \hline
\end{tabular}
\begin{flushleft}

$^a$: Data taken from NASA/Extragalactic Database (NED).\\
$^b$:  $Z=0.02\times10^{[ \log({\rm O/H}) - \log({\rm O/H})_{\odot}]}$, assuming 12+log(O/H)$_{\odot}$ = 8.66, Asplund et al. (2005). \\ 
$^c$: Foreground extinction from Schlegel et al (1998). \\
$^d$: Parameters derived in this work.\\
$^e$: Luminosity distance. \\
$^f$: Projected distance, in kpc, of 1.2 arcmin at the distance of the galaxy.

\end{flushleft}

\end{table*}

\begin{figure}
\begin{center}
\includegraphics[scale=0.35]{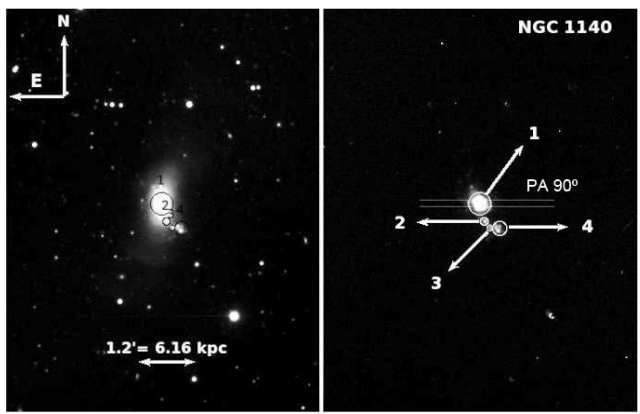} 
\caption{$R$-band (left) and  continuum-subtracted H$\alpha$ (right) images of NGC~1140. Both panels have the same dimensions. The slit position has been indicated on top of the continuum-subtracted H$\alpha$ image. \label{NGC1140image}}
\end{center}
\end{figure}

\section{Introduction}

Wolf-Rayet (WR)  galaxies   are   defined as those galaxies which  
show broad emission features associated to WR stars in their  integrated spectra. 
The presence of the Wolf-Rayet stars are reflected in 
galaxy spectrum as two important broad features, the blue WR bump  
(between 4650 - 4690~\AA,  mainly due to \ion{N}{iii}, \ion{N}{v} and \ion{He}{ii})
and the red WR bump (at $\sim$5808~\AA,  due to \ion{C}{iv} emission line). The
nebular \ion{He}{ii}~$\lambda$4686
is also associated to the  presence of these massive  stars, however 
 other ionization mechanisms  may also create this line, as discussed 
by \citet{Garnett91,G04,GIT00} and \citet{LSE10a}.
Wolf-Rayet galaxies were first cataloged 
 by \citet{C91} and later by \citet{SCP99}, but hundreds
 of these objects were found  using the {\it Sloan Digital Sky Survey},
 SDSS \citep{Zhang07,BKD08}.
 
The morphological type of WR galaxies  varies from the  low-mass
blue compact dwarf (BCD) irregular galaxies, to massive spirals and
luminous  merging  galaxies. WR features are often found in starburst   galaxies. 
The progenitors of the Wolf-Rayet stars are the most massive ($M\gtrsim25$~\Mo\ for \Zo),
luminous (10$^5$ to 10$^6$~$L_{\odot}$) and hot ($\sim$50,000~K) O stars,
and they finalize their days exploding as type Ib/Ic supernovae
\citep{MeynetMaeder05}.
Actually, the minimum stellar  mass that an O star  needs to reach the  WR phase and
its duration depends on  the metallicity. In general,  the  WR 
phenomenon is short lived, it  exists only for  $\leq 1$~Myr. 

 Hence,  the detection of WR features in the spectra
of a galaxy constrains the properties of the star-formation
processes. Because the first WR stars typically appear around
2 -- 3~Myr after the starburst is initiated and disappear within
some 5~Myr \citep{MeynetMaeder05}, their detection informs
about both the youth and strength of the burst, offering the opportunity
to study an approximately coeval sample of very young
starbursts \citep{SV98}
and the role that interaction with or between
dwarf galaxies and/or low surface brightness objects plays in
the triggering mechanism of the strong star-formation activity \citep{LSE08, LS10}.
Furthermore, the detection of WR stars also allows to 
study the formation and feedback of massive stars in
starburst galaxies \citep*{GIT00,FCCG04,Buckalew05,LSE10a,LSE10b}.

\begin{figure}
\begin{center}
\includegraphics[scale=0.35]{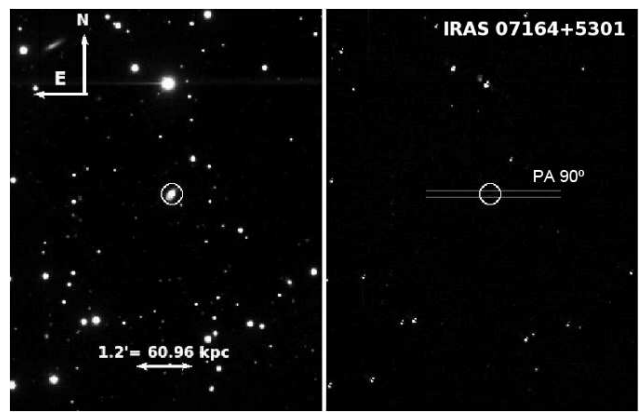} 
\caption{$R$-band (left) and  continuum-subtracted H$\alpha$ (right) images of  IRAS~07164+5301. Both panels have the same dimensions. The slit position has been indicated on top of the continuum-subtracted H$\alpha$ image. \label{IRAS07image}}
\end{center}
\end{figure}

\begin{table*}
\scriptsize
  \caption{Details of the optical broad-band, H$\alpha$ and low-resolution spectroscopical observations of WR galaxies. All spectroscopical observations were conducted at the 2m Himalayan Chandra  Telescope  (HCT).  \label{obs}}
  \begin{tabular}{l r r@{\hspace{6pt}}r@{\hspace{6pt}}r@{\hspace{6pt}}r@{\hspace{6pt}}rr @{\hspace{2pt}} r r@{\hspace{6pt}}r@{\hspace{6pt}}r  r@{\hspace{2pt}}  r@{\hspace{6pt}}r@{\hspace{6pt}}r@{\hspace{6pt}}r@{\hspace{6pt}}r}
  \hline
 
 &  \multicolumn{7}{c@{\hspace{4pt}}}{Broad-band Images} &  &
\multicolumn{3}{c@{\hspace{4pt}}}{H$\alpha$ images} & & 
\multicolumn{3}{c@{\hspace{4pt}}}{Optical Spectroscopy} \\
 \noalign{\smallskip}
\cline{2-8}
\cline{10-12}
\cline{14-17} 
  \noalign{\smallskip}
  
 Galaxy & 
 Date     & $U$    &  $B$  & $V$    & $R$    &  $I$   &  Teles. & & 
 Date     & H$\alpha$ &  $R$  & & 
 Date      &Exp. T. &  PA  & Airmass \\ 
 
 & & [s] & [s] & [s] & [s] & [s] & & & & [s] & [s] & & & [s]  & [$^{\circ}$] \\ 
 
 \hline    
 
NGC 1140       &14/12/2007 &3900 &3600 &2700 &3000 &3300 &ARIES & &17/12/2006&2100       &900  & & 05/12/2007&1900 &  90      &  1.38  \\        
IRAS\,07164+5301&16/02/2010 &7x900&5x600&5x300&5x240&5x240&HCT &  &16/02/2010&4x900      &3x180 & &15/02/2010 &900&   90     & 1.09    \\
NGC 3738     &17/02/2010 &6x720&7x480&7x300&6x240&5x240&HCT &  &12/03/2008&1200       &300  & & 15/02/2010 &1500&   90     &  1.10 \\       
UM 311         &31/10/2008 &3000 &1200 & 900 & 360 & 260 & HCT&  &31/10/2008 &4200      &360  & & 16/02/2010 &1800&   135      & 2.20 \\
NGC 6764      &31/10/2008 &3200 &650& 360& 200& 200& HCT     &  &31/10/2008 & 2x900    &900  & & 11/07/2011$^{a}$ &1200 &  70       & 1.08  \\
" knot~\#1   &  \nodata   & \nodata     &  \nodata  & \nodata    &  \nodata  &   \nodata &  \nodata      & & \nodata    &  \nodata          & \nodata    & &  31/10/2008$^{b}$ &3000 &     80      & 1.15  \\
NGC 4861       &03/05/2011 &2x600 &2x450 &2x300 &3x150 &3x150& HCT & &26/01/2011 &1210 & 120 & & 01/02/2011&1800 &                90 & 1.13 \\
NGC 3003       &26/03/2009 &2700 &1450 &1300 & 960 & 940& HCT & & 17/12/2006 & 2x600 & 300 & & 20/01/2012&2x1200&            70   & 1.02 \\
\hline   
\end{tabular}
\begin{flushleft}
$^a$: The spectrum was taken over the nuclear region (knot~\#2), which shows a clear AGN feature, and in region~\#3.\\
$^b$: The spectrum was taken over the region \#1 (see Fig.~\ref{NGC6764image}).
\end{flushleft}
\end{table*}

\begin{figure}
\begin{center}
\includegraphics[scale=0.35]{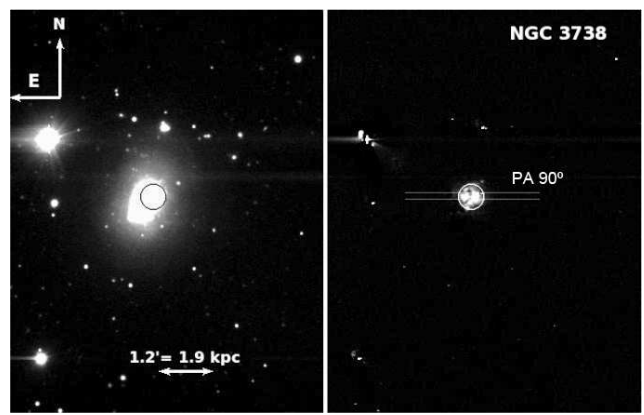} 
\caption{$R$-band (left) and  continuum-subtracted H$\alpha$ (right) images of  NGC~3738. Both panels have the same dimensions. The slit position has been indicated on top of the continuum-subtracted H$\alpha$ image. \label{NGC3738image}}
\end{center}
\end{figure}

With the aim of getting a better understanding of the properties of the WR galaxies, 
we have performed a detailed analysis of a sample of these objects using 
broad-band $U, B, V, R, I$,  narrow-band H$\alpha$ imaging and  low-resolution optical spectroscopy.
In this paper we present the results of 7 of our analyzed WR galaxies. 
We compile the main properties of these objects in Table~\ref{Table1}.

The  paper is  organized  as follows:  observations 
and data reduction are  described in Sect.~2. 
The broad-band and narrow-band photometric results are presented in Sect.~3. This section includes
the identification of the star-forming regions, the determinations of their optical colours,
and the estimation of the both the SFR and the most recent star-forming episode  via the
analysis of the net-H$\alpha$ images.
Section~4 describes our results from optical spectroscopy. We here analyze the physical
conditions of the ionized gas (reddening, nature of the ionization, electron density and electron temperature
when possible) and compute the oxygen abundances of the ionized gas within our sample galaxies. 
Section~4 also includes the analysis of the WR features in NGC~1140 and NGG~4861.
We discuss our results in Sect~5, where we first describe how we determined the ages of
both the old and young stellar populations. We also present the analysis of the individual galaxies
in Sect.~5. Finally, we list our summary and conclusions in Sect.~6.

\section{Observations and data reduction}


Except for NGC~1140, broad-band ($UBVRI$), narrow-band  H$\alpha$ (6563\,\AA/100\,\AA) 
and  spectroscopic  observations  of  the  star-forming  knots within our WR galaxy sample  were obtained  
with HFOSC ({\it Hanle Faint Object Spectrograph Camera})  mounted  on the 2m Himalayan Chandra  Telescope  (HCT), 
of  the Indian  Astronomical Observatory (IAO). HFOSC  is  equipped   with  a
2k$\times$4k SITe CCD chip. The central 2k$\times$2k region with plate
scale  of 0.296''/pixel, provides  10$\times$10 arcmin  field of view.

Broad-band photometric observations of NGC~1140  were  obtained in $UBVRI$ 
bands  using  ARIES 1m telescope with  2k$\times$2k  CCD. The  plate  scale is  0.37''/pixel, covering an area of 12'.6$\times$ 12'.6. Observations   were   done   under   photometric   sky condition. To improve  the signal  to noise ratio (S/N)  on chip $2\times2$ binning  mode was used.  Apart  from the galaxy  frames, the typical calibration
frames --e.g.  bias, flats--  were also taken  for processing  the
science frames.  Photometric standard stars  from the list compiled by \citet{L92}
were observed for photometric calibration.

\begin{figure}
\begin{center}
\includegraphics[scale=0.35]{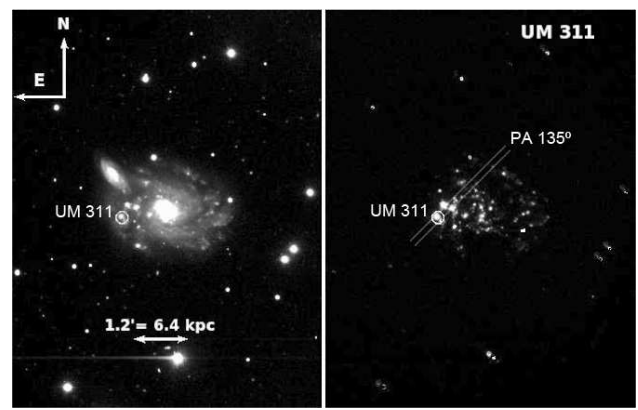} 
\caption{$R$-band (left) and  continuum-subtracted H$\alpha$ (right) images of UM~311. Both panels have the same dimensions. The slit position has been indicated on top of the continuum-subtracted H$\alpha$ image. \label{UM311image}}
\end{center}
\end{figure}

\begin{figure}
\begin{center}
\includegraphics[scale=0.35]{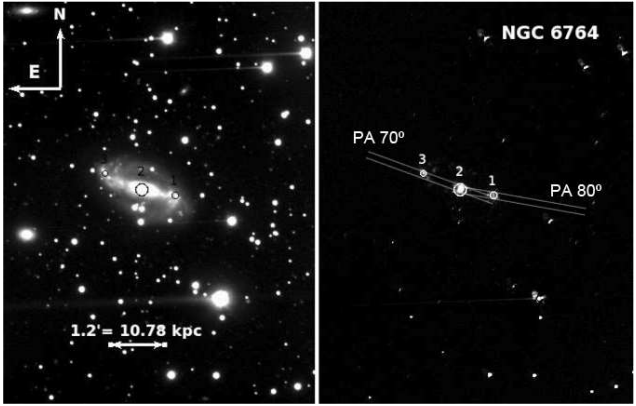} 
\caption{$R$-band (left) and  continuum-subtracted H$\alpha$ (right) images of  NGC~6764. Both panels have the same dimensions. The slit positions have been indicated on top of the continuum-subtracted H$\alpha$ image.\label{NGC6764image}}
\end{center}
\end{figure}

\begin{figure}
\begin{center}
\includegraphics[scale=0.35]{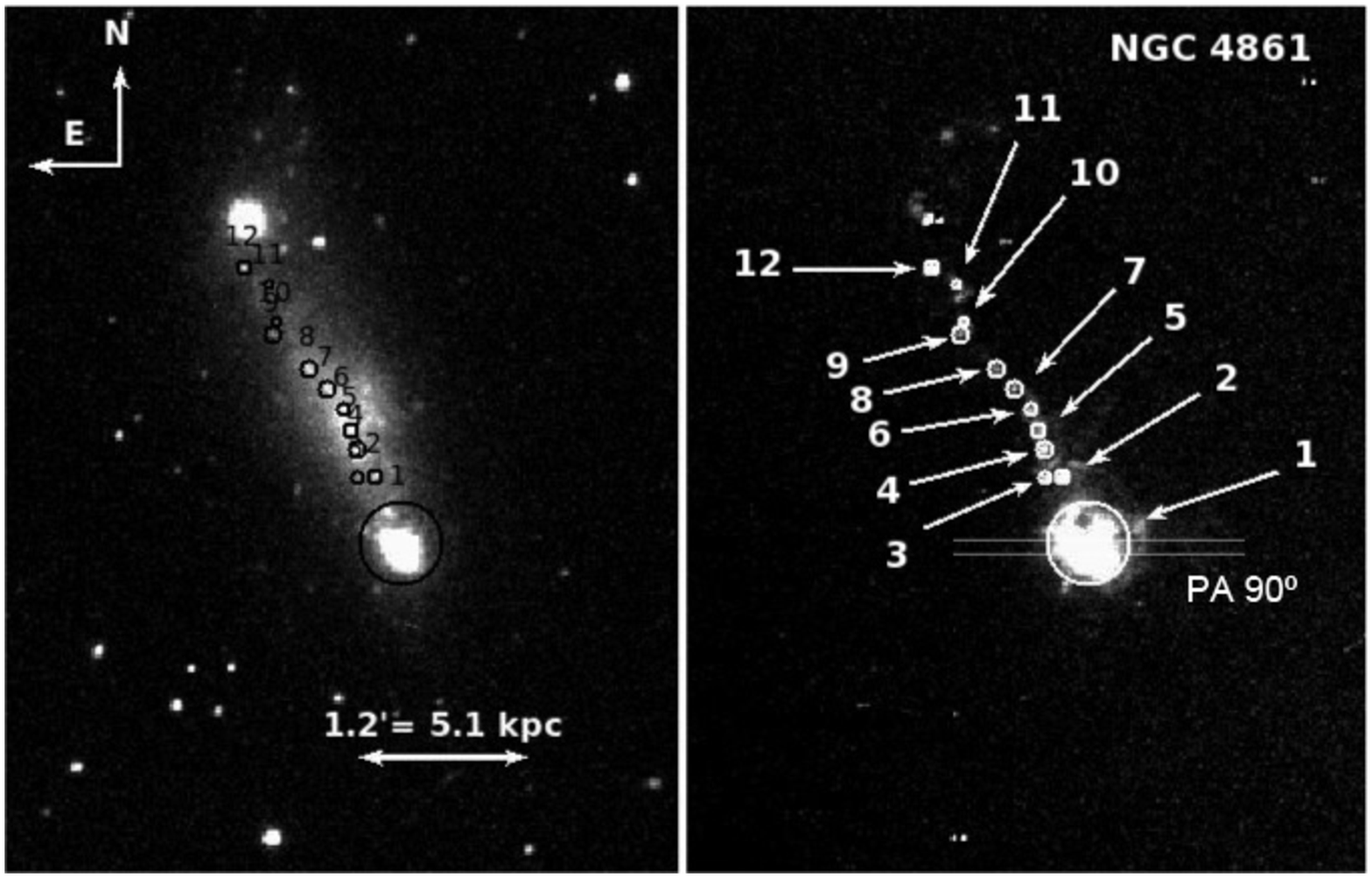} 
\caption{$R$-band (left) and  continuum-subtracted H$\alpha$ (right) images of NGC~4861. Both panels have the same dimensions. The slit position has been indicated on top of the continuum-subtracted H$\alpha$ image. \label{NGC4861image}}
\end{center}
\end{figure}

\begin{figure}
\begin{center}
\includegraphics[scale=0.35]{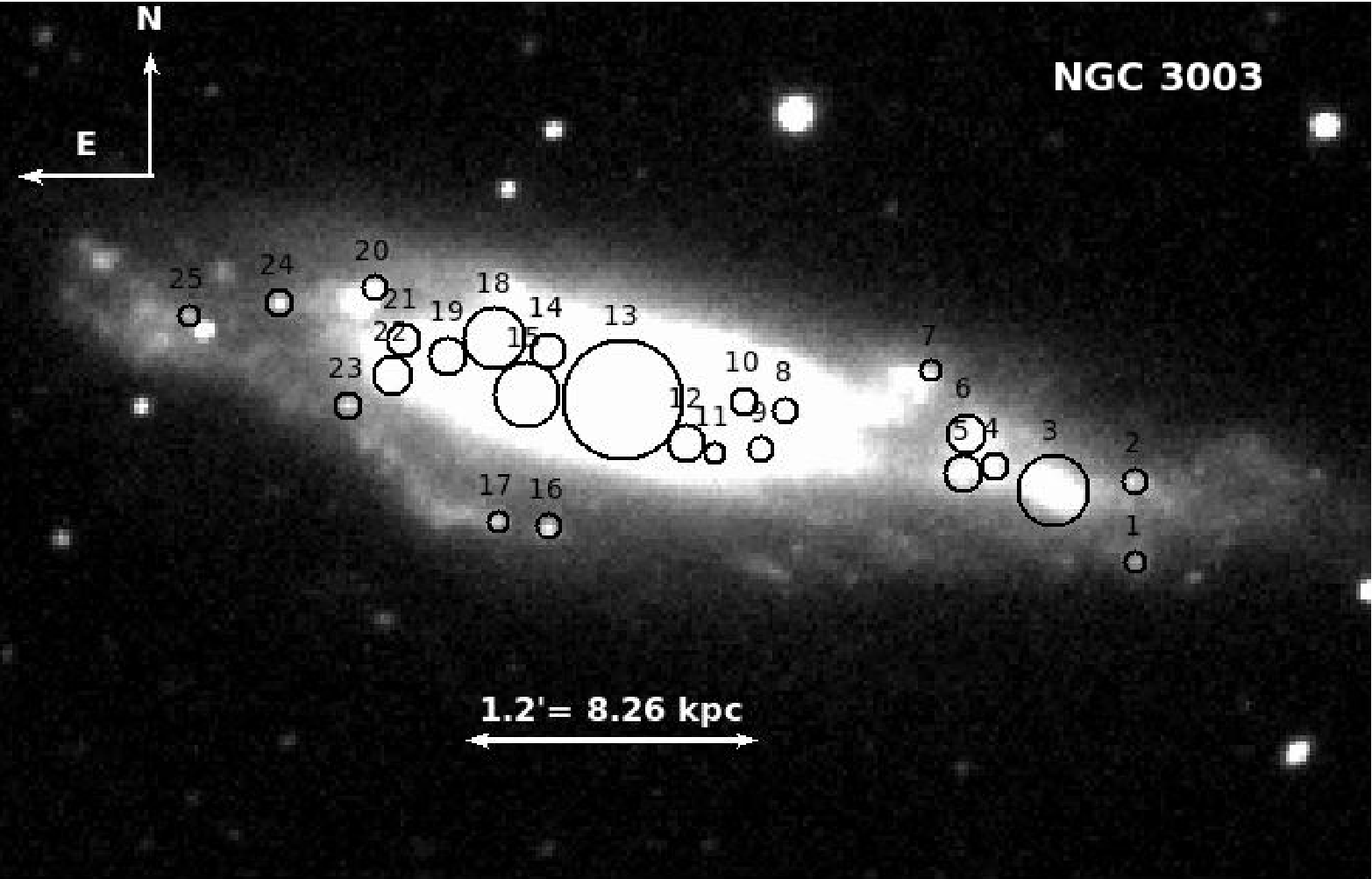} 
\includegraphics[scale=0.35]{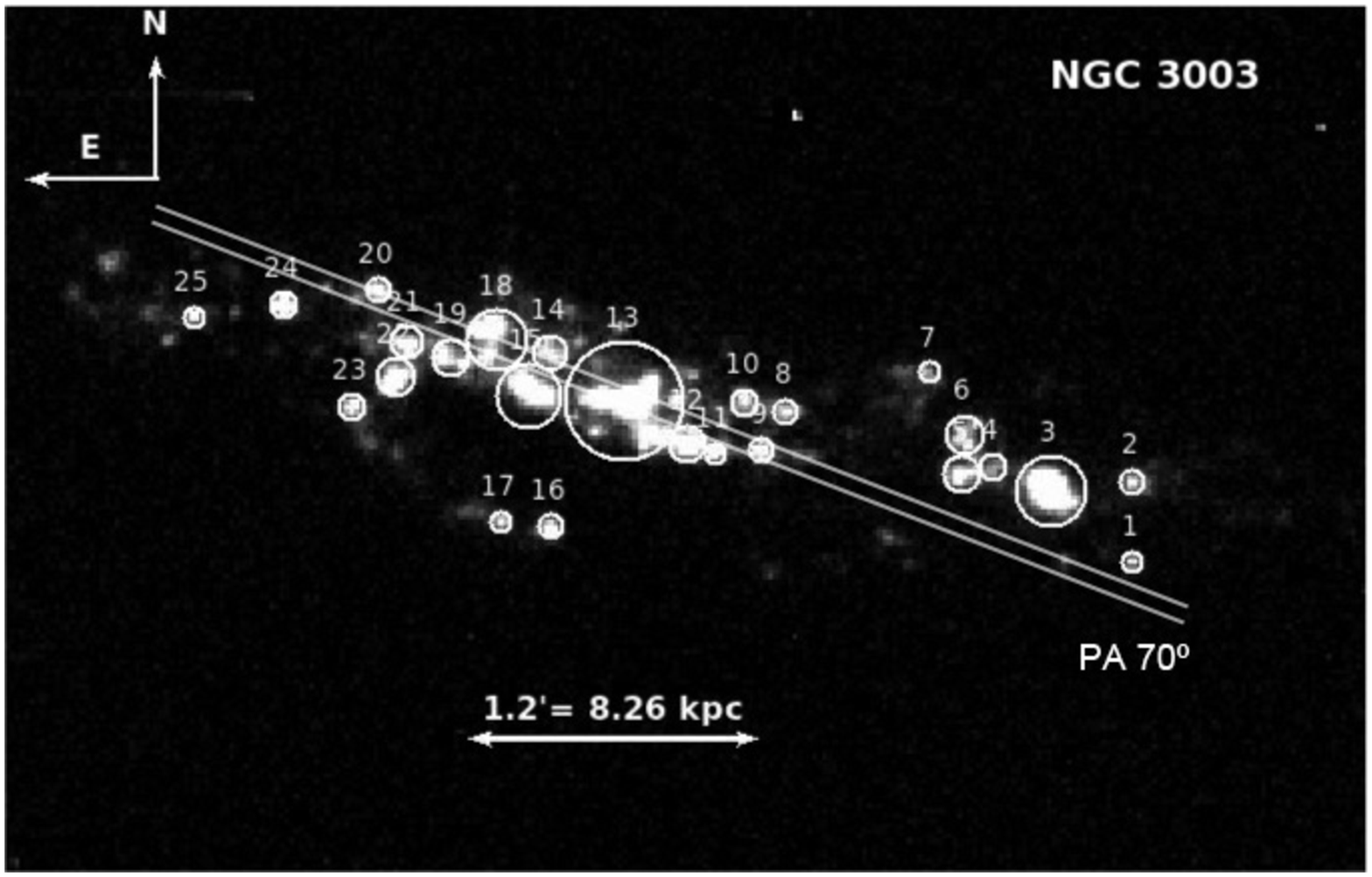} 
\caption{$R$-band (left) and  continuum-subtracted H$\alpha$ (right) images of NGC~3003. Both panels have the same dimensions. The slit position has been indicated on top of the continuum-subtracted H$\alpha$ image. The spectra we analyse here corresponds to knot~\#13. \label{iNGC3003}}
\end{center}
\end{figure}

Spectroscopic  observation of  the galaxies  were obtained  with   a  combination  of the slit  167l  
 (slit  width  1''.92$\times$11') and Grism 7, covering a wavelength 
 range  3500\,\AA -- 7500\,\AA. 
This  slit and grism combination gives a  dispersion of
1.5\,\AA/pixel  and a  spectral resolution  of $\sim$11\,\AA.  
As we used the HFOSC instrument at the 2m HCT for the majority of our spectroscopic observations,  
the spatial scale is 0.296''/pixel. In the case of our observations of NGC~1140, the spatial scale is 0.37''/pixel.
The star-forming  
regions   were  identified  using   short  exposures  through the
H$\alpha$ filter and it then they were centered through the slit. 
The position of the   slit  used in each case  is   marked   in   Figures~\ref{NGC1140image} to \ref{iNGC3003}.
The typical seeing was between 1.0 and 1.5~arcsec.
Because of the low air-mass at which the galaxies were usually observed ($m_X$ between 1.0 and 1.4), we should not expect problems coming from differential refraction. However, we did use the parallactic angle in the case of our observations of UM~311 because this galaxy was observed at an airmass of $m_X$=2.2. 
Spectrophotometric  standard star were observed for flux calibration of
the spectrum. In this case we used the same  Grism~7 but the
broad slit 1340 (15".41$\times$11'). A log of our observations is given in Table~\ref{obs}.


All  preprocessing and data reduction was  done in the
standard manner using various  tasks available with {\it Image Reduction and
Analysis  Facility}  (IRAF\footnote{IRAF is distributed by NOAO which is operated by AURA Inc.,
under cooperative agreement with NSF.}).  The {\it Munich Image Data Analysis System} (MIDAS) 
software was  used   for
identification  and removal  of  cosmic ray  events  from the  science
frames.  All images  were bias
subtracted and  flat-fielded using  master bias and  normalized master
flats. Dark frames were not needed in any case.
Multiple frames  of  the  galaxies taken  using
different filters were aligned and co-added in a final frame.

\begin{table*}
\scriptsize
  \caption{Results of the analysis of the broad-band photometry in the detected star-forming knots within our WR galaxy sample. These values have been corrected for Galactic extinction but not by intrinsic extinction. Last two columns compile the diameter and the size of each region.\label{phot}  }
 \begin{tabular}{ccccccc  cc}
 \hline

Object&Knot&      $m_V$     & $U-B$              & $B-V$             & $V-R$              &  $V-I$            &   Diameter    &    Size    \\
              & ID &    [ mag ]        &     [ mag ]              &     [ mag ]             &     [ mag ]              &        [ mag ]              &  [ " ]   &   [ kpc$^2$ ]  \\
      \hline

NGC 1140  	& \#1   &   12.10$\pm$0.03 &   -0.64$\pm$0.03 &    0.01$\pm$0.02 &    0.15$\pm$0.03 &	  0.28$\pm$0.03 &    28.8 &    4.77 \\ 
          	& \#2   &   16.99$\pm$0.14 &   -0.55$\pm$0.09 &    0.16$\pm$0.10 &    0.27$\pm$0.11 &	  0.57$\pm$0.15 &     9.6 &    0.53 \\ 
          	& \#3   &   18.67$\pm$0.06 &   -1.07$\pm$0.06 &   -0.15$\pm$0.05 &    0.22$\pm$0.05 &	  0.04$\pm$0.04 &     7.2 &    0.30 \\ 
          	& \#4   &   16.54$\pm$0.06 &   -1.12$\pm$0.06 &   -0.16$\pm$0.04 &    0.15$\pm$0.05 &	  0.11$\pm$0.06 &    19.2 &    2.12 \\ 
IRAS 07164+5301 & ...   &   14.06$\pm$0.03 &   -0.47$\pm$0.05 &    0.06$\pm$0.02 &    0.16$\pm$0.02 &	  0.30$\pm$0.02 &    27.6 &  429    \\ 
NGC 3738 	& ...   &   13.01$\pm$0.07 &   -0.54$\pm$0.04 &    0.20$\pm$0.05 &    0.19$\pm$0.06 &	  0.56$\pm$0.06 &    32.4 &    0.57 \\ 
  UM 311 	& ...   &   16.93$\pm$0.03 &   -1.12$\pm$0.02 &    0.07$\pm$0.02 &    1.67$\pm$0.03 &	  0.16$\pm$0.03 &    14.4 &    1.29 \\ 
NGC 6764 	& \#1   &   12.66$\pm$0.04 &   -0.92$\pm$0.08 &   -0.15$\pm$0.03 &   -0.10$\pm$0.03 &	  0.60$\pm$0.03 &     8.4 &    1.24 \\ 
         	& \#2   &   10.68$\pm$0.03 &   -0.81$\pm$0.07 &   -0.19$\pm$0.03 &    0.01$\pm$0.03 &	  0.30$\pm$0.02 &    15.6 &    4.28 \\ 
         	& \#3   &   13.74$\pm$0.05 &   -0.95$\pm$0.09 &   -0.18$\pm$0.04 &   -0.06$\pm$0.04 &	  0.63$\pm$0.03 &     7.2 &    0.91 \\ 

 NGC 4861 	& \#1   &   13.83$\pm$0.05 &   -0.81$\pm$0.08 &    0.33$\pm$0.02 &   -0.06$\pm$0.04 &	  0.11$\pm$0.04 &    35.1 &    4.85 \\ 
         	& \#2   &   19.92$\pm$0.09 &   -0.96$\pm$0.10 &    0.64$\pm$0.07 &   -0.09$\pm$0.08 &	  ...           &     7.0 &    0.194 \\ 
         	& \#3   &   21.67$\pm$0.29 &   -0.72$\pm$0.13 &   -0.52$\pm$0.20 &    0.13$\pm$0.22 &	  0.31$\pm$0.35 &     5.3 &    0.109 \\ 
         	& \#4   &   19.27$\pm$0.08 &   -0.92$\pm$0.08 &   -0.17$\pm$0.06 &   -0.01$\pm$0.07 &	  ...           &     7.9 &    0.246 \\ 
         	& \#5   &   20.49$\pm$0.14 &   -0.88$\pm$0.12 &   -0.08$\pm$0.12 &    0.25$\pm$0.10 &	  0.08$\pm$0.25 &     7.0 &    0.194 \\ 
         	& \#6   &   19.92$\pm$0.07 &   -0.63$\pm$0.07 &   -0.32$\pm$0.05 &    0.28$\pm$0.06 &	  0.47$\pm$0.21 &     5.3 &    0.109 \\ 
         	& \#7   &   19.74$\pm$0.11 &   -1.10$\pm$0.10 &   -0.03$\pm$0.08 &    0.16$\pm$0.09 &	  ...           &     7.9 &    0.246 \\ 
         	& \#8   &   19.28$\pm$0.07 &   -0.99$\pm$0.07 &   -0.11$\pm$0.05 &    0.35$\pm$0.06 &	  0.33$\pm$0.11 &     7.9 &    0.246 \\ 
         	& \#9   &   20.48$\pm$0.14 &   -0.38$\pm$0.13 &   -0.03$\pm$0.09 &    0.29$\pm$0.12 &	  0.21$\pm$0.23 &     7.9 &    0.246 \\ 
         	& \#10  &   21.20$\pm$0.13 &   -0.84$\pm$0.20 &    1.02$\pm$0.11 &    0.35$\pm$0.11 &	  ...           &     4.4 &    0.076 \\ 
         	& \#11  &   21.12$\pm$0.20 &   -0.71$\pm$0.14 &   -0.12$\pm$0.14 &    0.43$\pm$0.16 &	  ... 		&     4.4 &    0.076 \\ 
         	& \#12  &   20.18$\pm$0.07 &   -1.06$\pm$0.08 &    0.20$\pm$0.05 &    0.21$\pm$0.06 &	  0.41$\pm$0.15 &     7.0 &    0.194 \\ 

 NGC 3003 	& \#1   &   20.82$\pm$0.08 &   -1.16$\pm$0.08 &    0.23$\pm$0.06 &    0.49$\pm$0.07 &	  ...           &     5.0 &    0.258 \\ 
         	& \#2   &   20.44$\pm$0.13 &   -0.93$\pm$0.10 &    0.07$\pm$0.10 &    0.33$\pm$0.10 &	  0.74$\pm$0.18 &     5.5 &    0.313 \\ 
         	& \#3   &   16.94$\pm$0.02 &   -0.71$\pm$0.06 &    0.30$\pm$0.02 &    0.42$\pm$0.02 &	  ...           &    18.0 &    3.35 \\ 
         	& \#4   &   20.52$\pm$0.23 &   -0.85$\pm$0.12 &   -0.11$\pm$0.13 &    0.29$\pm$0.20 &	  0.80$\pm$0.14 &     5.5 &    0.313 \\ 
         	& \#5   &   20.88$\pm$0.27 &   -1.46$\pm$0.16 &    0.10$\pm$0.21 &    1.20$\pm$0.21 &	  1.45$\pm$0.21 &     9.0 &    0.837 \\ 
         	& \#6   &   19.18$\pm$0.11 &   -0.57$\pm$0.08 &   -0.06$\pm$0.06 &    0.16$\pm$0.09 &	  0.25$\pm$0.09 &     9.0 &    0.837 \\ 
         	& \#7   &   20.46$\pm$0.22 &   -0.19$\pm$0.20 &    0.23$\pm$0.15 &    0.31$\pm$0.19 &	  0.63$\pm$0.15 &     5.0 &    0.258 \\ 
         	& \#8   &   20.85$\pm$0.24 &   -1.17$\pm$0.10 &   -0.16$\pm$0.14 &    0.18$\pm$0.21 &	  0.52$\pm$0.24 &     5.5 &    0.313 \\ 
         	& \#9   &   19.89$\pm$0.24 &   -0.34$\pm$0.18 &    0.58$\pm$0.18 &    0.64$\pm$0.11 &	  0.88$\pm$0.17 &     5.5 &    0.313 \\ 
         	& \#10  &   22.27$\pm$0.71 &   -1.20$\pm$0.22 &   -0.58$\pm$0.50 &    ...           &	  ...           &     6.0 &    0.372 \\ 
         	& \#11  &   22.20$\pm$0.49 &   -0.98$\pm$0.25 &   -0.98$\pm$0.96 &    ...           &	  ...           &     5.0 &    0.258 \\ 
         	& \#12  &   18.39$\pm$0.22 &   -0.88$\pm$0.10 &    0.21$\pm$0.12 &    0.42$\pm$0.20 &	  0.90$\pm$0.18 &     9.0 &    0.837 \\ 
         	& \#13  &   14.45$\pm$0.02 &   -0.21$\pm$0.06 &    0.65$\pm$0.02 &    0.59$\pm$0.01 &	  1.11$\pm$0.01 &    29.0 &    8.69 \\ 
         	& \#14  &   19.14$\pm$0.13 &   -0.56$\pm$0.08 &   -0.33$\pm$0.09 &    0.56$\pm$0.10 &	  0.49$\pm$0.14 &     8.0 &    0.662 \\ 
         	& \#15  &   17.34$\pm$0.11 &   -0.53$\pm$0.09 &    0.29$\pm$0.08 &    0.48$\pm$0.09 &	  1.04$\pm$0.09 &    15.0 &    2.33 \\ 
         	& \#16  &   19.82$\pm$0.06 &   -0.96$\pm$0.08 &    0.18$\pm$0.03 &    0.34$\pm$0.05 &	  0.95$\pm$0.05 &     5.5 &    0.313 \\ 
         	& \#17  &   20.53$\pm$0.11 &   -0.35$\pm$0.10 &    0.08$\pm$0.07 &   -0.04$\pm$0.10 &	  0.45$\pm$0.09 &     5.0 &    0.258 \\ 
         	& \#18  &   16.71$\pm$0.04 &   -0.41$\pm$0.08 &    0.20$\pm$0.03 &    0.41$\pm$0.04 &	  0.85$\pm$0.04 &    15.0 &    2.33 \\ 
         	& \#19  &   19.56$\pm$0.19 &   -0.67$\pm$0.10 &   -0.45$\pm$0.14 &    0.65$\pm$0.14 &	  0.78$\pm$0.18 &     8.0 &    0.662 \\ 
         	& \#20  &   19.38$\pm$0.05 &   -0.75$\pm$0.06 &    0.12$\pm$0.03 &    0.35$\pm$0.05 &	  0.71$\pm$0.04 &     5.5 &    0.313 \\ 
         	& \#21  &   18.57$\pm$0.13 &   -0.26$\pm$0.12 &    0.20$\pm$0.10 &    0.35$\pm$0.10 &	  0.91$\pm$0.09 &     7.0 &    0.507 \\ 
         	& \#22  &   18.10$\pm$0.05 &   -0.63$\pm$0.06 &    0.28$\pm$0.03 &    0.44$\pm$0.04 &	  1.18$\pm$0.05 &     9.0 &    0.837 \\ 
         	& \#23  &   20.49$\pm$0.12 &   -1.11$\pm$0.08 &   -0.00$\pm$0.07 &    0.41$\pm$0.10 &	  ...           &     6.0 &    0.372 \\ 
         	& \#24  &   20.00$\pm$0.06 &   -1.08$\pm$0.09 &    0.11$\pm$0.04 &    0.44$\pm$0.05 &	  0.81$\pm$0.07 &     6.0 &    0.372 \\ 
         	& \#25  &   21.05$\pm$0.15 &   -1.23$\pm$0.15 &    0.48$\pm$0.13 &    0.28$\pm$0.13 &	  0.61$\pm$0.15 &     5.0 &    0.258 \\
          
         \hline

\end{tabular}
\end{table*}

Narrow-band H$\alpha$  line images were obtained  following the standard
procedure  described by \citet{Waller90}. $R$-band  image is  used for  continuum
subtraction. Spectrophotometry standard stars chosen from \citet{Oke90} were 
observed to calibrate the H$\alpha$ images.
A scale  factor between the H$\alpha$ and  $R$-band images
was  determined using  the  non-saturated field  stars  in the  galaxy
field,  after taking care  of the  difference in  the full-width-half-maximum
(FWHM) of  the stellar
profile.  The  $R$-band  continuum image was  scaled to  the H$\alpha$
image and  subtracted from H$\alpha$  image to get the  H$\alpha$ line
image  of  the galaxy. 
We then defined apertures to cover all the flux coming from different knots throughout each galaxy.
A nearby emission-free region was always considered to estimate
the surrounding background. The H$\alpha$-fluxes of the knots
were then flux calibrated using the results provided by the standard stars.

The spectroscopic data analysis  was also performed using standard routines
of the IRAF software. 
For each two-dimensional spectra we extracted an one-dimensional
spectrum integrating a particular region along the spatial direction, usually centering on the brightest point of the galaxy,
We carefully checked that the one-dimensional spectrum obtained for each object had the optimal signal-to-noise (SNR), i.e., that the emission lines are not diluted by the stellar continuum and that the SNR is high.
We list the size of the aperture used for each galaxy in Table~\ref{spectra}.
Each 1-D  spectrum was then wavelength calibrated using a FeAr arc spectrum. 
The absolute flux calibration of
the spectra was achieved by  observing  spectrophotometry standard stars chosen from \citet{Oke90}.
We also used IRAF to correct the spectra for foreground extinction using the Galactic $E(B-V)$ values listed in Table~\ref{Table1} and extracted from \citet{SFD98}.

\section{Photometry results}  

The  $R$-band  and the  continuum-subtracted
H$\alpha$  images  of all galaxies  are  shown in Fig.\,\ref{NGC1140image}\,to\,\ref{iNGC3003}.
We used the continuum-subtracted H$\alpha$  images to identify
the  star-forming regions  (knots)  within each   galaxy. The  knots  were
identified by  visual inspection. However, a  threshold of  4-5 times sigma of
background was taken to delimit the size of each knot.
We always consider circular apertures. The diameter of the regions (in ") and their corresponding physical size (in kpc$^{2}$) have been compiled in Table~\ref{phot}. 
The corresponding identity number for each star-forming region 
is included in Fig.\,\ref{NGC1140image}\,to\,\ref{iNGC3003}.

\begin{table*}
\scriptsize
  \caption{H$\alpha$ flux, H$\alpha$ luminosity (corrected for extinction and [\ion{N}{ii}] emission), derived star-formation rate (SFR) and SFR per area,  $-W$(H$\alpha$),  and age of the most recent star-formation event for the knots of the WR galaxies analyzed in this work. {Last column lists the [\ion{N}{ii}]~$\lambda\lambda$6548,6583/\Ha\ ratio derived for some regions using our optical spectroscopic data and used to correct the narrow-band  \Ha\ fluxes for [\ion{N}{ii}] emission.}\label{halpha}}
  \begin{tabular}{lccccc  ccc}
  \hline
Galaxy &  Metallicity          & Flux                             & Luminosity             & log SFR   &  log (SFR/Area)   &                       $-W$(H$\alpha$)& Age   & [\ion{N}{ii}]/\Ha \\
Knot ID  & $Z [Z_{\odot}]$          &[10$^{-14}$  erg\,s$^{-1}$\,cm$^{-2}$]   &[10$^{38}$  erg\,s$^{-1}$] &  [$M_{\odot}$\,yr$^{-1}$]&  [$M_{\odot}$\,yr$^{-1}$\,kpc$^{-2}$ ]&           [\AA] & [Myr] &    \\ 
 \hline
\textbf{NGC 1140}&0.008&&&&&&& 0.13 \\
\#  1 &&   204$\pm$10    &       777$\pm$38     &  -0.21$\pm$0.02 & -0.89$\pm$0.02 &	 384$\pm$80 &   5.0$\pm$0.2  \\
\#  2 &&  2.81$\pm$0.25  &     10.72$\pm$0.96   &  -2.07$\pm$0.04 & -1.79$\pm$0.04 &	 432$\pm$80 &   4.8$\pm$0.1  \\
\#  3 &&  1.18$\pm$0.11  &      4.52$\pm$0.41   &  -2.45$\pm$0.04 & -1.92$\pm$0.04 &	 888$\pm$240 &   4.2$\pm$0.6   \\
\#  4 &&  9.00$\pm$0.27  &        33$\pm$2      &  -1.58$\pm$0.03 & -1.91$\pm$0.03 &	 976$\pm$240 &   3.5$\pm$0.6    \\

 \noalign{\smallskip}   
\textbf{IRAS 07164+5301$^a$ }&0.008& 109   & 9650    & 0.88 & -1.27 &     82$\pm$10 &   6.5$\pm$0.2 & \nodata \\  

 \noalign{\smallskip}   
\textbf{NGC 3738} & 0.008& 119$\pm$5     &     44$\pm$2    &     -1.46$\pm$0.02 & -1.21$\pm$0.02 &    124$\pm$24 &   6.2$\pm$0.2 &0.18\\

 \noalign{\smallskip}   
\textbf{UM 311} &0.008 &  20$\pm$1     &      82$\pm$5 &   -1.19$\pm$0.03 & -1.30$\pm$0.03 &   1300$\pm$80 &   3.2$\pm$0.1 & 0.09\\ 

 \noalign{\smallskip}   
\textbf{NGC 6764}&0.02&&&&&& \\ 
\#  1 &&   12.0$\pm$0.4  &     130$\pm$9  & -0.99$\pm$0.03 & -1.08$\pm$0.03 &    150$\pm$40 &   6.0$\pm$0.1 & 0.32  \\
\#  2$^b$ &&  152$\pm$8    &    1776$\pm$90  &  0.15$\pm$0.02 & -0.48$\pm$0.02 &    450$\pm$40 &   4.8$\pm$0.3& 1.40$^b$\\
\#  3 &&   10.0$\pm$0.4  &     107$\pm$9  & -1.07$\pm$0.04 & -1.03$\pm$0.04 &    300$\pm$40 &   5.6$\pm$0.1& 0.32  \\
 \noalign{\smallskip}   

\textbf{NGC 4861}&0.004&&&&&&&0.03\\ 
\#1  &&    223$\pm$11	  &	 583$\pm$29    &   -0.33$\pm$0.02 & -1.02$\pm$0.02 &	845$\pm$150 &	4.6$\pm$0.2 \\
\#2  &&   1.12$\pm$0.10   &	2.93$\pm$0.26  &   -2.63$\pm$0.04 & -1.92$\pm$0.04 &   1417$\pm$470 &	3.0$\pm$0.8 \\
\#3  &&   1.15$\pm$0.10   &	3.01$\pm$0.27  &   -2.62$\pm$0.04 & -1.66$\pm$0.04 &   2878$\pm$950 &	1.5: \\
\#4  &&   0.37$\pm$0.03   &	0.97$\pm$0.09  &   -3.11$\pm$0.04 & -2.50$\pm$0.04 &	202$\pm$60 &	6.1$\pm$0.5 \\
\#5  &&   0.40$\pm$0.04   &	1.04$\pm$0.09  &   -3.08$\pm$0.04 & -2.37$\pm$0.04 &	218$\pm$70 &	6.0$\pm$0.6 \\
\#6  &&   0.35$\pm$0.03   &	0.90$\pm$0.08  &   -3.15$\pm$0.04 & -2.18$\pm$0.04 &	267$\pm$90 &	5.6$\pm$0.6 \\
\#7  &&   0.81$\pm$0.07   &	2.11$\pm$0.19  &   -2.78$\pm$0.04 & -2.17$\pm$0.04 &	585$\pm$190 &	4.9$\pm$0.2 \\
\#8  &&   0.64$\pm$0.06   &	1.68$\pm$0.15  &   -2.87$\pm$0.04 & -2.27$\pm$0.04 &	480$\pm$160 &	5.0$\pm$0.2 \\
\#9  &&   0.44$\pm$0.04   &	1.15$\pm$0.10  &   -3.04$\pm$0.04 & -2.43$\pm$0.04 &	556$\pm$180 &	5.0$\pm$0.2 \\
\#10 &&   0.24$\pm$0.02   &	0.62$\pm$0.06  &   -3.31$\pm$0.04 & -2.19$\pm$0.04 &	666$\pm$220 &	4.8$\pm$0.2 \\
\#11 &&   0.31$\pm$0.03   &	0.82$\pm$0.07  &   -3.19$\pm$0.04 & -2.07$\pm$0.04 &	615$\pm$210 &	4.9$\pm$0.2 \\
\#12 &&   1.10$\pm$0.10   &	2.87$\pm$0.26  &   -2.64$\pm$0.04 & -1.93$\pm$0.04 &   1285$\pm$430 &	3.7$\pm$0.9 \\

 \noalign{\smallskip}   
\textbf{NGC 3003}&0.02&&&&&&&0.36 \\   

\#  1 &&   0.23$\pm$0.02   &	 1.58$\pm$0.14 &  -2.90$\pm$0.04 & -2.31$\pm$0.04 &    319$\pm$90 &   5.6$\pm$0.2	\\
\#  2 &&   0.31$\pm$0.03   &	 2.12$\pm$0.19 &  -2.77$\pm$0.04 & -2.27$\pm$0.04 &    411$\pm$120 &   5.3$\pm$0.9	\\
\#  3 &&   11.0$\pm$0.2    &	   72$\pm$5    &  -1.24$\pm$0.03 & -1.77$\pm$0.03 &    119$\pm$30 &   6.1$\pm$0.1	\\
\#  4 &&   0.17$\pm$0.02   &	 1.17$\pm$0.11 &  -3.03$\pm$0.04 & -2.53$\pm$0.04 &    215$\pm$70 &   5.8$\pm$0.2	\\
\#  5 &&   0.90$\pm$0.08   &	 6.2$\pm$0.6   &  -2.31$\pm$0.04 & -2.23$\pm$0.04 &    984$\pm$230 &   3.3$\pm$0.1	\\
\#  6 &&   0.77$\pm$0.07   &	 5.3$\pm$0.5   &  -2.38$\pm$0.04 & -2.30$\pm$0.04 &    364$\pm$110 &   5.5$\pm$0.5	\\
\#  7 &&   0.16$\pm$0.01   &	 1.13$\pm$0.10 &  -3.05$\pm$0.04 & -2.46$\pm$0.04 &    304$\pm$90 &   5.6$\pm$0.2	\\
\#  8 &&   0.33$\pm$0.03   &	 2.3$\pm$0.2   &  -2.75$\pm$0.04 & -2.24$\pm$0.04 &    874$\pm$220 &   3.4$\pm$0.2	\\
\#  9 &&   0.47$\pm$0.04   &	 3.2$\pm$0.3   &  -2.59$\pm$0.04 & -2.09$\pm$0.04 &    568$\pm$150 &   3.8$\pm$0.9	\\
\# 10 &&   0.38$\pm$0.03   &	 2.7$\pm$0.2   &  -2.68$\pm$0.04 & -2.25$\pm$0.04 &   1274$\pm$280 &   3.1$\pm$0.2	\\
\# 11 &&   0.36$\pm$0.03   &	 2.5$\pm$0.2   &  -2.71$\pm$0.04 & -2.12$\pm$0.04 &    480$\pm$130 &   4.7$\pm$0.9	\\
\# 12 &&   1.30$\pm$0.12   &	 8.9$\pm$0.8   &  -2.15$\pm$0.04 & -2.07$\pm$0.04 &    391$\pm$110 &   5.3$\pm$0.5	\\
\# 13 &&   14.0$\pm$1.0    &	  101$\pm$7    &  -1.10$\pm$0.03 & -2.03$\pm$0.03 &	87$\pm$20 &   6.3$\pm$0.1	\\
\# 14 &&   0.45$\pm$0.04   &	 3.1$\pm$0.3   &  -2.61$\pm$0.04 & -2.43$\pm$0.04 &    107$\pm$40 &   6.2$\pm$0.2     \\
\# 15 &&    2.9$\pm$0.3    &	   20$\pm$1    &  -1.80$\pm$0.02 & -2.17$\pm$0.02 &    152$\pm$50 &   6.0$\pm$0.2	\\
\# 16 &&   0.54$\pm$0.05   &	 3.7$\pm$0.3   &  -2.53$\pm$0.04 & -2.03$\pm$0.04 &    443$\pm$120 &   4.8$\pm$0.9	\\
\# 17 &&   0.24$\pm$0.02   &	 1.67$\pm$0.15 &  -2.88$\pm$0.04 & -2.29$\pm$0.04 &    513$\pm$140 &   4.0$\pm$0.9	\\
\# 18 &&    3.2$\pm$0.3    &	   22$\pm$2    &  -1.76$\pm$0.04 & -2.13$\pm$0.04 &	90$\pm$30 &   6.2$\pm$0.2      \\
\# 19 &&   0.66$\pm$0.06   &	 4.5$\pm$0.4   &  -2.44$\pm$0.04 & -2.26$\pm$0.04 &    205$\pm$70 &   5.8$\pm$0.2  \\
\# 20 &&   0.45$\pm$0.04   &	 3.1$\pm$0.3   &  -2.61$\pm$0.04 & -2.11$\pm$0.04 &    205$\pm$70 &   5.8$\pm$0.2  \\
\# 21 &&   0.73$\pm$0.07   &	 5.0$\pm$0.5   &  -2.40$\pm$0.04 & -2.11$\pm$0.04 &    146$\pm$50 &   6.0$\pm$0.2  \\
\# 22 &&   1.83$\pm$0.16   &	   13$\pm$1    &  -1.99$\pm$0.03 & -1.91$\pm$0.03 &    319$\pm$90 &   5.6$\pm$0.2   \\
\# 23 &&   0.57$\pm$0.05   &	 3.9$\pm$0.4   &  -2.50$\pm$0.04 & -2.08$\pm$0.04 &    869$\pm$210 &   3.4$\pm$0.1   \\
\# 24 &&   0.86$\pm$0.08   &	 5.9$\pm$0.5   &  -2.33$\pm$0.04 & -1.90$\pm$0.04 &    799$\pm$200 &   3.4$\pm$0.2   \\
\# 25 &&   0.25$\pm$0.02   &	 1.71$\pm$0.15 &  -2.87$\pm$0.04 & -2.28$\pm$0.04 &    686$\pm$180 &   3.5$\pm$0.6   \\

\hline
\end{tabular}

\begin{flushleft}

$^a$The H$\alpha$ data for IRAS 07164+5301 have been derived using the spectroscopic data. See text. \\
$^b$The center of NGC~6764 (knot~\#2) is a LINER (see Sect.~4.1), and hence the emission coming from the [\ion{N}{ii}] lines is larger (1.4 times) than the emission coming from \Ha. See text for details.

\end{flushleft}

\end{table*}

\subsection{Broad-band photometry\label{photo}}
  
Aperture photometry of the star-forming regions 
was carried out  within the circular apertures drawn in Figs.\,\ref{NGC1140image}\,to\,\ref{iNGC3003}.  
We then computed the apparent $V$-magnitude, $m_V$, 
and the colors $U-B$,  $B-V$, $V-R$, and $V-I$,  of all regions. 

Broad-band colours were corrected 
for foreground extinction --i.e., Galactic extinction-- using the $E(B-V)$ data provided by 
\citet*{SFD98} and listed in Table~\ref{Table1}. 
All derived magnitudes and colors are compiled in Table~\ref{phot}. 

Table~\ref{phot} does not include a correction for internal extinction. However, for many knots we
are able to determine this correction by the comparison of the broad-band colors 
with the predictions given by the Starburst99 \citep{L99} evolutionary synthesis models, 
as we will explain in Sect.~\ref{models}. 
In any case, for the majority of the knots the correction 
for internal extinction seems to be small, even almost zero. 

No correction for emission lines from the ionized gas has been considered, although it may be important in some 
bright, intense, star-forming knots \citep[e.g.,][]{Salzer89a,LSE08,Reines10}, as we will explain in Sect.~5.2.

Uncertainties quoted in Table~\ref{phot} consider only the photometric error.

\subsection{Narrow-band H$\alpha$ photometry}

We used the continuum-subtracted H$\alpha$  image (see Figs.\,\ref{NGC1140image}\,to\,\ref{iNGC3003})  
to estimate the H$\alpha$ flux 
of every star-forming knot within each galaxy. 
H$\alpha$ fluxes have been  corrected for extinction assuming:
\begin{eqnarray}
A_{\rm H\alpha} = 1.758/0.692 E(B-V) = 
2.54 E(B-V)
\end{eqnarray}
following \citet{LSE08}, who considered a Milky Way reddening law with $R_V=3.1$ following \citet{Cardelli89}.
Here $E(B-V)$ considers both the foreground and the intrinsic extinction, as listed in Table~\ref{Table1} (Galactic extinction) and
Table~\ref{ages} (internal extinction), i.e., 
\begin{eqnarray}
E(B-V) = E(B-V)_{Galactic} + E(B-V)_{internal}. 
\end{eqnarray}
The measured  H$\alpha$ flux  is contaminated by [\ion{N}{ii}] $\lambda\lambda$6548,6583 emission.
The estimation of the [\ion{N}{ii}]  contribution was derived using the  [\ion{N}{ii}]~$\lambda\lambda$6548,6583/\Ha\ ratio derived from our optical spectroscopic data (which is listed in last column of Table~\ref{halpha}) and applying
   \begin{eqnarray}
    F_{\rm H\alpha,cor} =   F_{\rm H\alpha} \times \Big(1+\frac{\rm [N\,\textsc{ii}]~\lambda\lambda6548,6583}{\rm H\alpha}\Big)^{-1},
  \end{eqnarray}
  where  $F_{\rm H\alpha,cor}$ and $F_{\rm H\alpha}$ are, respectively, the corrected and uncorrected H$\alpha$ flux.
  %
  We did not consider the transmittance of the narrow-band filter in the position of the [\ion{N}{ii}] lines --as it is explained in \citet{LSE08}-- because of the relatively large FWHM ($\gtrsim100$~\AA) of the narrow-band filter used.
The corresponding  [\ion{N}{ii}]/\Ha\ ratio  was typically found between  0.03 (NGC~4861) and 0.36 (NGC~3003). 
Except for the case of NGC~6764, we considered that all knots within a galaxy have the same [\ion{N}{ii}] contribution we estimated for the region we have spectroscopic data.
Knot~2 within NGC~6764 is the center of the galaxy, which is a low-ionization nuclear emission-line region (LINER, see Sect.~4.1), and hence  the emission coming from the [\ion{N}{ii}] lines is 1.4 times larger than the emission coming from \Ha.

The final H$\alpha$ fluxes corrected for both extinction and [\ion{N}{ii}] contribution derived for each star-forming region are compiled in Table~\ref{halpha}.

Using the H$\alpha$ fluxes 
 and considering the distance
to the galaxies listed in Table~\ref{Table1}, we derived the total H$\alpha$  luminosity 
for the analyzed star-forming regions within our WR galaxy sample. The results are also listed in Table~\ref{halpha}.

The H$\alpha$ luminosity can be used as a good tracer of star formation activity in 
galaxies \citep[e.g.,][]{K98,Calzetti07}. 
The radiation from the young  stars ionizes the surrounding 
hydrogen gas, giving rise to H$\alpha$ emission by recombination.
Since the H$\alpha$ luminosity is proportional to  the number of ionizing photons 
produced by the hot stars (which is also proportional to their birth rate), 
the star-formation rate (SFR) can  be easily derived from the H$\alpha$ luminosity.
Only stars with masses  $>10$ $M_{\odot}$ and life time $<20$\,Myr 
contribute  significantly to the integrated ionizing  flux, so the
emission  lines provide  a nearly  instantaneous measure  of  the SFR,
independent of the previous star formation history.

We adopted the standard \citet{K98} relation to derive the SFR from the \Ha\ luminosity.

We  compile the H$\alpha$-based SFR values for star-forming knots observed in our WR galaxy sample  in Table~\ref{halpha}.

\begin{figure}
\centering
\includegraphics[scale=0.47,angle=90]{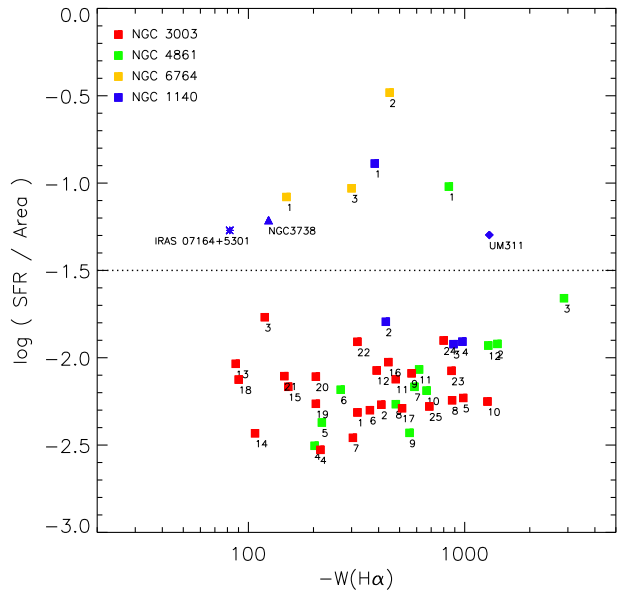}
  \caption{SFR per area vs.  $W$(H$\alpha$) for the knots analyzed in this work. The dotted line at $\log(SFR/A)=-1.5$ has been included for guiding the eye. Knots located over this line are experiencing a relatively high SFR per area and hence they might be considered as starburst regions or starburst galaxies. Note that knot~\#2 in NGC~6764 hosts a LINER, see text. \label{sfrawha}}
\end{figure}

Since this galaxy IRAS~07164+5301 has a larger redshift than the other objects (it has a radial velocity of 12,981~km\,s$^{-1}$, which means a redshift of $z\sim0.0433$),
 the continuum subtraction results with null. In this case, we estimated its H$\alpha$ luminosity through our spectroscopy
data. Considering the size and width of the slit used to get the spectrum of this galaxy  (1.92"$\times$5.2", see Table~\ref{spectra}) 
and the corresponding angular size of the galaxy (10"$\times$21"), we estimate that the total H$\alpha$ flux of the galaxy is $\sim21$ times the extinction-corrected H$\alpha$ flux we measured using our spectrum. Taking into account the values for Galactic (Table~\ref{Table1}) and internal (Table~\ref{spectra}) reddening, the H$\alpha$/H$\beta$ ratio (2.79) and the H$\beta$ flux (1.27$\times$10$^{-14}$~erg\,s$^{-1}$\,cm$^{-2}$), we estimate a total H$\alpha$ flux of $\sim$1.09$\times$10$^{-12}$~erg\,s$^{-1}$\,cm$^{-2}$ for IRAS~07164+5301, which is the value tabulated in Table~\ref{halpha}. Hence, its total SFR is SFR$_{\rm H\alpha}\sim$7.6~$M_{\odot}$\,yr$^{-1}$.

The H$\alpha$  equivalent width, $W$(H$\alpha$), of the star  forming knots was
calculated using  the expression  by \citet{Waller90}. As it will be explained in Sect.~\ref{ageHa}, 
we will use $W$(H$\alpha$) 
to estimate the age of the most recent star-formation event. Both quantities are also listed
in Table~\ref{halpha}  for each region.

We have also quantified the strength of the star-formation activity in each knot by computing their SFR per area. Table~\ref{halpha} also lists these values, in units of $M_{\odot}$\,yr$^{-1}$\,kpc$^{-2}$. Figure~\ref{sfrawha} compares the SFR per area with the H$\alpha$ equivalent width derived for each knot. The dotted line at $\log(SFR/A)=-1.5$ has been included for guiding the eye. Knots located over this line are experiencing a relatively high SFR per area and hence they might be considered as starburst regions (knot~\#1 in NGC~1140 and NGC~4861 and knots~\#1 and \#3 in NGC~6764) or starburst galaxies (IRAS~07164+5301, UM~311, NGC~3738). Note that knot~\#2 in NGC~6764 hosts a LINER, as we discuss in the next section. In the case of NGC~3003 all knots are located well below the dotted line in Figure~\ref{sfrawha}, suggesting that this is just a normal star-forming galaxy.

\begin{figure*}
\centering
 (a){\includegraphics[scale=0.38]{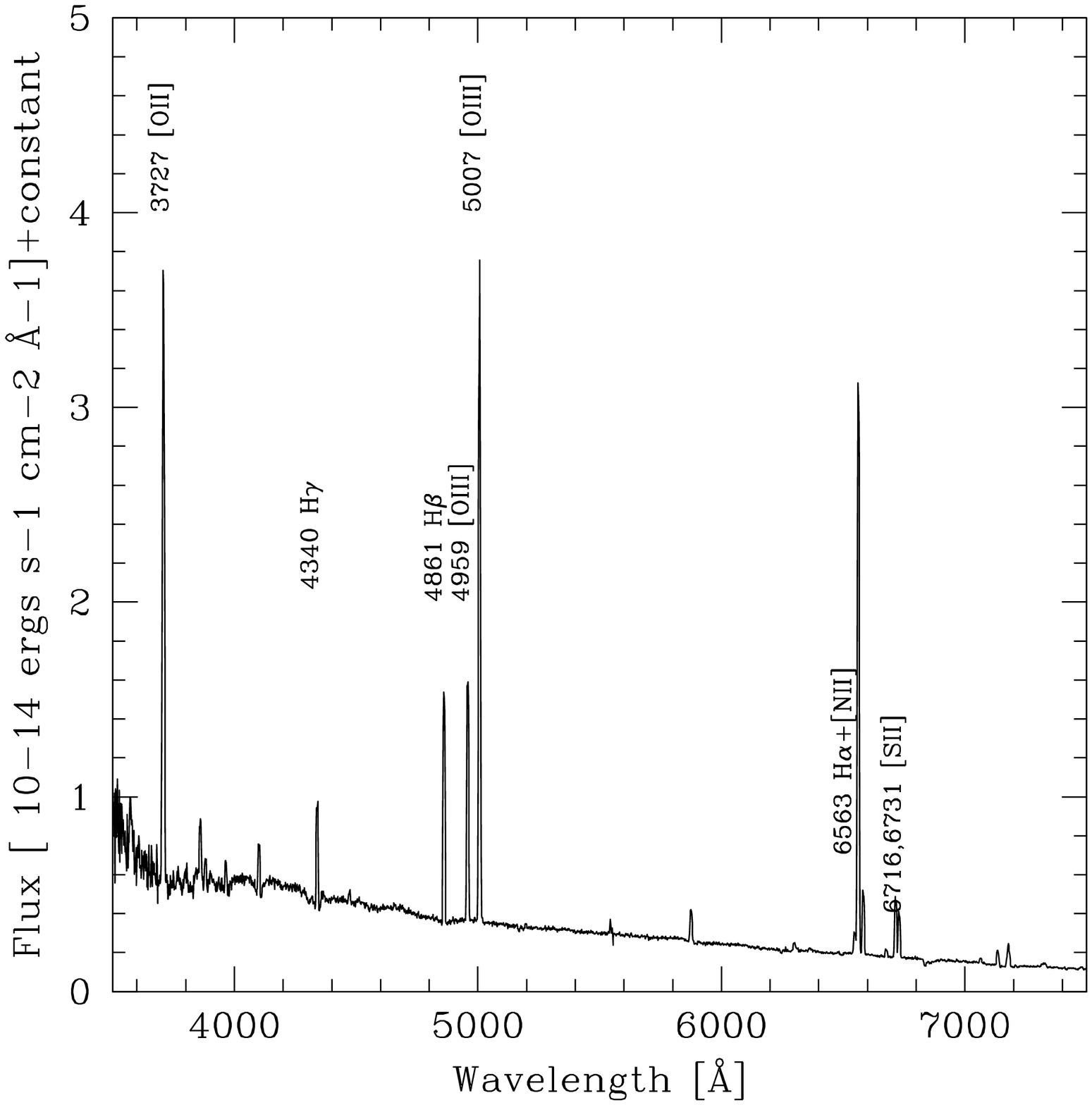}}
 (b){\includegraphics[scale=0.38]{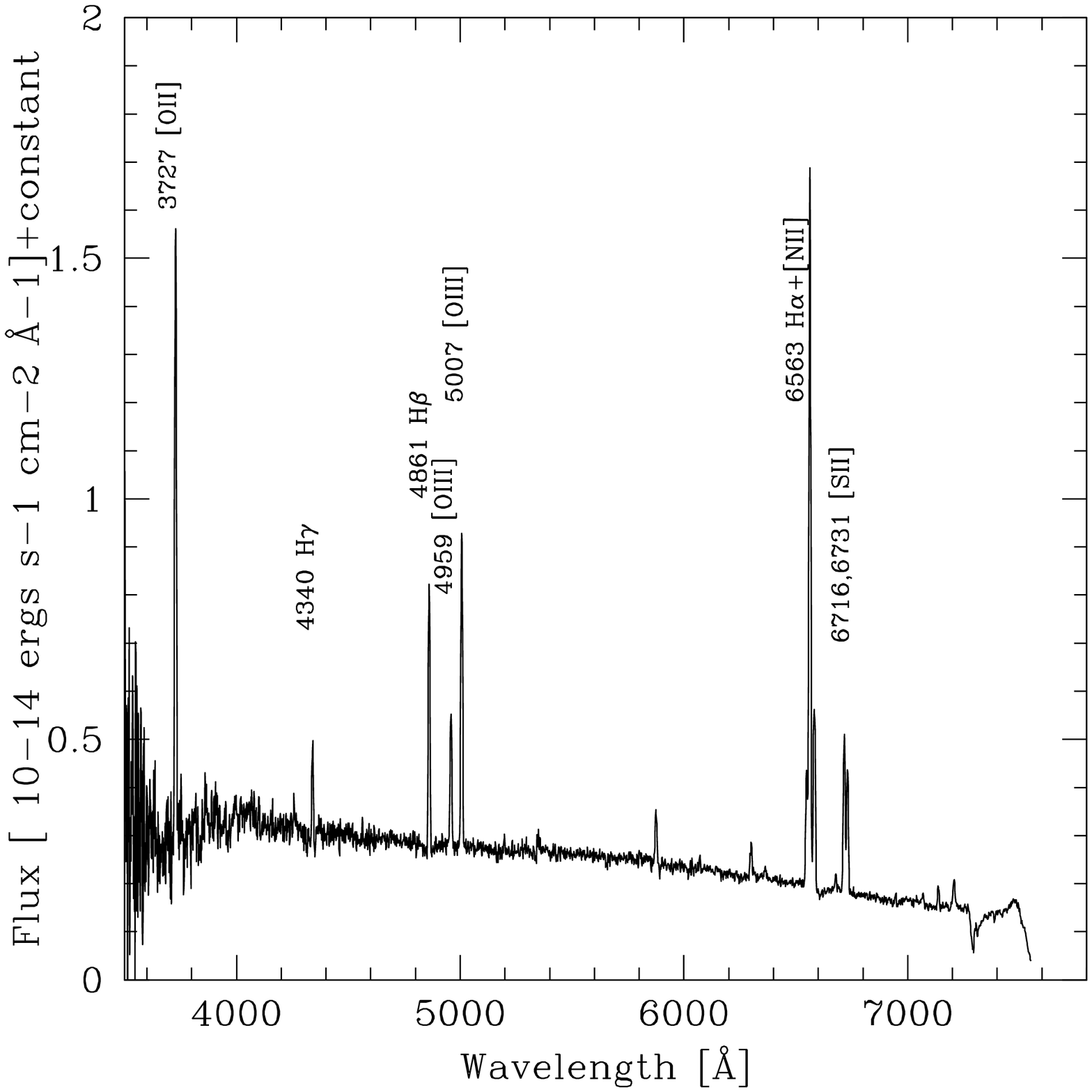}}
 (c){\includegraphics[scale=0.38]{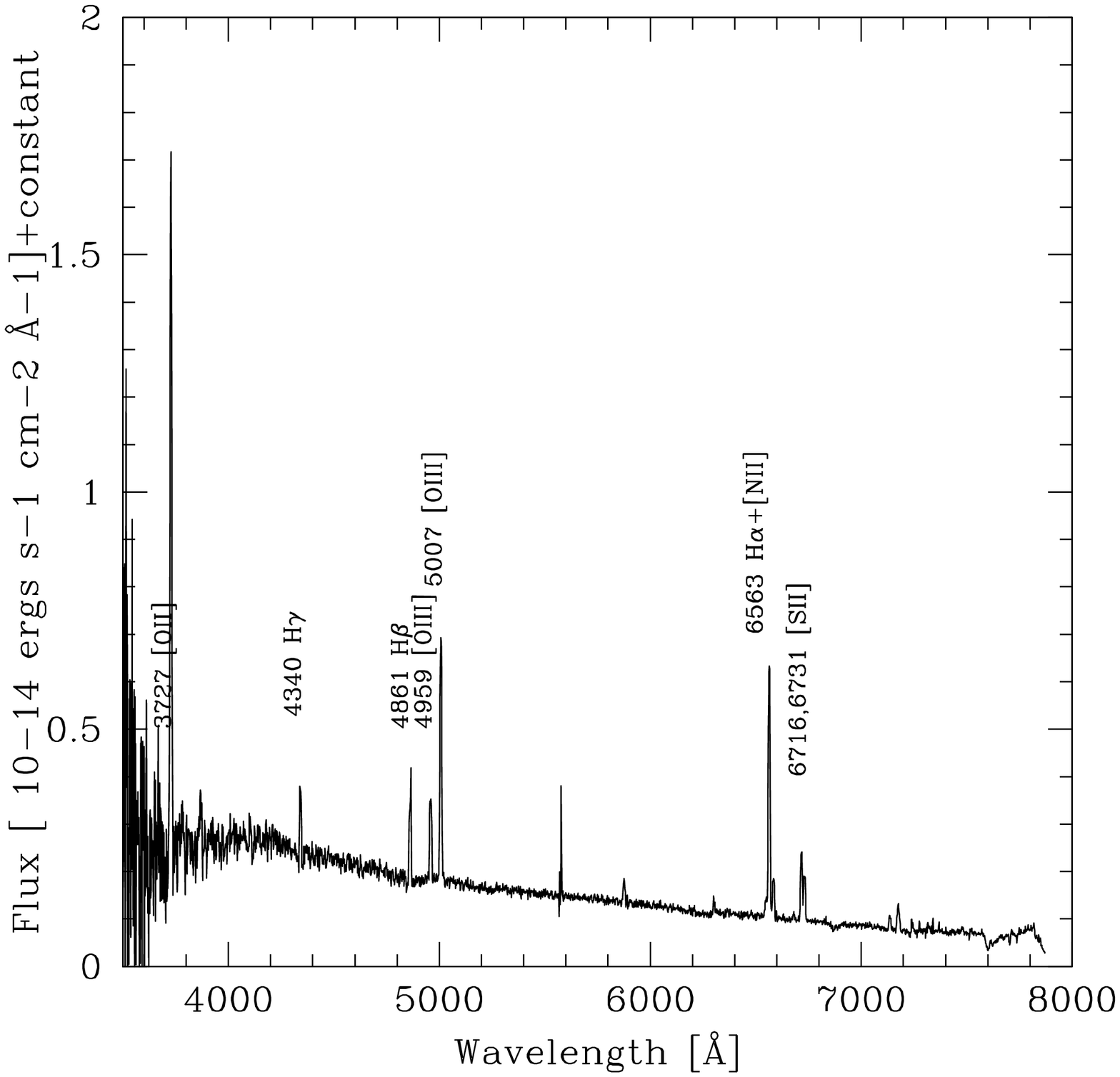}}
 (d){\includegraphics[scale=0.38]{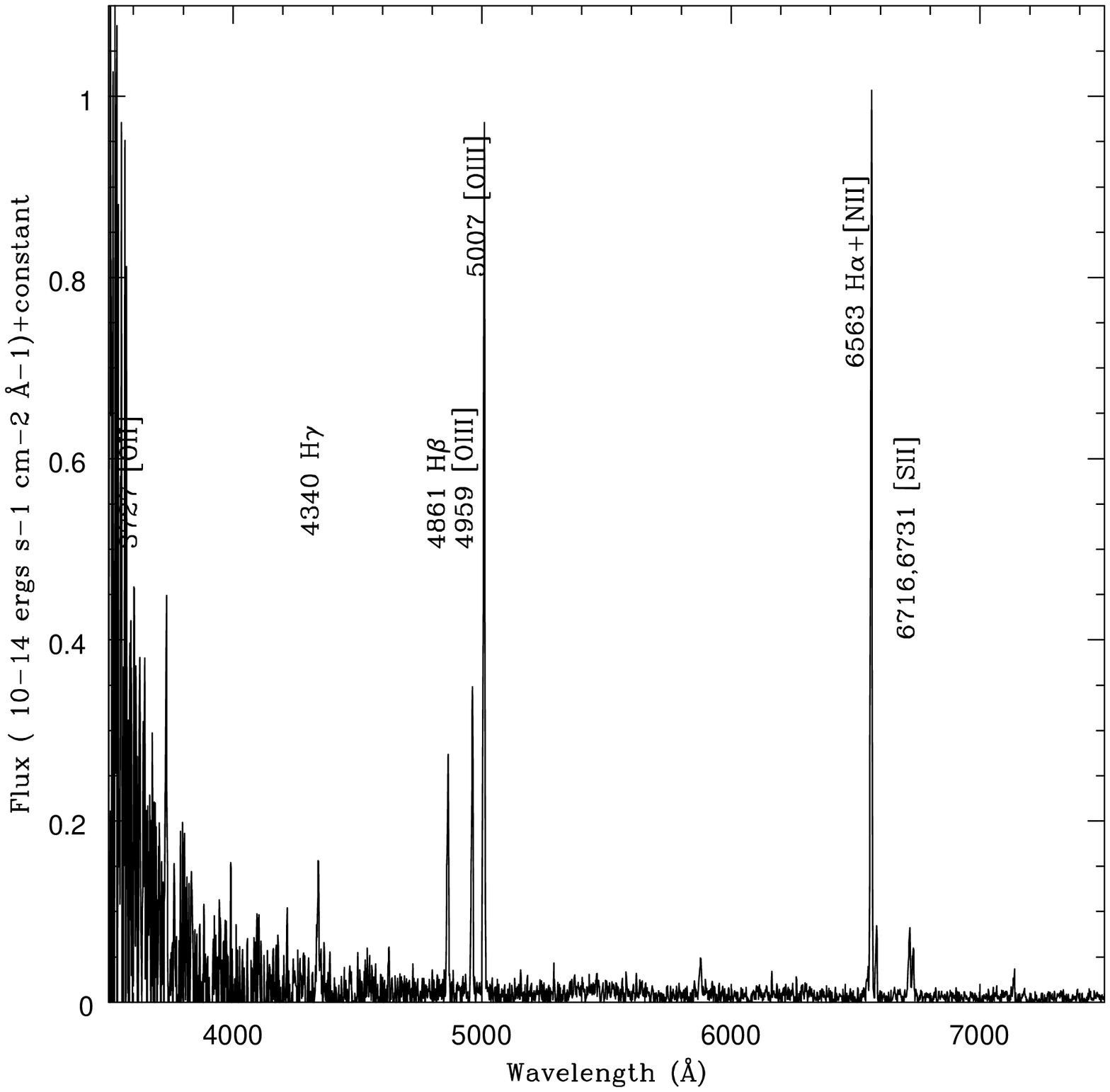}}
 (e){\includegraphics[scale=0.38]{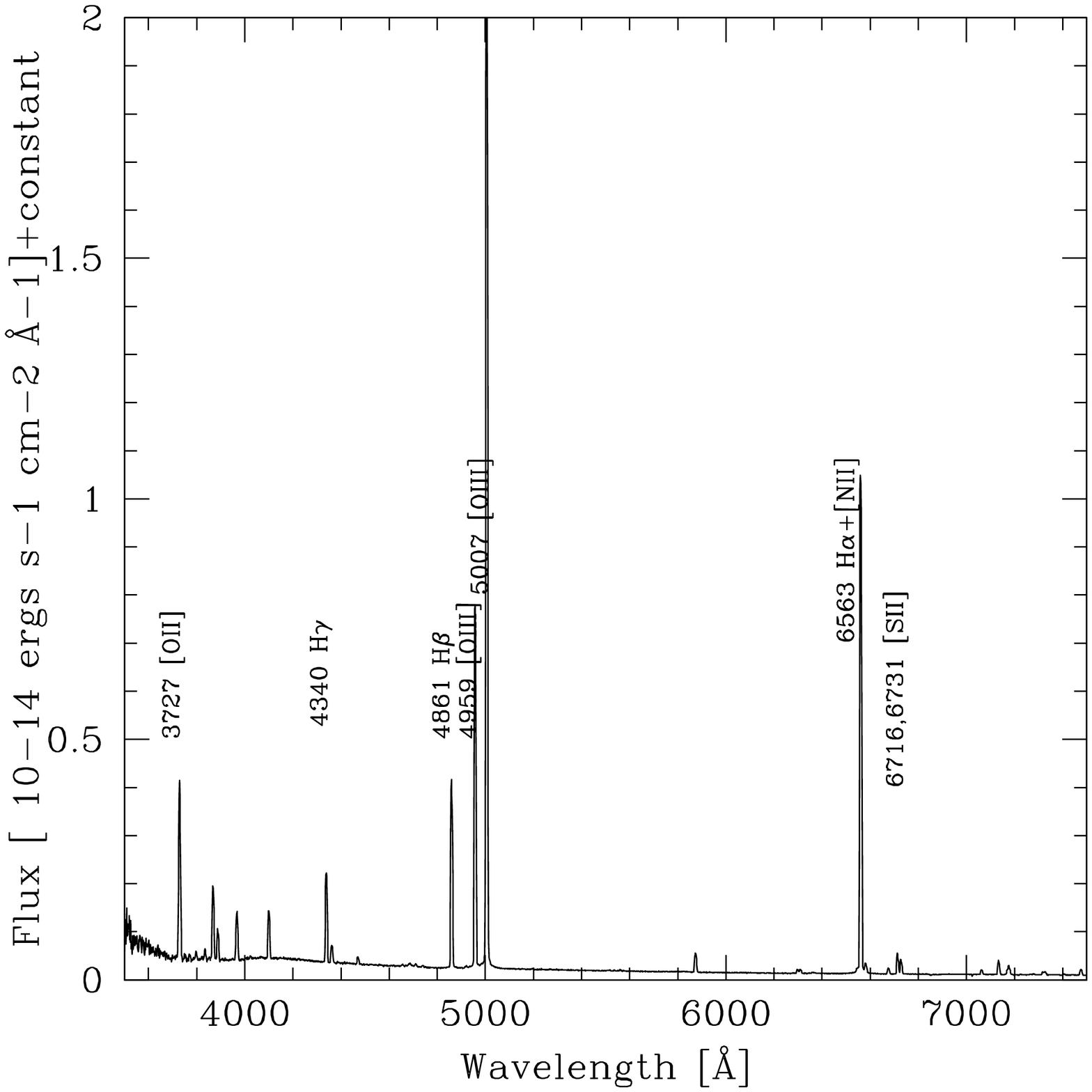}} 
 (f){\includegraphics[scale=0.38]{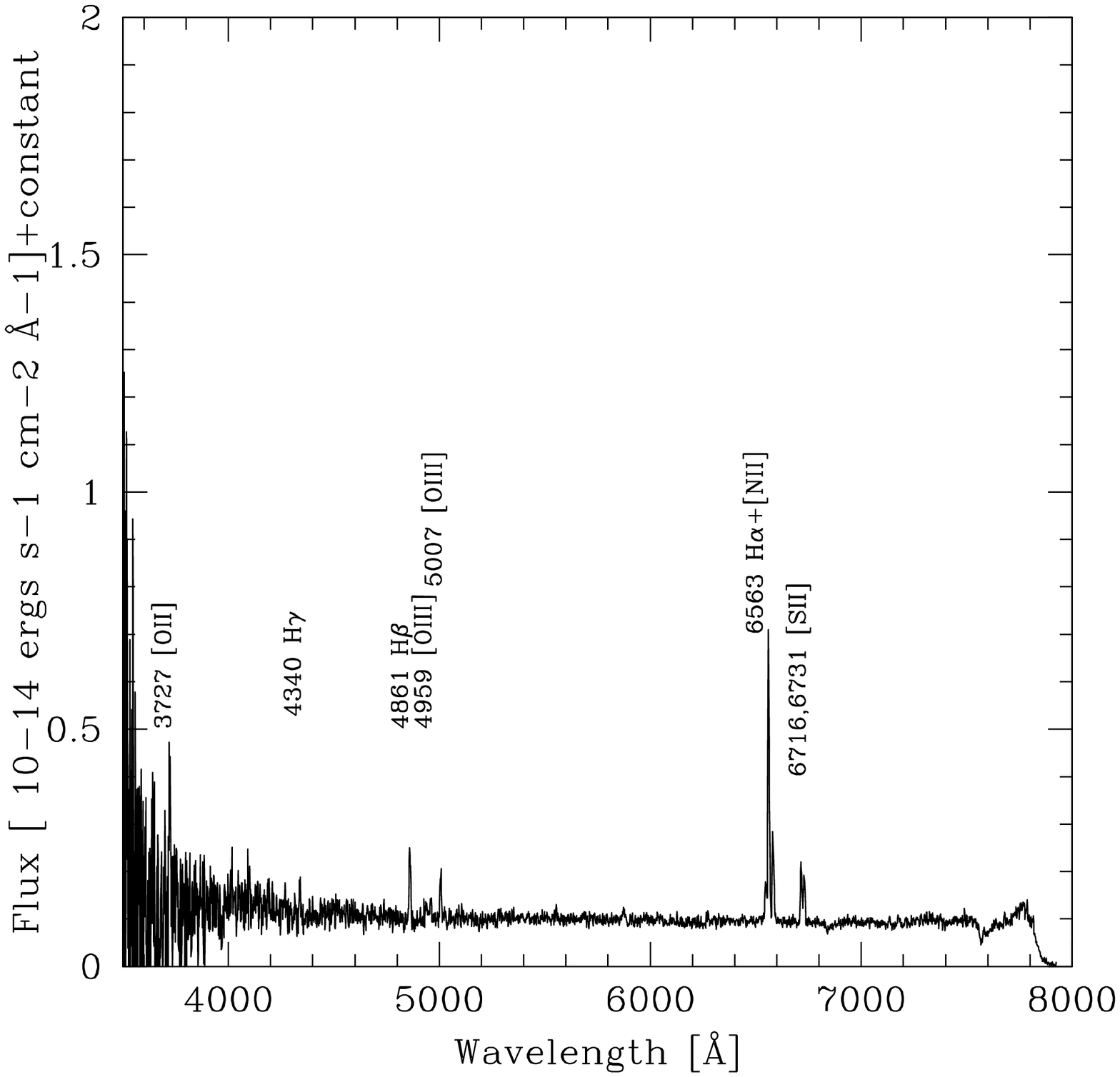}} 

  \caption{Optical spectra of the analyzed WR galaxies: (a) NGC~1140, (b) IRAS~07164+5301, (c) NGC~3738, (d) UM~311, (e) NGC~4861 and (f) NGC~3003.  \label{spectrafig1}. }
\end{figure*}

\begin{figure*}
\centering
(a)\includegraphics[scale=0.38]{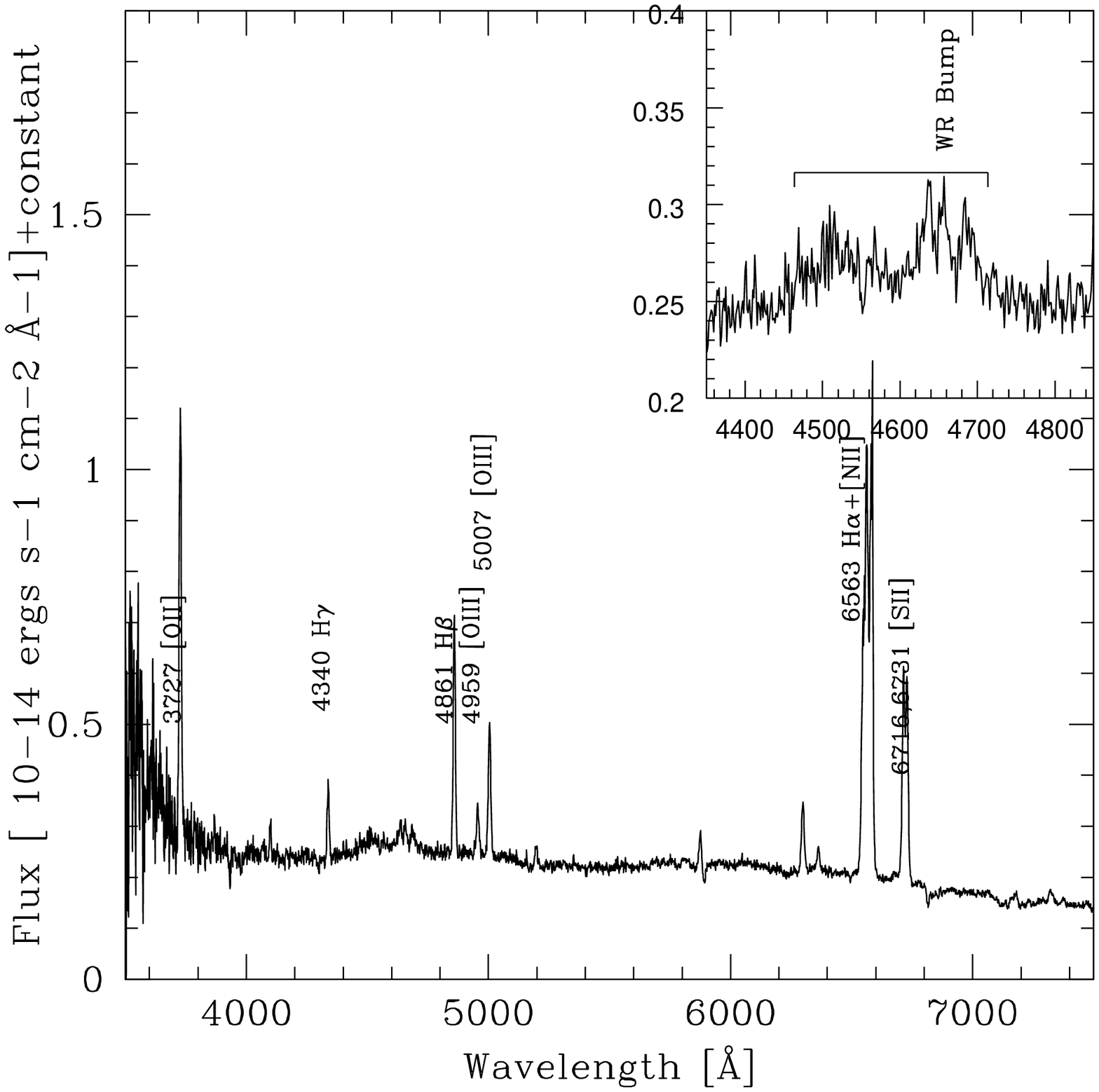}
(b)\includegraphics[scale=0.38]{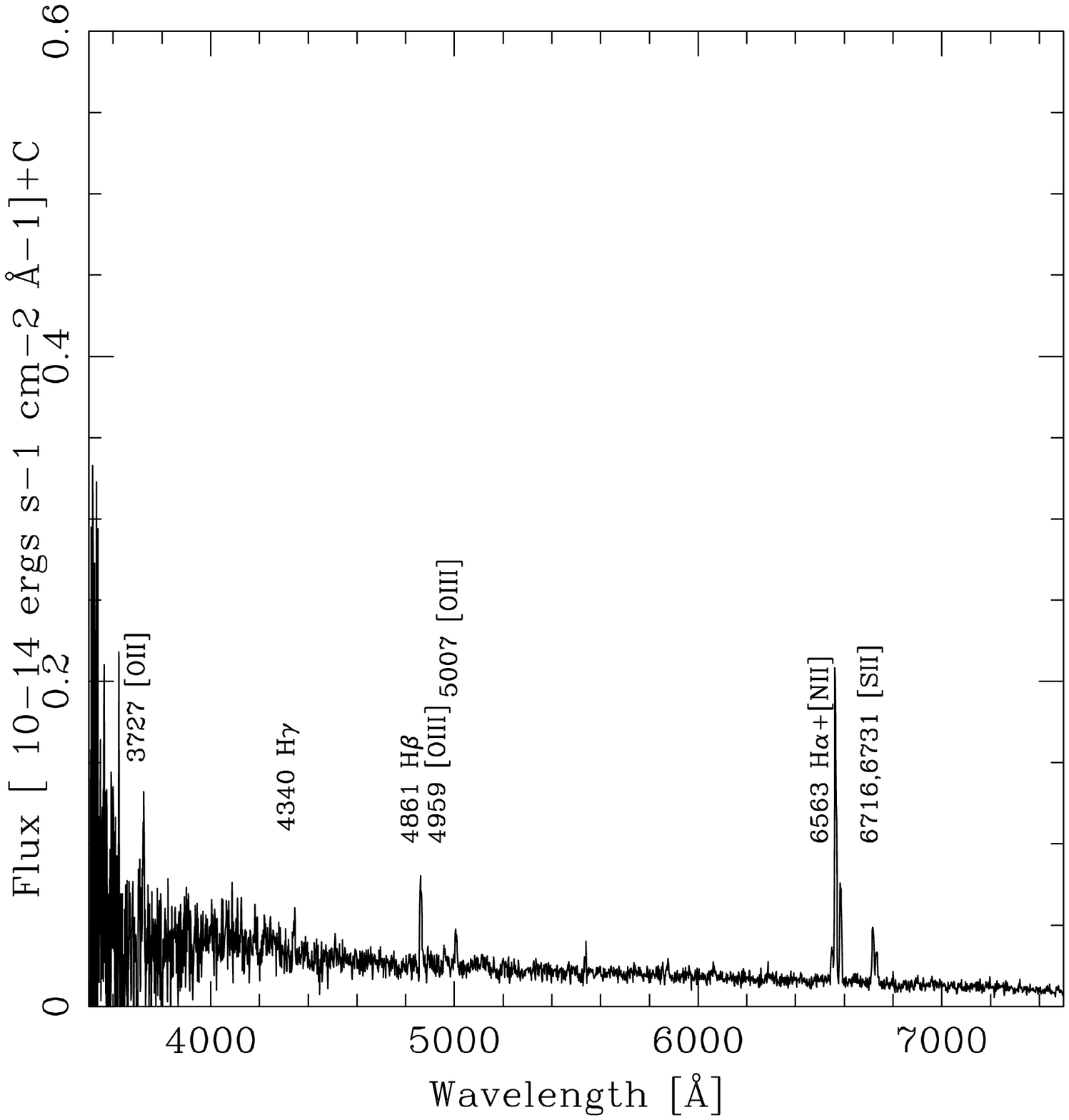} 
  \caption{Optical spectra of NGC~6764. (a) Spectrum of its center. (b) Spectrum of knot \#~1. \label{spectrafig2}}
\end{figure*}

\section{Results from optical spectroscopy}   

The one-dimensional spectra obtained for our WR galaxy sample are plotted in Figs.~\ref{spectrafig1} and \ref{spectrafig2}.
The prominent lines have been identified and marked in these plots.
These are emission lines from the hydrogen and helium Balmer series, and forbidden emission lines  due to
oxygen, nitrogen and sulphur. 
These spectra have been corrected for foreground (Galactic) extinction using the corresponding $E(B-V)$ values extracted from
\citet*{SFD98} 
(see Table~\ref{Table1}) and by making use of the IRAF {\it deredden} task. We then directly analyzed these dereddened spectra.

Line intensities and equivalent widths
were measured by integrating all the flux in the line between
two given limits and over a local continuum estimated by
eye. In the cases of line blending (usually, in the H$\alpha$+[\ion{N}{ii}] region), a multiple Gaussian profile 
plotting procedure was applied to obtain the line flux
of each individual line. We used the standard assumption,
i.e., that $I$(H$\beta$)=100, to compute the line intensity ratios.

\begin{table*}
\scriptsize
\caption{Dereddened line intensity ratios with respect to $I$(H$\beta$)=100 for the galaxies analyzed in this work. We also compile the H$\beta$ flux, the size of the extracted area, the  reddening coefficient, $c$(H$\beta$), and the equivalent widths of the absorption in the hydrogen lines, $W_{abs}$, used to correct the spectra for internal reddening, and the equivalent widths of the emission \ion{H}{i} Balmer lines. \label{spectra}}
\begin{tabular} {l r c c c c c c c}
\hline
 Line   & $f(\lambda)$     & NGC\,1140 \#1 & IRAS\,07164+5301  &NGC\,3738 & UM\,311    & NGC\,6764 \#1 & NGC\,4861 \#1 & NGC\,3003 \#13\\
\hline

 [\ion{O}{ii}]~3728   & 0.322    & 241$\pm$17  & 238$\pm$28   & 680$\pm$110    &    257$\pm$90   & 142$\pm$44 & 88.3$\pm$5.8  &  209$\pm$65   \\   
   
 [\ion{Ne}{iii}]~3869 & 0.291   &  21.9$\pm$3.9   &    \nodata    &  41:          &  \nodata           &  \nodata        & 37.0$\pm$2.5 & \nodata   \\
 
 [\ion{Ne}{iii}]~3969+H7&0.267&   10.2$\pm$1.1    &   \nodata   &  \nodata         &   \nodata          &    \nodata      & 26.5$\pm$2.1  & \nodata  \\
 H$\delta$~4101 & 0.230   &   26.2$\pm$2.1     &   \nodata       &   27:         &   \nodata          &  \nodata       &  26.0$\pm$1.7   & \nodata    \\
 H$\gamma$~4340 & 0.157  &   47.0$\pm$3.1   &  43.9$\pm$7.4   &     49$\pm$13    &  35$\pm$11    & 50$\pm$14     &   47.2$\pm$2.5   &  47$\pm$17  \\ 

 [\ion{O}{iii}]~4363  & 0.150   &   1.53$\pm$0.49 &   \nodata    & \nodata    &   \nodata          &   \nodata       & 9.79$\pm$0.77   & \nodata \\ 
 \ion{He}{i}~4471   & 0.116  &     3.63$\pm$0.64  &     \nodata    &  \nodata &   \nodata & \nodata         & 3.28$\pm$0.35 & \nodata  \\

 [\ion{Fe}{iii}]~4658& 0.059   &   \nodata       &      \nodata  & \nodata  &   \nodata    &   \nodata       & 0.93$\pm$0.26  & \nodata  \\
 Broad \ion{He}{ii}~4686 & 0.049    &     3.3$\pm$1.4         &      \nodata   & \nodata  &   \nodata  &    \nodata      & 2.62$\pm$0.78  &\nodata  \\
 
 [\ion{Ar}{iv}]~4711 &0.043   &   \nodata           &    \nodata    &  \nodata  &  \nodata  &  \nodata        & 1.09$\pm$0.23  &\nodata  \\
 
 [\ion{Ar}{iv}]~4740 &0.034    &  \nodata         &     \nodata  &  \nodata   &   \nodata   &  \nodata        & 0.56$\pm$0.20 & \nodata  \\ 
 
 H$\beta$~4861    & 0.000 &100.0$\pm$3.8 & 100.0$\pm$5.5     & 100$\pm$10 &100$\pm$11    &100$\pm$12            & 100.0$\pm$3.3  & 100$\pm$14 \\  
 
 [\ion{O}{iii}]~4959 &$-0.025$   & 107.4$\pm$5.7 & 50.2$\pm$5.3	   & 96$\pm$14 & 107$\pm$13    & 18.2$\pm$3.1 & 197.7$\pm$9.1    &   21.8$\pm$4.1 \\      
 
 [\ion{O}{iii}]~5007 &$-0.037$   & 274$\pm$13 & 120.2$\pm$9.8	   & 257$\pm$31 & 270$\pm$16    & 38.6$\pm$4.6 & 596$\pm$26   &  58.2$\pm$6.5 \\      
 
 Broad \ion{C}{iv}~5808  & $-0.191$   &  3.34$\pm$0.35 & \nodata & \nodata & \nodata & \nodata &  1.23$\pm$0.37 & \nodata \\
 
 \ion{He}{i}~5876   & $-0.203$   &  16.8$\pm$1.3 &     43.9$\pm$4.8    &  28.0$\pm$5.2    &  16.1$\pm$5.6   &   13.8:  & 10.28$\pm$0.59  & 12.8: \\
 
 [\ion{O}{i}]~6300  & $-0.262$     &  4.48$\pm$0.41 &   11.0$\pm$1.9     &   12.4$\pm$3.5   &  \nodata  &    \nodata      &    1.78$\pm$0.15  & \nodata \\ 
 
 [\ion{S}{iii}]~6312  & $-0.264$    &  \nodata &  \nodata  &   \nodata        &   \nodata  &   \nodata       &    2.13$\pm$0.15 & \nodata \\

[\ion{N}{ii}]~6548   & $-0.295$  & 10.5$\pm$1.1  & 43.9$\pm$4.8	   & 15.5$\pm$4.9 & 5.5:   & 33.8$\pm$3.9  & 2.55$\pm$0.32 & 52$\pm$11 \\      

H$\alpha$ 6563    & $-0.297$ & 282$\pm$14 & 279$\pm$19   & 281$\pm$31 &  281$\pm$25  & 284$\pm$23 & 282$\pm$13 &  281$\pm$29 \\      

[\ion{N}{ii}]~6583  & $-0.300$   & 30.6$\pm$1.8 & 75.4$\pm$6.0 & 44.2$\pm$6.5 &   28.1$\pm$2.6   & 102$\pm$9 & 7.06$\pm$0.41 & 103$\pm$12 \\      

 \ion{He}{i}~6678   & $-0.312$   &  3.49$\pm$0.46 & 4.2$\pm$1.4  &   4.5:        &   \nodata       &     \nodata     & 3.23$\pm$0.24 & \nodata \\
  
 [\ion{S}{ii}]~6716  & $-0.318$   & 29.2$\pm$1.7 & 62.0$\pm$5.1	   & 59.2$\pm$8.4 & 35.9$\pm$3.9    & 60.2$\pm$8.5 & 9.90$\pm$0.53  & 69$\pm$11 \\      

 [\ion{S}{ii}]~6731   & $-0.319$  & 21.0$\pm$1.3 & 46.0$\pm$4.0	   & 38.3$\pm$7.1 & 22.0$\pm$3.0    & 37.9$\pm$7.8 & 7.33$\pm$0.42  & 53$\pm$10 \\
 
 \ion{He}{i}~7065   & $-0.364$   &   2.44$\pm$0.60   &     \nodata       &  \nodata          &   1.7:          &     \nodata     & 2.32$\pm$0.31 & \nodata \\

 [\ion{Ar}{iii}]~7135 & $-0.373$   &  7.44$\pm$0.58     &    7.4$\pm$1.5    &  14.6$\pm$3.1   &   10.8$\pm$2.4     &   \nodata       &    7.20$\pm$0.41 & 5.8:  \\

 \hline
$F_{\rm H\beta}$ [10$^{-14}$ erg\,cm$^{-2}$\,s$^{-1}$] 
        &     & 10.0$\pm$0.4  &   1.27$\pm$0.07       &   2.1$\pm$0.2  &  1.89$\pm$0.15     &   0.66$\pm$0.06  &  33.8$\pm$1.1  & 1.31$\pm$0.16 \\
                    
 \noalign{\smallskip}                   
                    
$-W$(H$\alpha$) [\AA] & &      212$\pm$15    & 82$\pm$10 &    54$\pm$5    &  1100$\pm$100     &  109$\pm$11  &  $757\pm35$ & 51$\pm$5 \\
$-W$(H$\beta$) [\AA]  & &       90$\pm$8       & 9.3$\pm$1.1    &   13.1$\pm$1.4  &    376$\pm$50     &  22.8$\pm$2.9 &  $136\pm15$  & 14.0$\pm$1.6\\
$-W$(H$\gamma$) [\AA] & &   12.9$\pm$1.2     &  6.9$\pm$0.6  & 4.8$\pm$0.6   &     68$\pm$20       & 7.6$\pm$2.2  &  $45.5\pm2.8$ & 5.1$\pm$2.8 \\
$-W$(H$\delta$) [\AA] & &       4.9$\pm$0.5      &   \nodata &    2.2:     & \nodata & \nodata  &  $19.1\pm1.6$  &  \nodata\\

\noalign{\smallskip}
Aperture size [arcsec] & &     1.92$\times$8.18   & 1.92$\times$5.2  & 1.92$\times$12.3    & 1.92$\times$3.4    &    1.92$\times$17.6 &  1.92$\times$6.2      &1.92$\times$4.1 \\
$c$(H$\beta$)$_{\rm internal} $  & &  0    &     0.06$\pm$0.02       &    0       &   0.17$\pm$0.03    &     0.02$\pm$0.01    &  0.04$\pm$0.01  & 0.04$\pm$0.01\\
$E$(B-V)$_{\rm internal} $  & &  0    &     0.041$\pm$0.014       &    0       &   0.12$\pm$0.02    &     0.014$\pm$0.007    &  0.028$\pm$0.007  & 0.028$\pm$0.014\\

$W_{abs}$   [\AA]   & &         2.2        &     1.6$\pm$0.2              &       0    &      1.1$\pm$0.2       &   1.2$\pm$0.2       &  0.5$\pm$0.1 &   0.5$\pm$0.2 \\

\hline
   
\end{tabular}

\end{table*}

The results obtained for each galaxy are compiled in Table~\ref{spectra}. 
These data have been corrected for both internal reddening and underlying stellar absorption
following the method described in \citet{LSE09}.
This is an iterative procedure to derive simultaneously the reddening coefficient, $c$(H$\beta$), and the equivalent 
widths of the absorption in the hydrogen lines, $W_{abs}$, to correct the observed line intensities for both effects. The method also
assumes that $W_{abs}$ is the same for all the Balmer lines and uses the
relation given by \citet{MB93} to perform the absorption correction,
\begin{eqnarray}
c(H\beta)=\frac{1}{f(\lambda)} \log\Bigg[\frac{\frac{I(\lambda)}{I(H\beta)}\times
\Big(1+\frac{W_{abs}}{W_{H\beta}}\Big)} {\frac{F(\lambda)}{F(H\beta)}\times 
\Big(1+\frac{W_{abs}}{W_{\lambda}}\Big)}\Bigg],
\end{eqnarray}
for each detected hydrogen Balmer line, where $F(\lambda)$ and $I(\lambda)$ are the observed and the theoretical fluxes (unaffected by 
reddening or absorption), $W_{abs},\ W_{\lambda}$, and $W_{H\beta}$ are the equivalent widths of the underlying stellar absorption, the considered Balmer line and H$\beta$, respectively, and $f(\lambda)$ is the reddening curve normalized to H$\beta$ using the \citet*{Cardelli89} extinction law. All these values are compiled in Table~\ref{spectra}.
We always considered the
theoretical ratios of the pairs of \ion{H}{i} Balmer lines expected
for case B recombination given by \citet{SH95} 
assuming the derived electron temperature (only for the case of NGC~1140 and NGC~4861) or using the oxygen abundances derived from empirical calibrations (as we explain below) and the expected electron temperature for these values, following Tables~2 and 3 in \citet{LSE10b}.
As we already corrected the spectra for foreground reddening, the obtained values of $c$(H$\beta$)$_{\rm internal}$ 
represent the intrinsic extinction within each galaxy.
We note that for two galaxies, NGC~1140 and NGC~3738, we assumed $c$(H$\beta$)$_{\rm internal}$=0 because 
the method gave slightly negative values for the reddening correction.
Table~\ref{spectra} also includes the  $E(B-V)_{\rm internal}$ derived from the reddening coefficient, following 
$E(B-V)_{\rm internal}$ = 0.692$\times c$(H$\beta$)$_{\rm internal}$.

\subsection{Physical conditions of the ionized gas}

We first analyze the nature of the ionization using the so-called diagnostic diagrams,
as firstly proposed by \citet{BPT81} and \citet{VO87}. 
These diagnostic diagrams  plot two different excitation line ratios for
classifying the excitation
mechanism of ionized nebulae. \ion{H}{ii} regions (or \ion{H}{ii} or starburst galaxies) 
lie within a narrow band within these diagrams, 
but when the gas is ionized by shocks, accretion
disks, or cooling flows (in the case of AGNs or LINERs)
its position in the diagram is away from the locus of \ion{H}{ii}
regions.

\begin{figure}
\centering
\includegraphics[scale=0.38,angle=90]{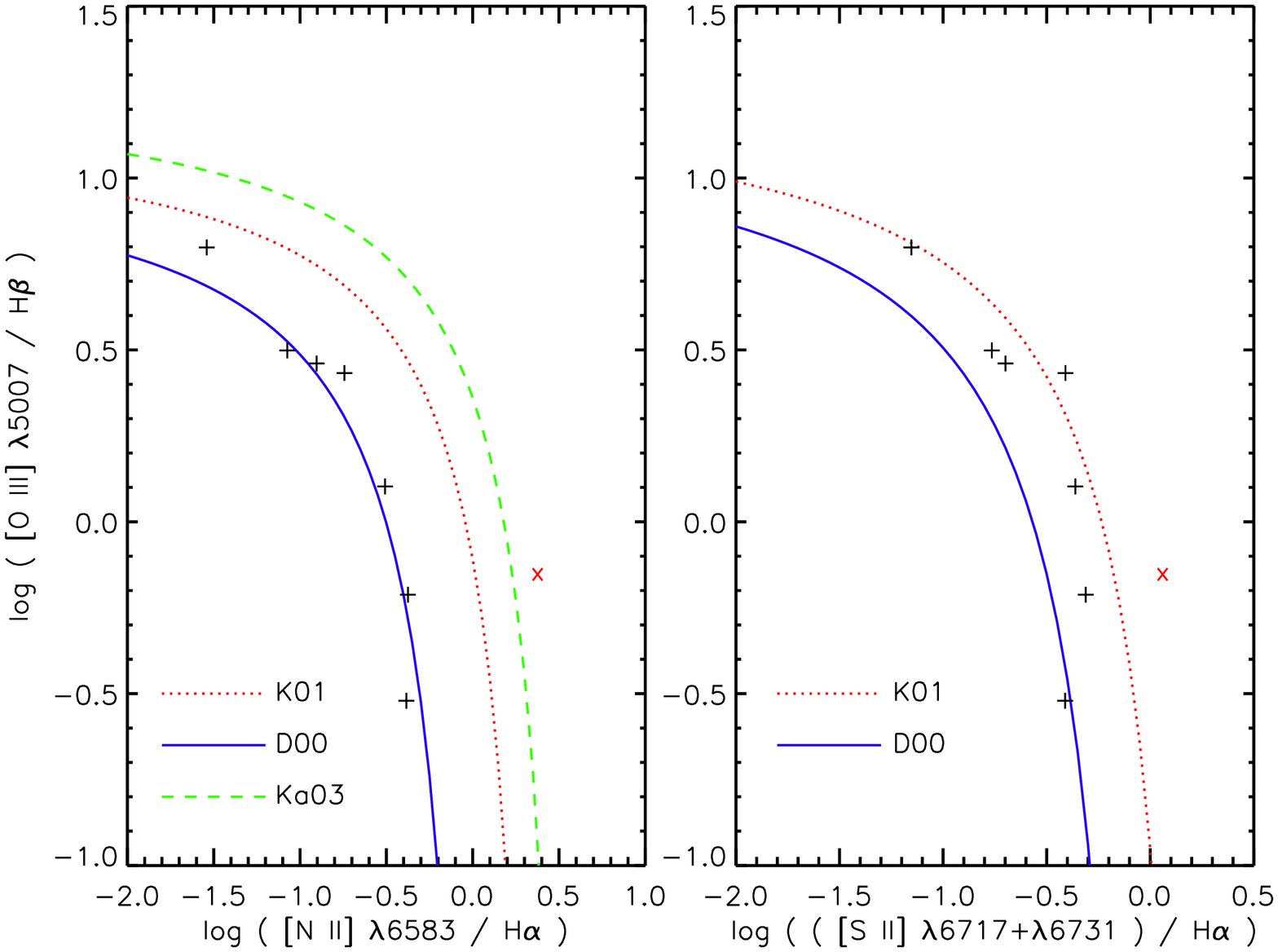}
  \caption{Comparison of some observational flux ratios obtained
for all analyzed WR galaxies (black crosses) 
with the diagnostic diagrams proposed by \citet{Do00}, 
blue continuous line (D00), and \citet{KD01}, 
red discontinuous line (K01). The left panel also shows
the empirical relation provided by \citet{Kauffmann03} with
a dotted-dashed green line (Ka03). We also plot
the observed flux ratios for the center of NGC~6764 (red x),
which is classified as a LINER.  \label{diagnostic}}
\end{figure}

Figure~\ref{diagnostic} plots
the typical [\ion{O}{iii}] $\lambda$5007/H$\beta$ versus [\ion{N}{ii}] $\lambda$6583/H$\alpha$ and
 [\ion{O}{ii}] $\lambda$5007/H$\beta$ versus ([\ion{S}{ii}] $\lambda$6716+$\lambda$6731)/H$\alpha$ diagrams.
We used the analytic relations given by \citet{Do00} and \citet{KD01} 
between different line
ratios to check the nature of the excitation mechanism of
the ionized gas within the bursts. 
Actually, the dividing line
given by the \citet{KD01} models represents an upper
envelope of positions of star-forming galaxies. The left
panel of  Fig.~\ref{diagnostic} includes the empirical relation between the
[\ion{O}{iii}] $\lambda$5007/H$\beta$ and the [\ion{N}{ii}] $\lambda$6583/H$\alpha$ 
provided by \citet{Kauffmann03} analyzing a large data sample of star-forming
galaxies from the SDSS \citep{York00}. 
As we see, all analyzed regions lie below the \citet{KD01} 
theoretical line. This clearly indicates that photoionization
is the main excitation mechanism of the gas and
that there is very little evidence for a significant contribution
from shock excitation. However, this is not satisfied in the case of the center of NGC~6764 (red cross in Fig.~\ref{diagnostic}).
which lies in the region occupied for LINERs. Indeed, this starburst galaxy has been classified as 
a classical LINER galaxy in the past \citep{AlonsoHerrero00}. 

The electron density of the ionized gas, $n_e$, was computed via the [\ion{S}{ii}] $\lambda\lambda$6716,6731 doublet
by making use of the five-level program for the analysis
of emission-line nebulae included in IRAF NEBULAR task \citep{SD95}. 
All regions were found in the low-density limit, $n_e < 100$~cm$^{-3}$, and hence we adopt  $n_e$=100~cm$^{-3}$.

The electron temperature of the ionized gas was computed for only two galaxies: NGC~1140 and NGC~4861, as only in these two cases we detect the faint auroral [\ion{O}{iii}]~$\lambda$4363 emission line. We therefore inferred $T_e$([\ion{O}{iii}]) from the [\ion{O}{iii}] ($\lambda$4959+$\lambda$5507)/$\lambda$4363 ratio by making use of the IRAF NEBULAR task. As we assumed a two-zone approximation to define the temperature structure of the nebula, we used $T_e$([\ion{O}{iii}]) as representative of high-ionization potential ions. The electron temperature assumed for the low-ionization potential ions was derived from the linear relation between $T_e$([\ion{O}{iii}])  and $T_e$([\ion{O}{ii}])  
provided by \citet{G92}. 
The results are listed in Table~\ref{abunda}.

\begin{table}
\scriptsize
\caption{Physical conditions and chemical abundances of the ionized gas of the regions analyzed in NGC\,1140 and NGC\,4861. \label{abunda}}
\centering
\begin{tabular} {l c c }
\hline
          &     NGC\,1140~\#1     &    NGC\,4861~\#1  \\
\hline

$T_e$[\ion{O}{iii}]  [K]  &  9500$\pm$900    &     14000$\pm$600 \\
$T_e$[\ion{O}{ii}]  [K]   & 9700$\pm$700 &   12800$\pm$500 \\
$n_e$ [cm$^{-3}$]      &     $<$100  &     $<$100 \\
\noalign{\smallskip}

12+log(O$^+$/H$^+$)       &  8.06$\pm$0.12  &     7.14$\pm$0.08    \\
12+log(O$^{++}$/H$^+$)  &  8.09$\pm$0.09 &    7.87$\pm$0.04          \\
{\bf 12+log(O/H)   }           & 8.38$\pm$0.10  &  7.95$\pm$0.05             \\
log(O$^{++}$/O$^+$)  & 0.03$\pm$0.14  &    0.73$\pm$0.09 \\

\noalign{\smallskip}

12+log(N$^+$/H$^+$)      & 6.82$\pm$0.06 & 5.90$\pm$0.05    \\
12+log(N/H)                      & 7.13$\pm$0.10 & 6.70$\pm$0.09          \\
$icf$(N)                            & 2.08$\pm$0.21 &  6.38$\pm$0.95      \\
log(N/O)                            & -1.24$\pm$0.07 & -1.25$\pm$0.08  \\ 

\noalign{\smallskip}

12+log(S$^+$/H$^+$)      & 6.10$\pm$0.06 & 5.36$\pm$0.04    \\
12+log(S$^{++}$/H$^+$)   & \nodata &6.16$\pm$0.09    \\
$icf$(S)          & \nodata  &  1.8$\pm$0.7   \\
12+log(S/H)                      & \nodata & 6.35$\pm$0.08         \\
log(S/O)                            & \nodata & -1.59$\pm$0.12   \\

\noalign{\smallskip}

12+log(Ne$^{++}$/H$^+$)      & 7.48$\pm$0.15  & 7.01$\pm$0.06    \\
$icf$(Ne)                  &   1.93$\pm$ 0.65  &  1.19$\pm$0.22 \\
12+log(Ne/H)                      & 7.16$\pm$0.19 &   7.09$\pm$0.10          \\
log(Ne/O)                         & -0.61$\pm$0.16 &   -0.86$\pm$0.06 \\

\noalign{\smallskip}
12+log(Ar$^{++}$/H$^+$)      & 5.89$\pm$0.10 & 5.52$\pm$0.06    \\
12+log(Ar$^{+3}$/H$^+$)      & \nodata &   4.78$\pm$0.11    \\
$icf$(Ar)                                & 0.69$\pm$0.23  &   0.46$\pm$0.02  \\
12+log(Ar/H)                         & 5.73$\pm$0.18 & 5.60$\pm$0.06         \\
log(Ar/O)                      &    -2.65$\pm$0.21 &  -2.35$\pm$0.11 \\

\hline
   
\end{tabular}
\end{table}

\subsection{Estimation of the chemical abundances}

The preferred method for
determining oxygen abundances in galaxies using \ion{H}{ii} regions is through electron temperature
sensitive lines such as the [\ion{O}{iii}]~$\lambda$4363 line, the so-called $T_e$ method 
\citep{PC69,S78,Esteban04}.
However, in absence of [\ion{O}{iii}]~$\lambda$4363 line, the alternative empirical relations, such
as $R_{23}$ method \citep[][hereafter M91]{McGaugh91},  
$R_{23}-P$ method  \citep{P01a,P01b,PT05,PVT10}, 
$N_2O_2$ method \citep[][hereafter KD02]{KD02}, 
which use the strong emission lines, can be used for determining oxygen abundance. 
However,  the use of these empirical methods must be done carefully, see recent reviews by \citet{LSE10b} and \citet{LSD12}.

In our case, we can only derive the oxygen abundances of the ionized gas following the $T_e$ method for two galaxies: NGC~1140 and NGC~4861. We followed the very same prescriptions and {\it ionization correction factors, icf}, indicated by \citet{LSE09} to compute the O, N, S, Ar and Ne abundances, and the N/O, S/O, Ar/O and Ne/O ratios, for the ionized gas within these two galaxies. 
In particular, we  assumed a two-zone scheme for deriving the ionic abundances and considered the \citet{G92} relation between $T_e$[\ion{O}{iii}] and  $T_e$[\ion{O}{ii}]. We then used the IRAF {\it nebular} task \citep{SD95} to compute the ionic abundances from the intensity of collisionally excited lines. 
We finally assumed the standard $icf$ of \citet{PC69} to derive the total N and Ne abundances. For NGC~4861 we have estimations of both  S$^+$/H$^+$ and S$^{++}$/H$^+$ ratios, and hence we used the $icf$ given by the photoionization models by \citet{S78} to derive the total S. The total Ar abundance was  calculated by considering the $icf$ proposed by \citet{ITL94}.
The results are compiled in Table~\ref{abunda}. We note that we used an updated atomic dataset for O$^+$, S$^+$, and S$^{+2}$ for {\it nebular}. The references of these of these updated values are indicated in Table~4 of  \citet{GRE05}. 

\begin{table}
  \caption{Parameters used to derive the oxygen abundance following the strong emission-line methods.  \label{parameters}}
\scriptsize
\begin{tabular}{l r@{\hspace{8pt}}  r@{\hspace{8pt}}r@{\hspace{8pt}}r@{\hspace{8pt}}r@{\hspace{8pt}}r}

\hline
Galaxy Name / Knot         & $N_2$     &  $N_2O_2$   & $O_3N_2$ & $R_{23}$  & $P$    &  $y$    \\    

\hline

NGC\,1140~\#1             & $-$0.971 & $-$0.897 & 1.409   &   6.23  &  0.613 & 0.200     \\ 
IRAS\,07164+5301   & $-$0.580 & $-$0.499 &  0.659 & 4.08  & 0.418 & $-$0.144 \\
NGC\,3738             & $-$0.810  & $-$1.19  &  1.220  &   10.33  &   0.342   &  $-$0.285  \\ 
UM\,311                 & $-$1.002 & $-$0.962 & 1.433  &  6.34 & 0.593 & 0.164 \\
NGC\,6764~\#1             &  $-$0.446 & $-$0.141 & 0.0332 & 1.985 & 0.286 & $-$0.397 \\
NGC\,4861~\#1              & $-$1.602 &  $-$1.10   &   2.37  & 8.82 & 0.900 &  0.953    \\
NGC\,3003~\#13            & $-$0.444  & $-$0.308  & 0.209  &  2.89 & 0.277 & $-$0.417 \\
\hline

\end{tabular}

\begin{flushleft}
 
The definition of the parameters are: \\
$N_2$ = log([\ion{N}{ii}]~$\lambda$6583/H$\alpha$), \\
$O_3N_2$ = log [([\ion{O}{iii}]~$\lambda$5007/H$\beta$)/log([\ion{N}{ii}]~$\lambda$6583/H$\alpha$)], \\
 $N_2O_2$ = log([\ion{N}{ii}]~$\lambda$6583/[\ion{O}{ii}]~$\lambda$3727), \\ 
 $R_{23}$ = log [ (  [\ion{O}{ii}]~$\lambda$3727 + [\ion{O}{iii}]~$\lambda$4959 + [\ion{O}{iii}]~$\lambda$5007)/H$\beta$], \\
  $P$ =  ([\ion{O}{iii}]~$\lambda$4959 + [\ion{O}{iii}]~$\lambda$5007)/$R_{23}$,\\
   $y$=log [ ([\ion{O}{iii}]~$\lambda$4959 + [\ion{O}{iii}]~$\lambda$5007) /  [\ion{O}{ii}]~$\lambda$3727 ].

\end{flushleft}

\end{table}

We derive oxygen abundances --in units of 12+log(O/H)-- of 8.38$\pm$0.10 and 7.95$\pm$0.05 for NGC\,1140 and NGC\,4861, respectively, being their associated N/O ratios --in units of log(N/O)-- $-$1.24$\pm$0.07 and $-$1.25$\pm$0.08, respectively. The N/O ratio computed in NGC\,1140 agrees with that expected for its O/H \citep[e.g.,][]{IT99,Izotov04,LSE10b}. 
However, the N/O ratio derived in NGC~4861 is clearly higher than that expected for its oxygen abundance.
Following the data compiled by these authors, an object with 12+log(O/H) between 7.9 and 8.0 should have an N/O ratio between $-$1.6 and $-$1.4.
Hence, we estimate this excess in the N/O ratio in NGC~4861 on $\sim$0.25--0.35~dex. 
The fact that WR stars are clearly detected in this galaxy suggests that the excess of nitrogen has been released by the ejecta of these massive stars \citep*{Kobulnicky97,Pustilnik04,BKD08,LSE10b}. 
However, only few observations confirming the localized N enrichment are nowadays available 
\citep{Kobulnicky97,LSEGRPR07,James09,Monreal-Ibero+10,LS+IC10+11},
and in some cases the chemical pollution produced by the WR stars detected 
are not able to explain the observed N excess \citep{PerezMontero11,Amorin12}. 

For all galaxies (even those two for which we have a direct estimation of $T_e$) we estimate their oxygen abundances using the most-common empirical calibrations. Table~\ref{parameters} compiles all parameters (and their definitions) used for computing the oxygen abundances following these strong-line methods, while Table~\ref{empirical} lists the results. For more information about this methods and their equations, see Appendix~A in \citet{LSE10b}. The KD02 method using the $N_2O_2$ parameter can only be used for objects with 12+log(O/H)$\gtrsim$8.60, and hence Table~\ref{empirical} only gives its value for such objects.

\begin{table*}

\scriptsize

  \caption{Oxygen abundances derived for our WR galaxy sample using the most commonly used strong emission-line methods. Last two columns compile the adopted oxygen abundance and the branch (lower, medium or upper) used. The third column also lists previous estimations of the O/H ratio in the literature. The strong emission-line calibrations are: 
 M91: \citet{McGaugh91};
 KD02: \citet{KD02}; 
 PT05: \citet{PT05};  
 P01: \citet{P01a,P01b}; 
 PP04a: \citet{PP04}, 
 using a linear fit to the $N_2$ parameter;  
 PP04c:  \citet{PP04}, using the $O_3N_2$ parameter.
  \label{empirical} }

 \begin{tabular}{l   cc    cccc  cc  cc  cc c   }
\hline

Galaxy Name     & $T_e$  & Lit.  &  M91   & KD02    & KD02    &  &     PT05  &  P01   &     PP04a   &    PP04c        &  \multicolumn{2}{c@{\hspace{4pt}}}{Adopted$^{a}$}   &Branch \\

\noalign{\smallskip}
\cline{12-13}
\noalign{\smallskip}

knot                      &  &   &    $R_{23}$, $y$  & $R_{23}$, $y$ &   $N_2O_2$        & &       $R_{23}$, $P$  & $R_{23}$, $P$ &
                          $N_2$    &  $N_2O_3$   &             MKD     &    PPP   \\

\hline 
NGC\,1140~\#1         & 8.38$\pm$0.10 &  8.29$\pm 0.09^b$     &   8.61 &  8.40 & 8.66     & &   8.36 & 8.42  &  8.35 &  8.30 &  8.56 &  {\bf  8.36}   & Upper   \\
 
IRAS\,07164+5301   & \nodata & 8.96$^c$  & 8.77 & 8.99 & 8.91   &&  8.42  &  8.50  & 8.57  & 8.52  & 8.89 & {\bf 8.50}  &Upper  \\ 

NGC\,3738              & \nodata & 8.23$^d$     & 8.53 & 8.57 &  ...    && 8.18   & 8.26  & 8.44 & 8.35                    &   8.55 &   {\bf 8.31}  & Med   \\

UM\,311                   & \nodata &  8.31$\pm$0.04$^e$ & 8.31  & 8.43 &  \nodata    &&  8.12 & 8.15   &   8.33     &    8.28    & 8.37    &   {\bf 8.22}  & Med \\

NGC\,6764~\#1              & \nodata & \nodata  &  8.97  & 9.12    & 9.07  &  &   8.53   & 8.68    & 8.65   &     8.72    & 9.05 & {\bf 8.65}  & Upper  \\

NGC\,4861~\#1              & 7.95$\pm$0.05 & 8.05$\pm$0.04$^f$   & 8.03 & 8.22 &  \nodata && 7.77  &  7.83   &   7.99     &  7.97   & 8.13 & {\bf 7.89}   & Lower\\

NGC\,3003~\#13          & \nodata & \nodata &   8.87 & 9.06 &   9.00    && 8.41 & 8.56 &       8.65 &   8.66   &  8.98 & {\bf 8.57} & Upper   \\

\hline 
\end{tabular}
\begin{flushleft}

$^a$: Average abundance value using all the empirical methods, the $T_e$ method is not considered here. We provide two results: PPP, which considers the average value obtained with the  PT05, P01, PP04a  and  PP04c calibrations and MKD, which assumes the average value of the M91 and KD02 calibrations. The KD02 method using the $N_2O_2$ parameter is only considered for objects with  12+log(O/H)$\gtrsim$8.60 dex (NGC\,3738, NGC\,6764 and NGC\,3003). The uncertainty in these values is $\sim$0.10~dex.\\
$^b$: \citet{Moll07} through the  $T_e$ method.\\
$^c$: Taken from the literature \citep{Huang96} through $T_e$ method, although we consider this value is not correct, see Sect.~5.4.2. \\
$^d$: \citet{Martin97}. \\ 
$^e$: \citet{IT98}. \\ 
$^f$: \citet{Esteban09}. 

\end{flushleft}

\end{table*}

The final adopted value for the oxygen abundance in each galaxy has been computed by averaging all the results provided by these calibrations, and it is compiled in Table~\ref{empirical}.
As it has been already noticed by several authors \citep{Peimbert07,Bresolin09,LSE10b,Moustakas+10,RO11,LSD12}, those empirical calibrations which assume strong emission-lines based on photoionization models
\citep{McGaugh91,KD02} tend to overpredict the observed oxygen abundances derived using the $T_e$ method and the empirical calibrations based on it \citep{P01a,P01b,PT05,PP04,PVT10} by 0.2--0.4~dex.
Here we also observe that behavior in the majority of the analyzed regions. 
Hence, Table~\ref{empirical} compiles two sets of average values: MKD, which is the average value of the oxygen abundance provided by the empirical calibrations based on photoionization models \citep{McGaugh91,KD02}, and PPP, which considers the average value obtained with those calibrations based on the $T_e$ method  \citep{P01a,P01b,PT05,PP04}. 
The adopted oxygen abundance following the empirical calibrations for the two galaxies for which we have a direct estimation of $T_e$ agrees well within the errors with the adopted PPP value, so we consider that the oxygen abundances derived for the rest of the objects are reliable besides the uncertainties. 

Assuming that the solar abundance is 12+log(O/H)$_{\odot}$=8.66 \citep{ASP05}, we finally derived the metallicity, $Z$, for each galaxy using the PPP results. These values have been listed in Table~\ref{Table1}.

\begin{figure*}
\centering
(a)\includegraphics[scale=0.43,angle=90]{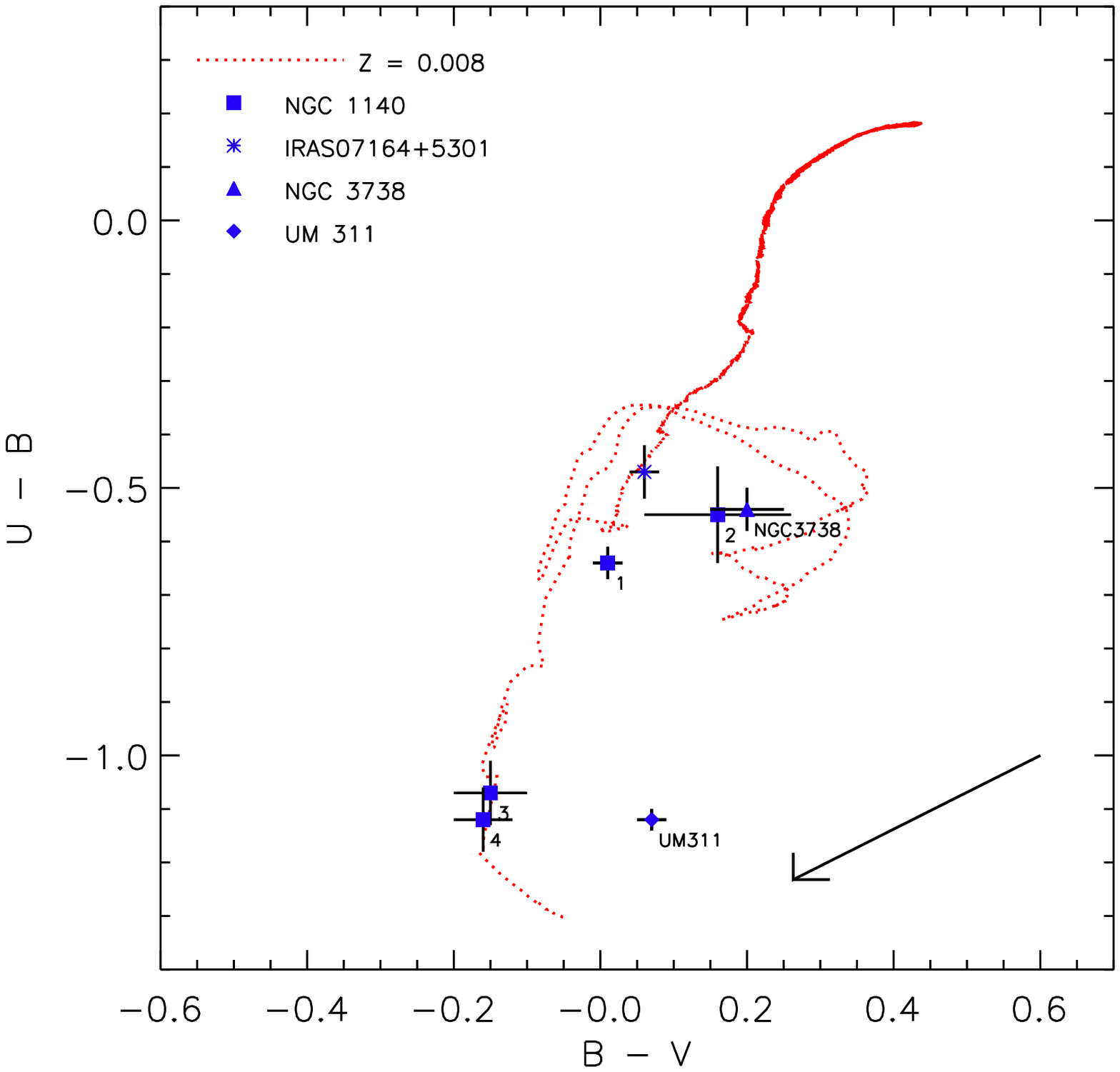} 
(b)\includegraphics[scale=0.43,angle=90]{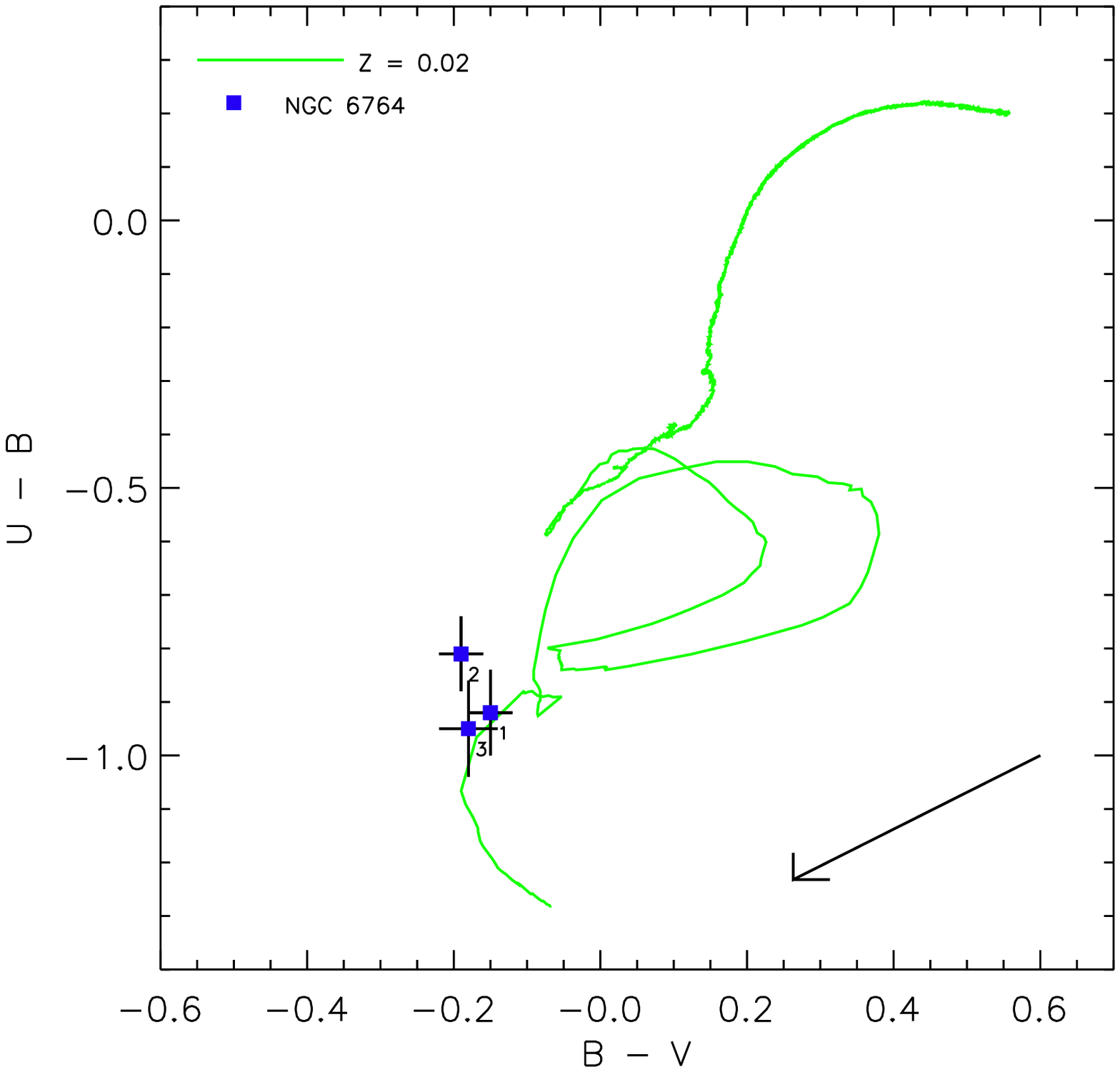} 
(c)\includegraphics[scale=0.43,angle=90]{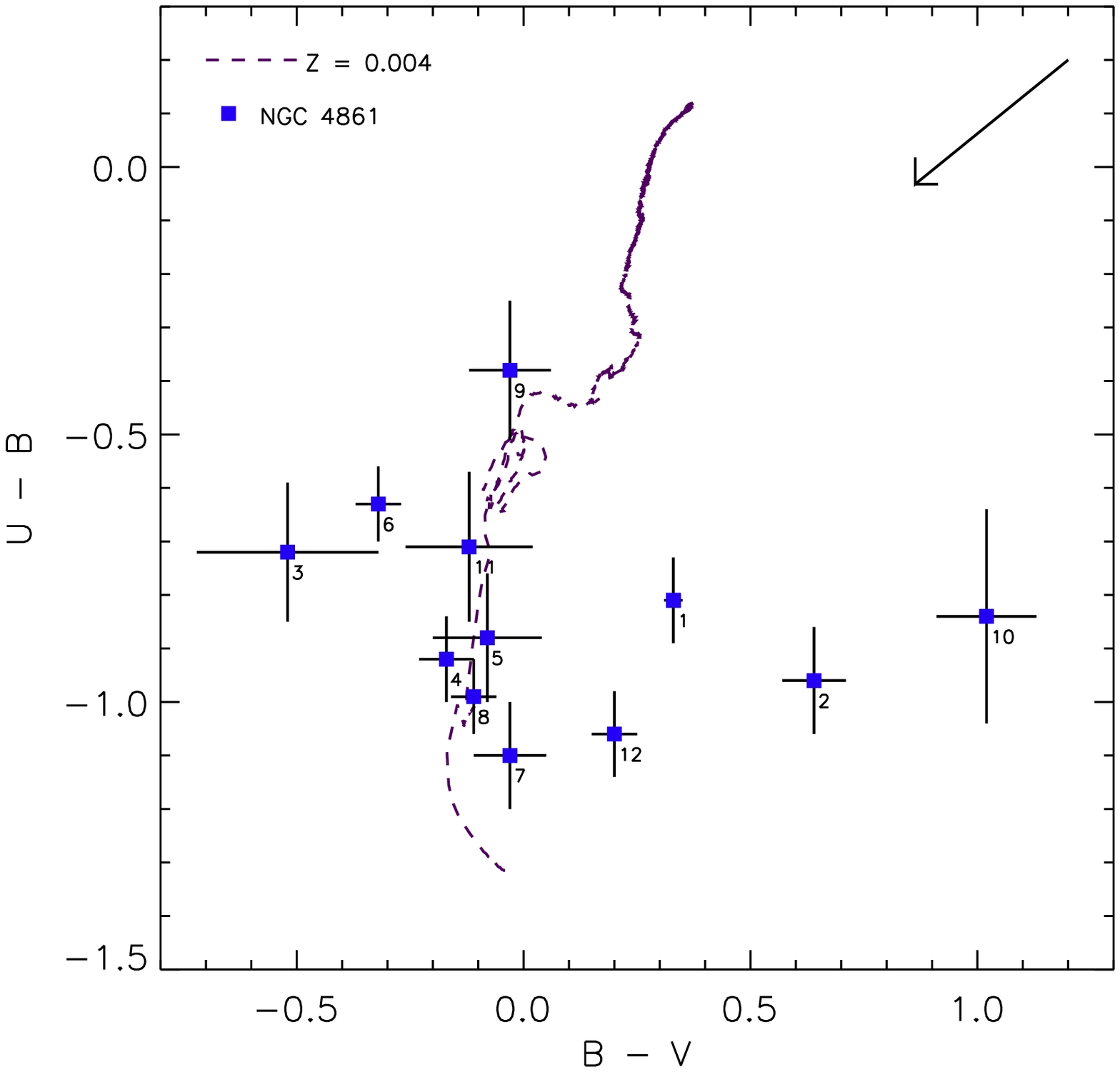} 
(d)\includegraphics[scale=0.43,angle=90]{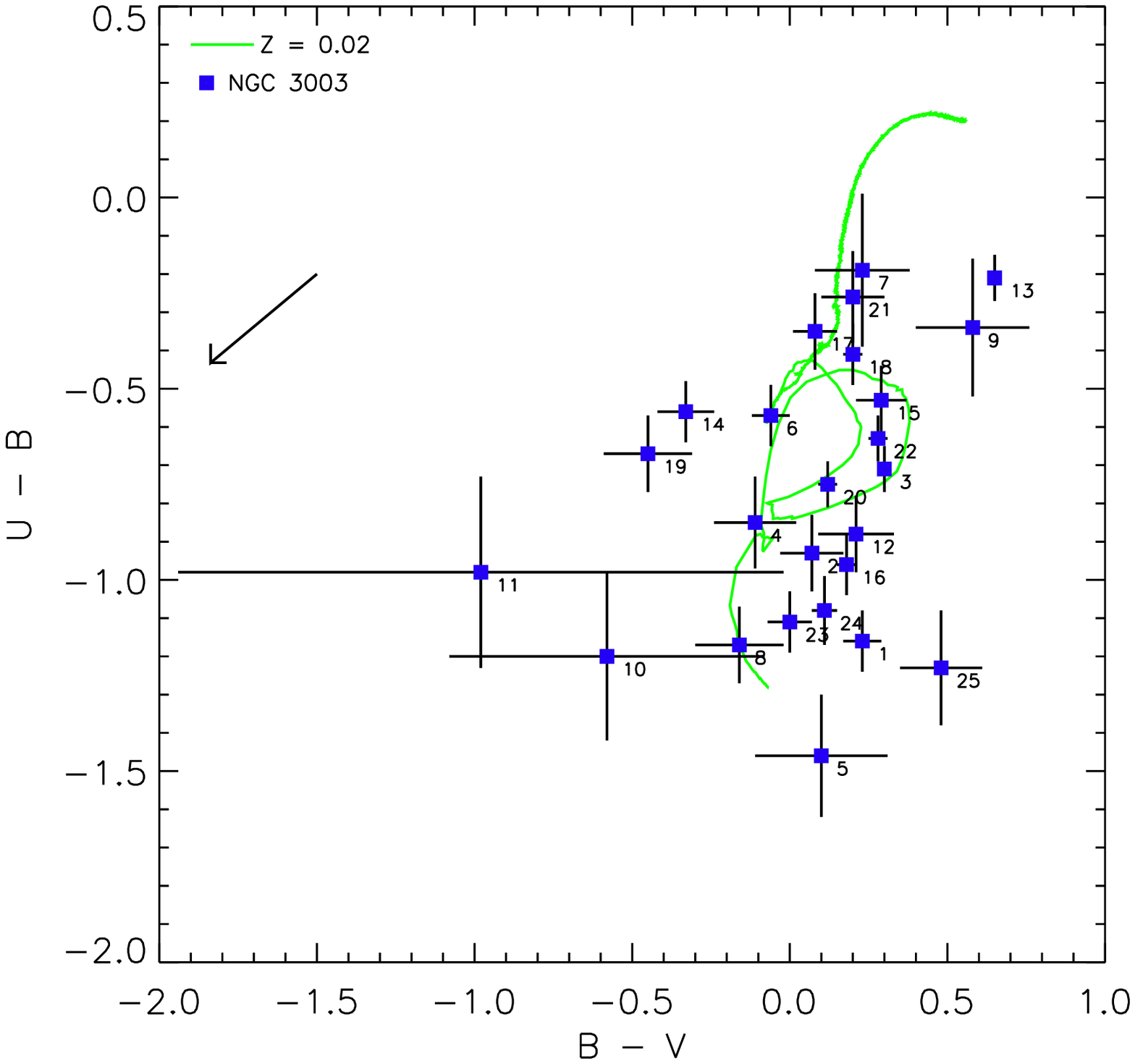} 
  \caption{ $U-B$ vs. $B-V$ diagrams comparing the predictions given by
evolutionary synthesis models provided by the Starburst~99 code \citep{L99}
   with the colors obtained from the star-forming regions within
the WR galaxies of our sample. The metallicity of each model is also shown, 
   and it is $Z$= 0.008 in panel (a) NGC~1140, IRAS~07164+5301, NGC~3738 and UM~311;
   $Z$=0.02 in panel (b) NGC~6764; $Z$=0.04 in panel   (c) NGC~4861; and $Z$=0.02 in 
   panel (d) NGC~3003. The vector $E(B-V)=0.323$\,mag  [$A_V=1$,  $(U-B)=-0.232$, $(B-V)=-0.337$]  used for dereddening the data is also shown. \label{sb99models}}
\end{figure*}

\begin{table*}
\scriptsize
  \caption{Estimation of the internal reddening,  $E(B-V)_{internal}$, and the age of the young and old stellar populations of the analyzed knots using the broad-band colors. The second last column compiles our best estimation for the age of the most recent star-formation event, and considers the data derived using both the H$\alpha$ images and the spectroscopy.
  The uncertainty for the ages of the young stellar populations typically is  $\sim$1~Myr. The uncertainties of the age of the underlying stellar population vary from $\sim$10 - 20~Myr (ages up to 100~Myr) to $\sim$100 - 300~Myr (ages up to 100~Myr).
  The typical uncertainty for $E(B-V)_{internal}$ is 0.05~mag.
 \label{ages}}
 \begin{tabular}{l ccccccc  ccc}
 \hline

   &    &    &     &  
   \multicolumn{4}{c@{\hspace{4pt}}}{Ages estimated from broad-band colors.   } &  & \multicolumn{2}{c@{\hspace{4pt}}}{Age Adopted} \\  

\noalign{\smallskip}
\cline{5-8}\cline{10-11}
\noalign{\smallskip}

Galaxy &  $Z$ &  Knot & $ E(B-V)_{internal}$  &   $U-B$    &       $B-V$         &     $V-R$     &        $V-I$         &  &  Burst     &     Underlying     \\

\noalign{\smallskip}
  &[$Z_{\odot}$]&        & [mag]  	  &     [Myr]    &    [Myr]&     [Myr] &        [Myr]    &&  [Myr] &        [Myr]               \\
\hline	                 	 	  											         
NGC~1140 &0.008&  1       & 0.1    & 5.2 - 5.4       & 6 - 23         &     1 - 25          & 7 - 20	       &&   5.0   &   20		       \\        
                            &&  2       & 0.2    & 5.2 - 6.2       & 5 - 40         &	 1 - 45          &  5 - 25  	       &&    5.5  &  25  \\	       
                            &&  3       & 0.0$^d$    & 3.3 - 3.8       & 2.3 - 4.8      &     20 - 100      &     2 - 5           &&   3.5  &   50 \\   
                            &&  4       & 0.0$^d$    & 3.2 - 3.3        & 2.3 - 4.8	    &	1 - 20 	      &	1 - 7	       &&   3.2  &  ...   \\	         
\noalign{\smallskip}	                 	 	  	    		       			    				           
IRAS~07164+5301         &0.008&  1        & 0.0     & 5.9 - 6.3           & 34 - 44	       & 15 - 25	& 15 - 25	    &&   6.0 &    25	      \\	         
\noalign{\smallskip}	                  	  	    		       			    				           
NGC~3738&0.008&  1$^a$      & 0.00 - 0.25  &    5.0 -  7.0       & 16 - 470        &	  10 - 100             &     10 - 800	   & &   6.0   &  600	       \\  
\noalign{\smallskip}	                 	 	  	    		       			    				           
UM~311   &0.008&  1$^b$      & 0.00 - 0.17 	  &    3.2 - 3.7	     &	    2 - 45       &	  ... 	& 5 - 10  &&	3.2   &   ... 		       \\	      
\noalign{\smallskip}	                 	 	  	    		       			    				           
NGC~6764&0.02&  1       & 0.0    &    3.5 - 9.0    	     &	    2.0 - 3.5           &	 ...	        &		30 - 500	&&  6.0 &   400       \\	      
            &&  2$^c$       & 0.0   &    4.0 - 10  	              &	    2.5 - 3.5           &	2.5 - 6.0	& 		6 - 20	&&  6.0 &    15       \\	      
            &&  3       & 0.0    &    3.4 - 6.0   	                     &	    2.5 - 3.5          &  ... 	        &		40 - 500  	&&  5.0 &  400      \\  
\noalign{\smallskip}	         				    	               			                                       
NGC~4861&0.004 &  1$^b$  &  0.0 - 0.5  &   3.5 - 5.0  &	...		&	   ...            	& 2.0 - 7.0                  && 4.6 & ...	                       \\          
            &&  2$^b$        &  ...    &  3.1 -   4.8  	   &	...	           	&	 ...  			&   \nodata		&& 3.0 & ...			   \\		   
      	    &&  3$^b$        &  ...    &  4.8 -   6.4 	   &	      ...          	&      1 - 100    		& 7 - 25	  		 & & 5.0  & 25		    \\  
      	    &&  4       & 0.0  	  &  4.1 -   4.8  	 	   &    1.9 - 6.8	  	&      2.5 - 7.0		&  ...    			&&  4.5  & ..			    \\  	    
      	    &&  5       & 0.0  	  &  4.1 -   5.0  		  &        1 - 20    	&	$>$20  		&   1 - 40			 &&   5.0 & 30			    \\  
      	    &&  6       & ...	  &  5.2 -   6.8  		  &    \nodata  	   	&    $>$20		       &     20 - 100 		& & 5.6 & 50		    \\  
      	    &&  7       & 0.1  	  &  2.6 -   4.1    	  &  1 - 20          	&	  7 - 25  	 	&    ...   			& &  4.0  & 15		    \\   
      	    &&  8       & 0.0  	  &  3.1 -   4.6    	  &  1.5 - 6.9	   	&      $>$30		 &    8 - 30    		 &&  4.5 & 30	    \\  	   
      	    &&  9       &  0.0      &  7.2 -   10      	  & 3 - 25            	&    2.5 - 7.5		&     5 - 30 		&& 5.0 & 20		    \\  	      
      	    && 10       &    ...     &  3.1 -   5.4      	  &	  ...	    	        &	   $>$30	 	&    ...  			&& 4.5 &  $>$30		    \\  	   
      	    && 11       & 0.0  	  &  4.8 -   6.6                &  1 - 30 	   	&       $>$30		 &    \nodata 		&  & 5.0 & $>$30		    \\  	   
      	    && 12$^a$ & 0.00 - 0.25  &  2.5 -   4.0      & 4 - 88	     	&	  20 - 150	   	 &    10 - 200  		& & 3.7 & 100		    \\     
\\	                 	 	  	    		       			    				           
NGC~3003&0.02 
  &  1        & ...      &      2.0 -  3.0              &	200 - 460  	 & $>$1000		& \nodata	 		&&  3.0  & $>$1000	  \\	     
 &&    2     & 0.2     &      3.0 -   5.0   		&	   2 - 20	 	&    14 - 400 		&   20 - 600  		 && 5.0 & 500		    \\   
 &&  3       & 0.4     &      4.5 -   10   		&	   1 - 20    		& 	7 - 21	  	&   \nodata	        & & 8.0 & 20			    \\  	   
 &&  4       & 0.0     &      3.4 -   6.1   		&	 100		 	&  $>$20    	       	&   $>$500	  	&& 5.5 & $>$500  		 \\	 
 &&  5$^b$   &   ... &      ...	     			&	...       		&      ...        	 	&    ...         		&& 3.3 & ...		    \\  	   
 &&  6       &  0.0    &      5.8 -   6.5   		&	 2 - 22       	&      14 - 60	  	& 15 - 25 			& & 5.5 & 20			    \\      
 &&  7       &   0.0   &      60 - 120      	&    40 - 600	  	& $>$30	      		&   20 - 900 		& & 5.6 & 600		    \\     
 &&  8       &   0.0   &      3.2 - 3.4     		&	1 - 20 		&  10 - 800    		& 50 - 200 		&& 3.4 & 200		\\   
 &&  9       &   0.5   &      4.5 -   10    		&   10  - 400	 	&  20 - 800	 	&  20 - 900 		&&  4.5 & 500		    \\  
 && 10$^b$  &   ... &      3.2 -   3.4   		&	     \nodata  	&     ...			&    ...	 		&& 3.1 &  ... 		  \\	   
 && 11       &  ...     &      1.9 -   6.1   		&	   \nodata  	&  ...				&    ...	 		&& 4.7 &  ... 		    \\  	   
 && 12       &  0.3    &      3.0 -   5.5   	&	      1 - 30  	&  20 - 800	 	&   30 - 400 		&& 5.3 & 300		    \\  	   
 && 13  & 0.0 - 0.6$^a$ &      100 - 175	&	      $>$1000    &  $>$1000		&   $>$1000 		 && 6.7 & $>$1000  		    \\  	   
 && 14       &   ...    &      6.3 -   6.6   		&  \nodata  	   	& $>$1000		 &	20 - 600    		 && 6.2 &  500			   \\	 
 && 15       &   0.4   &      5.5 -  9.5   		&  1 - 20	       		& 20 - 400       		&	20 - 800	 	& & 6.0 &  500			    \\    
 && 16       &   0.3   &      3.4 -   3.5  	 	&	2 - 6      		&   15 - 25		 &  30 - 200   		& & 3.5 & 100		    \\     
 && 17       &   0.0   &      15 -   50   		&  23 - 78	  		&  \nodata		   	&  20 - 300  		& & 4.0 & 200				    \\     
 && 18       &   0.2   &      5.8 -  9.6   		& 6 - 20	 		&   20 - 470		&   20 - 300  		 & & 6.2 & 200			    \\        
 && 19       &  ...     &      6.1 -   6.4   		& \nodata			&	$>$1000 		&  $>$400	        &   & 5.8 & $>$1000			    \\    
 && 20       & 0.2     &      5.6 -   6.0   	&	4 - 20 		&   20 - 40  	  	&   15 - 25 		& & 5.8 & 20		      \\     
 && 21       & 0.0     &      55 - 150   		&  40 - 490   		&     $>$100	        &      $>$1000          && 6.0 & $>$1000 	   \\	   
 && 22       & 0.4     &      5.5 -   6.2   	&	1 - 8	    		&	10 - 45	       &      $>$20	       &   &     5.6 & 40		    \\     
 && 23$^b$& 0.0 - 0.1&      2.4 -   3.6   	&	     3 - 35		 & $>$30    		&   \nodata		 & & 3.4 & $>$30		    \\  	   
 && 24       & 0.2     &      2.2 -   3.2      	&       2 - 20  	   	&    30 - 600    		&  30 - 600   		&  & 3.4 & 500			    \\  
 && 25       &  ...    &      3.3 -   3.4      	&       600 - 1000     &    $>$30		        &    20 - 900 		&& 3.4 & 800			    \\     
 
  \hline			 	  											      
\end{tabular}

\begin{flushleft}

$^a$ The derived $E(B-V)_{internal}$ for this object is very probably overestimated because of the effect of an important old stellar population underlying the burst to the $B-V$ colour. See text for details.\\
$^b$ The broad-band colours of these knots, specially $B-V$, are probably strongly affected by the emission of the ionized gas. See text for details.\\
$^c$ The center of NGC~6764 (knot~\#2) is a LINER, and not a star-forming region. Hence, these values are meanless. See text for details. \\
$^d$ As these knots have a relatively large $W$(\Ha),  perhaps the contribution of the emission of the gas cannot be neglected. If this is true, the real value of the intrinsic reddening may be up to $E(B-V)_{internal}\sim$0.4~mag. See text for details.

\end{flushleft}

\end{table*}

\section{Discussion}

\subsection {Age of the most recent star-forming event\label{ageHa}} 

The estimated  age of  the most-recent star forming event can be derived from
the  H$\alpha$ equivalent  width, $W$(H$\alpha$), of  the star-forming
knots, as it decreases with time \citep[e.g.,][]{LH95,JC00}.
We used the predictions provided by the Starburst99 code  \citep{L99}, 
which are AGB phase corrected. 
We considered an instantaneous  burst  with Salpeter initial  mass
function  (IMF) of 2.35. The  lower and  upper mass  range was  taken from  1 to
100\,\Mo. We assumed a total mass of 10$^6$\,\Mo, and a
metallicity of $Z/Z_{\odot}$ = 0.2, 0.4 and 1 which was chosen depending on the 
oxygen abundance of the galaxy derived from our spectroscopic
data. The advantage of using $W$(H$\alpha$) for deriving the age of the most 
recent starburst event is that they provide a very small error, only between 0.1 and 0.5~Myr. 
Table~\ref{halpha} compiles our results. Following our $W$(H$\alpha$) data, the majority of the analyzed 
regions have experienced a strong star-formation event between 3 and 6~Myr ago. 
In particular, the starbursts observed in regions \#1 of NGC~1140 and NGC~4861 have an age of 5.0 and 4.6~Myr, respectively, in 
agreement with the fact we observe WR stars in them.

\subsection {Age of the underlying stellar populations\label{models}} 

We used the same Starburst99 models to estimate the age of the stellar populations underlying each 
\ion{H}{ii} region. We compared Starburst99 model tracks  for  various colours and metallicities 
with the colors we derived in each star-forming region and determine the age which best fits the observing
data with the predictions given by the models. Although the ages derived using this method have larger uncertainties
(typically, between 1 and 3 Myr for young stellar populations, and between 100 and 300 for Myr for old stellar populations)
than the ages determined from the H$\alpha$ equivalent widths, this exercise is
useful for discriminating between young ($\lesssim$25 Myr),
intermediate (100--300 Myr), and old ($>$500 Myr) stellar populations
\citep[e.g.][]{J99,ME00,Buckalew05,LSEGR06}.
However, the analysis has to be done with care, as the integrated optical colours of a star-forming region also depend on other factors.
\begin{enumerate}

\item {\bf Older stellar populations}. We should expect some disagreement between the models and the observational data, as
the star-forming regions would actually host mixed stellar populations, with stars which were created
in the last starburst event on top of an older stellar population. In the case of having a
bright or very recent star-formation event in a region which already has an old underlying population
we will get high  $W$(H$\alpha$) values, low $U-B$ colours but higher $B-V$ or $V-R$ colours than those expected using the models.
That is because 
absorptions in the H$\alpha$ line are expected not to be important in this case (the H$\alpha$ emission line is much stronger than the H$\alpha$ absorption line) and the old, reddish stellar population is not affecting much the emission in the $U$ filter but it considerably
affects $B$, $V$, $R$ and $I$ data, being specially important in the infrared filters. Indeed, $NIR$ data are usually needed to disentangle
the effect of the mixed (young/old) stellar populations in strong star-forming galaxies 
\citep[e.g.][]{Vanzi00,Vanzi02,N03,Noeske05,LSE08}.

\item {\bf Emission from the ionized gas}.  Theoretical models, as those used here, only consider the emission of the stellar continuum to obtain the broad-band colours, but the contribution of the emission lines coming from the ionized gas may be important in some cases.
The value of these corrections depend on several factors (position of the emission lines within the broad-band filters, metallicity and ionization of the gas, ratio between the region occupied  by the \ion{H}{ii} region and the area covered by the slit) and it is not easy to provide an {\it average} number. Nevertheless, we  used the values for $\sim$40 independent regions analyzed by \citet{LSE08} --see their Appendix~A-- and tabulated in their Table~6 to get some estimations of the correction of the broad-band colours for the emission of the gas according to the $W$(H$\alpha$) value. For $W$(H$\alpha$)$\sim-50$~\AA\ we find  $\Delta(U-B)=-0.01$, $\Delta(B-V)=0.03$ and $\Delta(V-R)=-0.02$~mag and hence for these cases the contribution should be within the errors. For $W$(H$\alpha$)$\sim-200$~\AA,  $\Delta(U-B)=-0.05$, $\Delta(B-V)=0.10$ and $\Delta(V-R)=-0.10$~mag. For these cases we should take into account the contribution of the emission lines, we do so increasing the uncertainties of the ages derived using the broad-band filters for these knots. Finally, in the case of $W$(H$\alpha$)$\sim-1000$~\AA, $\Delta(U-B)=-0.1$,  $\Delta(B-V)=0.6$ and $\Delta(V-R)=-0.2$~mag. We have some few knots with $W$(H$\alpha$)$\lesssim-1000$~\AA\ (see Table~\ref{halpha}).

\item {\bf Extinction}, which will move the broad-band colours to redder values. 
The intrinsic extinction can be determined from the $c$(H$\beta$) derived from spectroscopy, 
however here we use a different approach to estimate the internal extinction on each knot using the $U-B$ vs. $B-V$ diagram.

\end{enumerate}

\begin{figure*}
\centering
(a)\includegraphics[scale=0.44,angle=90]{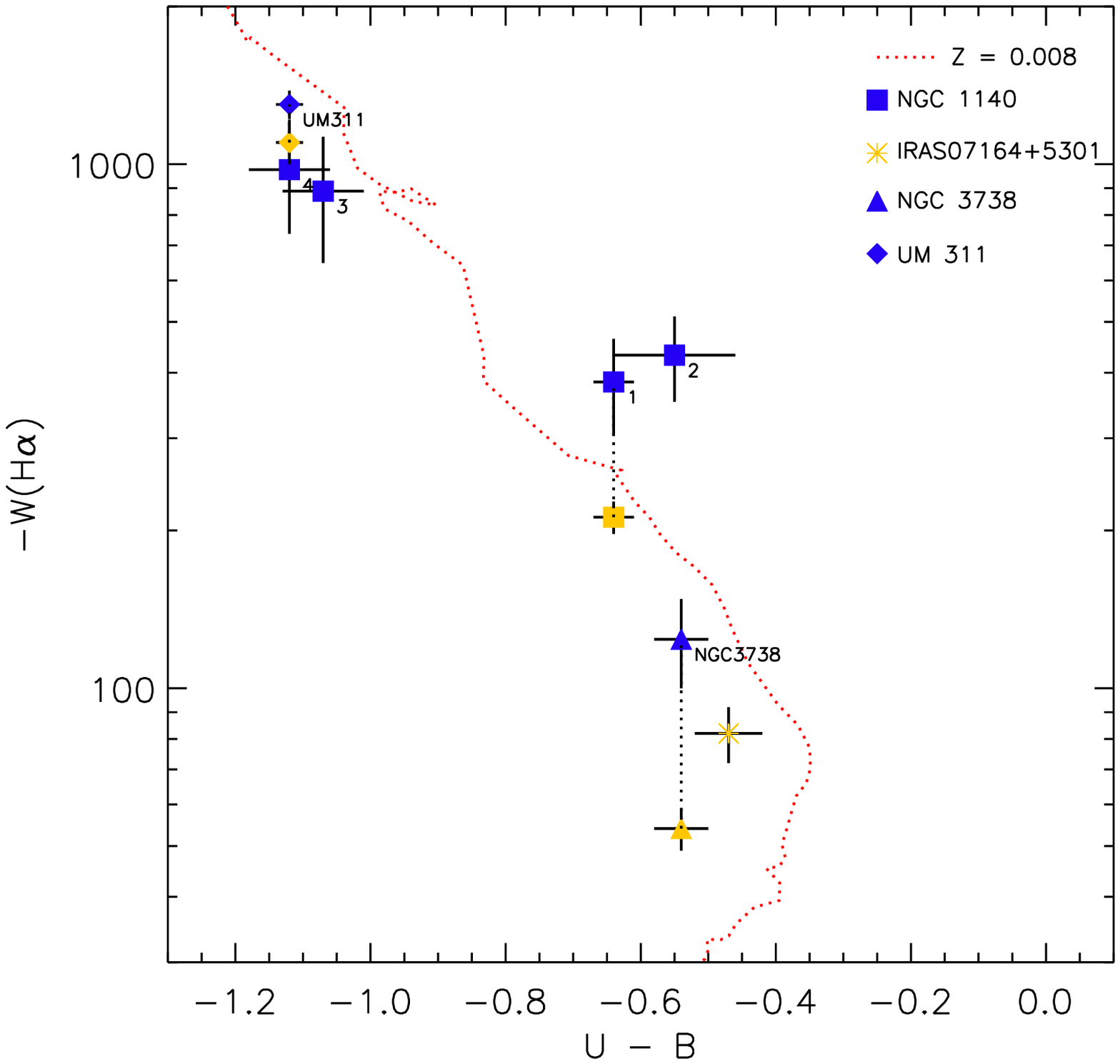}
(b)\includegraphics[scale=0.44,angle=90]{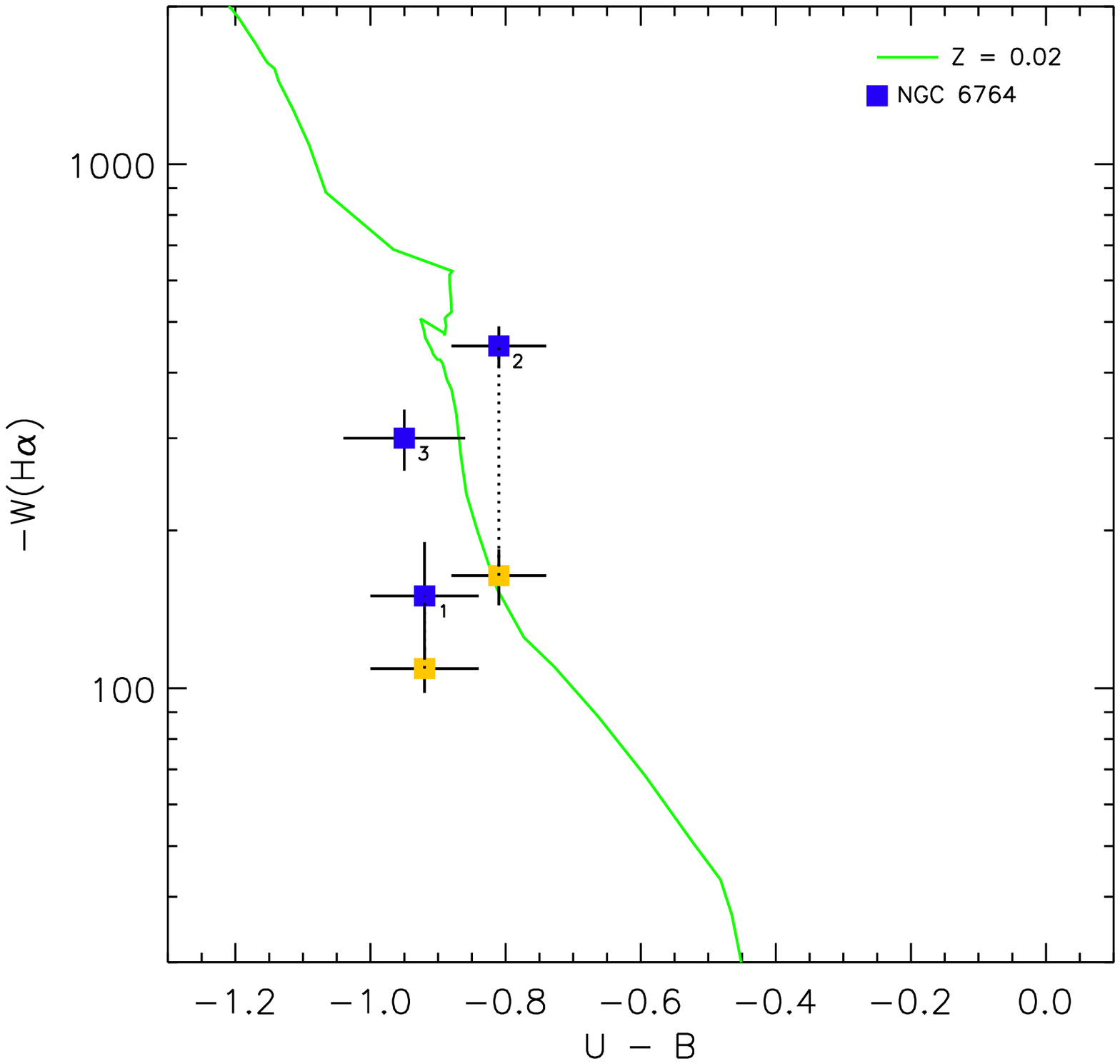}
(c)\includegraphics[scale=0.44,angle=90]{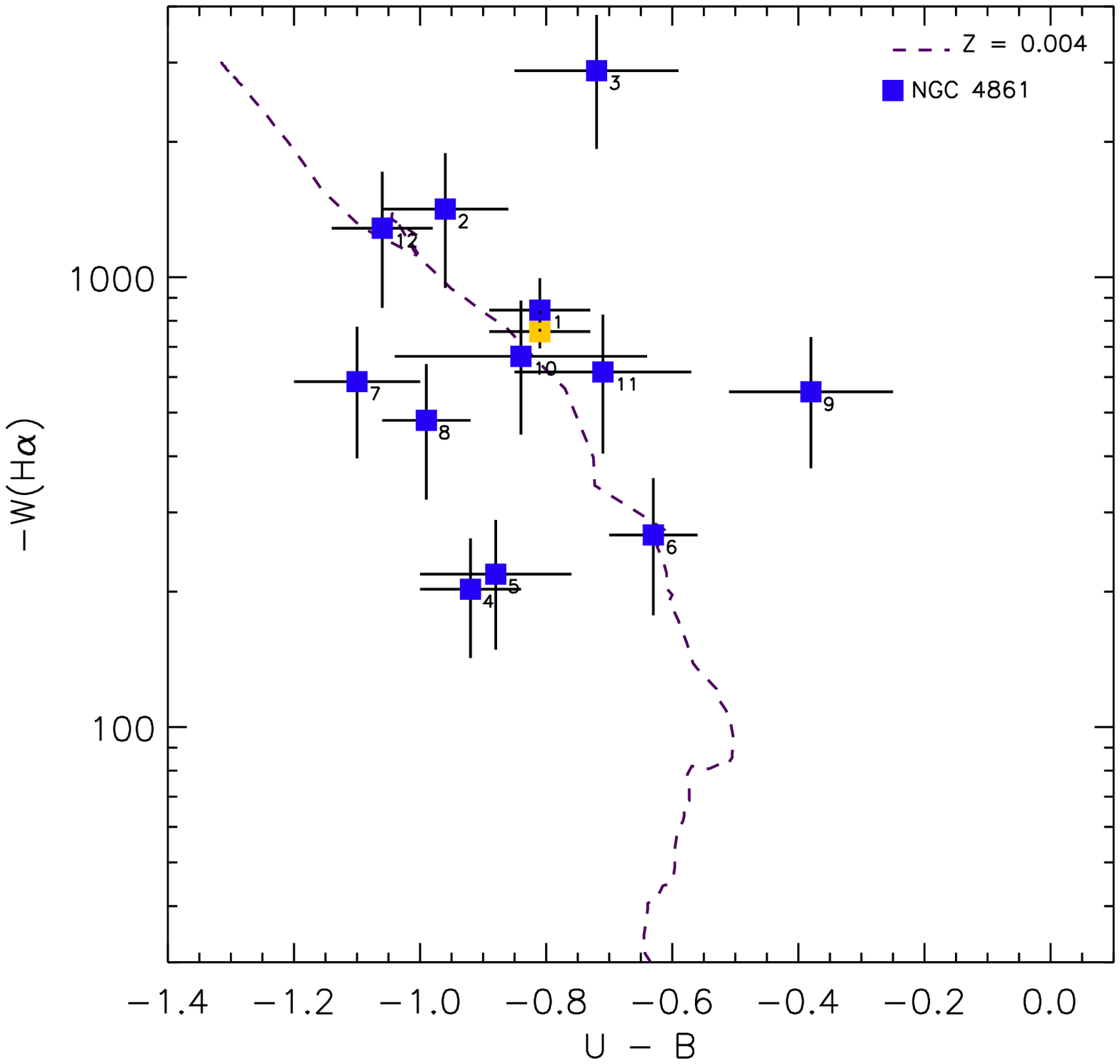}
(d)\includegraphics[scale=0.44,angle=90]{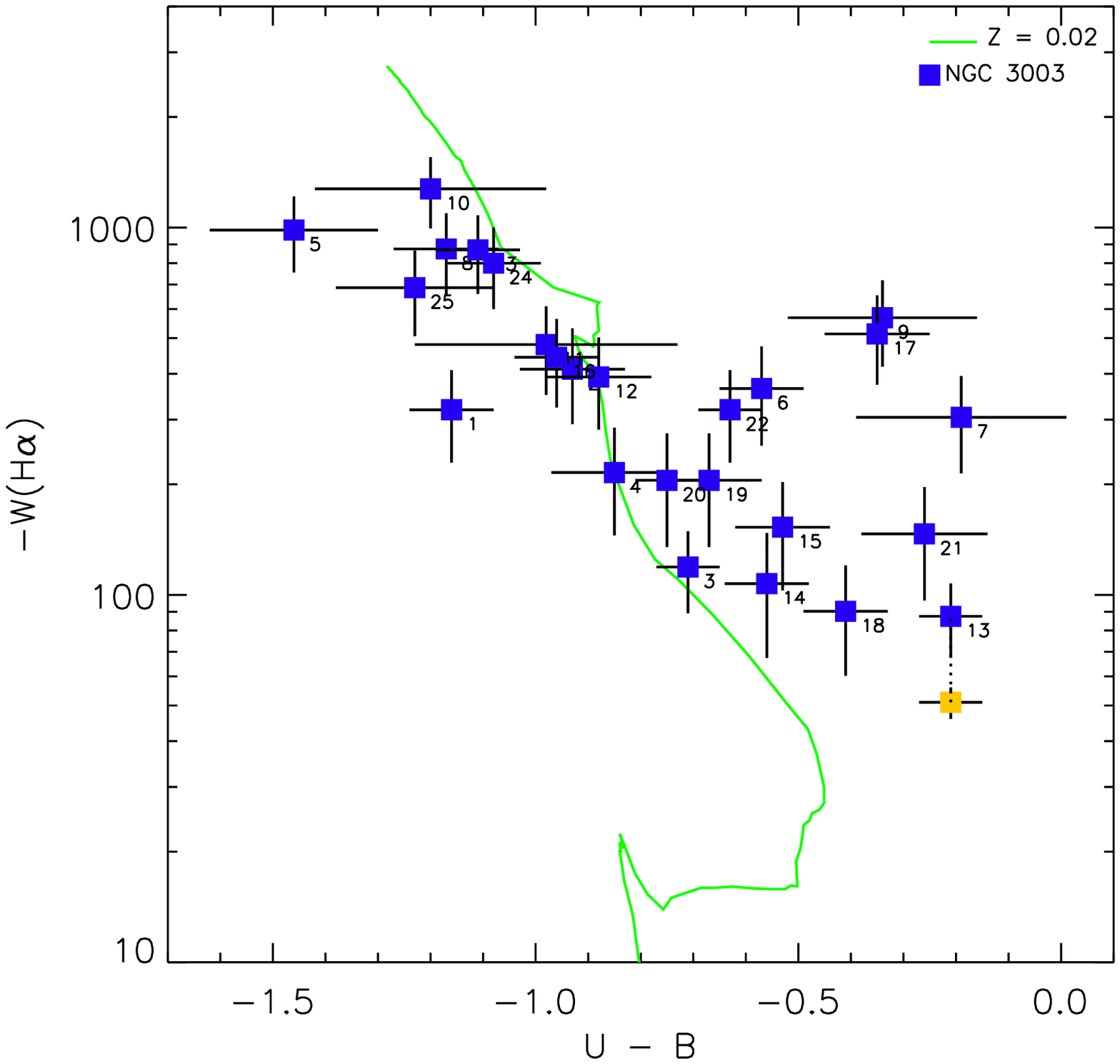}
  \caption{$W$(H$\alpha$)  vs. $U-B$ diagrams comparing the predictions given by
evolutionary synthesis models provided by the Starburst~99 code \citep{L99}
   with the colors obtained from the star-forming regions within
the WR galaxies of our sample. The metallicity of each model is shown.
   The yellow points indicate that the $W$(H$\alpha$) value is coming from our spectroscopic data (see Table~\ref{spectra}). Except for the case of IRAS~07164+5301, the yellow data points are connected to the values derived using our H$\alpha$ images for the same knot. \label{wha}}
  
\end{figure*}

Figure~\ref{sb99models} plots the Starburst99 model tracks  for the $U-B$ vs. $B-V$ color-color diagram; 
the  colours derived for our star-forming knots are  overlaid onto these model tracks.
As we previously specified, we have corrected our broad-band colors by Galactic extinction only using the \citet*{SFD98} data. 
Assuming that the offset between our data point and the models is due to the reddening, we can estimate the intrinsic extinction, $E(B-V)_{internal}$, of each knot. For this, we use
the reddening vector in the $U-B$ vs. $B-V$ diagram,  $(U-B)=-0.232$, $(B-V)=-0.337$, to move our data points to the color-color model track. 
The amount of the movement provides $E(B-V)_{internal}$, which we list 
in Table~\ref{ages}. Note that this cannot be done for all knots, as sometimes the movement of our data points following the reddening vector put them away from the theoretical track (e.g., knots \#5, \#10, \#11, \#14, \#19, and \#25 in NGC~3003). 
This is typically the case of faint regions where the uncertainties are large, but it may also be a consequence of the emission of the gas. 
As we described before, the effect of the emission lines in the $B-V$ color may be important, even +0.6 mag. 
This is what may be happening in UM~311, knots~\#1, \#2, \#3, and \#12 in NGC~4861 and  knots~\#5, \#10, and \#23 in NGC~3003, which show the lowest  $W$(H$\alpha$) values of our sample. In particular, it seems the case of knot~\#1 in NGC~4861, the strongest star-forming region in our sample (see Fig.~\ref{sfrawha}). The $E(B-V)_{internal}$ value we computed for this knot using our spectrum is $\sim$0.03~mag, but we need and offset of $\sim$0.5~mag to account its position in Fig.~\ref{sb99models}c for reddening.
On the other hand there are 2 regions with relatively low  $W$(H$\alpha$) values --knots \#3 and \#4 in NGC\,1140, which have  $W$(H$\alpha$)$\sim-900$\,\AA-- that show a good agreement with the models besides we do not correct the observed colors for the emission of the gas. Perhaps in these two very young star-forming regions there is a contribution of both internal reddening (which moves the $B-V$ color to redder values) and emission of the gas (which moves the $B-V$ color to bluer values), being null the net contribution of both effects. 
As the maximum variation of the $B-V$ color because of the emission of the gas is  $\Delta(B-V)\sim0.4$, we should expect that the highest value of the internal reddening in these two regions is $E(B-V)_{internal}\sim$0.4~mag. However, spectroscopic data are needed to confirm this hypothesis.

Furthermore, we consider that the offset between the position of knot~\#13 of NGC~3003 (the center of the galaxy) and the theoretical model is mainly because of the presence of a very important old stellar population underlying this star-forming region. Indeed, this knot has a high $W$(H$\alpha$) value, -90~\AA, hence we expect that the emission of the gas is negligible. On the other hand, the  $E(B-V)_{internal}$  value derived using our spectrum, $\sim$0.03~mag, is too low in comparison with the offset of $\sim$0.6~mag needed to match with the model following the reddening vector. Finally, the circular aperture used for this knot  (see Fig.~\ref{iNGC3003}) actually considers not only the starburst but a large non star-forming region probably dominated by older stellar populations. 

In the case of NGC~3738 we derive a $E(B-V)_{internal}$ = 0.25~mag following the $U-B$ vs. $B-V$ color-color diagram. However, for this galaxy we adopted $E(B-V)_{internal}$=0~mag using our spectroscopic data. The starburst is located in the center of this galaxy, which hosts an extended disk-like structure  (see Fig.~\ref{NGC3738image}) composed by older stars. Hence, we suspect there is also an important contribution of this old stellar population to the optical broad-band colours of the star-forming region.

The values for the intrinsic extinction calculated using our photometric data of knot \#1 of NGC 1140, IRAS~07164+5301, and knot \#1 of NGC 6764 (0.0$\pm$0.3 mag in all cases)
match well the values derived from our optical spectroscopy data (adopted 0~mag for the first object, and derived 0.041$\pm$0.014~mag and  0.014$\pm$0.007~mag for the second and third knot). 
This also suggests that the correction for the emission of the gas is not critical in these objects.

Once the intrinsic reddening is determined, we estimate the age of the young stellar population 
from the reddening-corrected $U-B$ color and compared with that derived from the $W$(H$\alpha$). Second last column in Table~\ref{ages} compiles the age adopted for each burst, which has a typical uncertainty of 1~Myr.

We further check the agreement between the data and ages derived using the narrow-band H$\alpha$ images and the broad-band optical colours. Figure~\ref{wha} compares the $W$(H$\alpha$) obtained using our H$\alpha$ images with the $U-B$ color derived for each knot. We also plot the predictions given by
evolutionary synthesis models provided by the Starburst~99 code for several metallicities, as well as the  $W$(H$\alpha$) obtained using our spectroscopic data (yellow points). We find a relatively good agreement between all measurements. However in the cases of NGC~4861 (knots \#3 and \#9) and NGC~3003 (knots \#7, \#13, \#17, and \#21) we find regions which are located somewhat far from the models. The reason of this discrepancy is the effect of an important intermediate-old stellar population underlying the bursts.

Finally, we use again the predictions given by the Starburst~99 models and the values obtained for the rest of colors to get an estimation of the age of the dominant stellar population underlying the starburst of each knot. For this we specially consider the results coming from the $V-R$ and the $V-I$ colors. The ages adopted for the underlying stellar populations are compiled in last column of Table~\ref{ages}. As we see, knots in NGC~3738, NGC~6764 and NGC~3003 usually have old underlying stellar populations, with ages even older than 500~Myr in many cases.  However, the rest of the galaxies (NGC~1140, IRAS~07164+5301, UM 311 and NGC~4861) host relatively young underlying stellar populations (20 - 50~Myr), which suggests that the strong star-formation activity observed in these galaxies has been ongoing for that period.

\begin{figure*}
\centering
(a){\includegraphics[scale=0.44, angle=90]{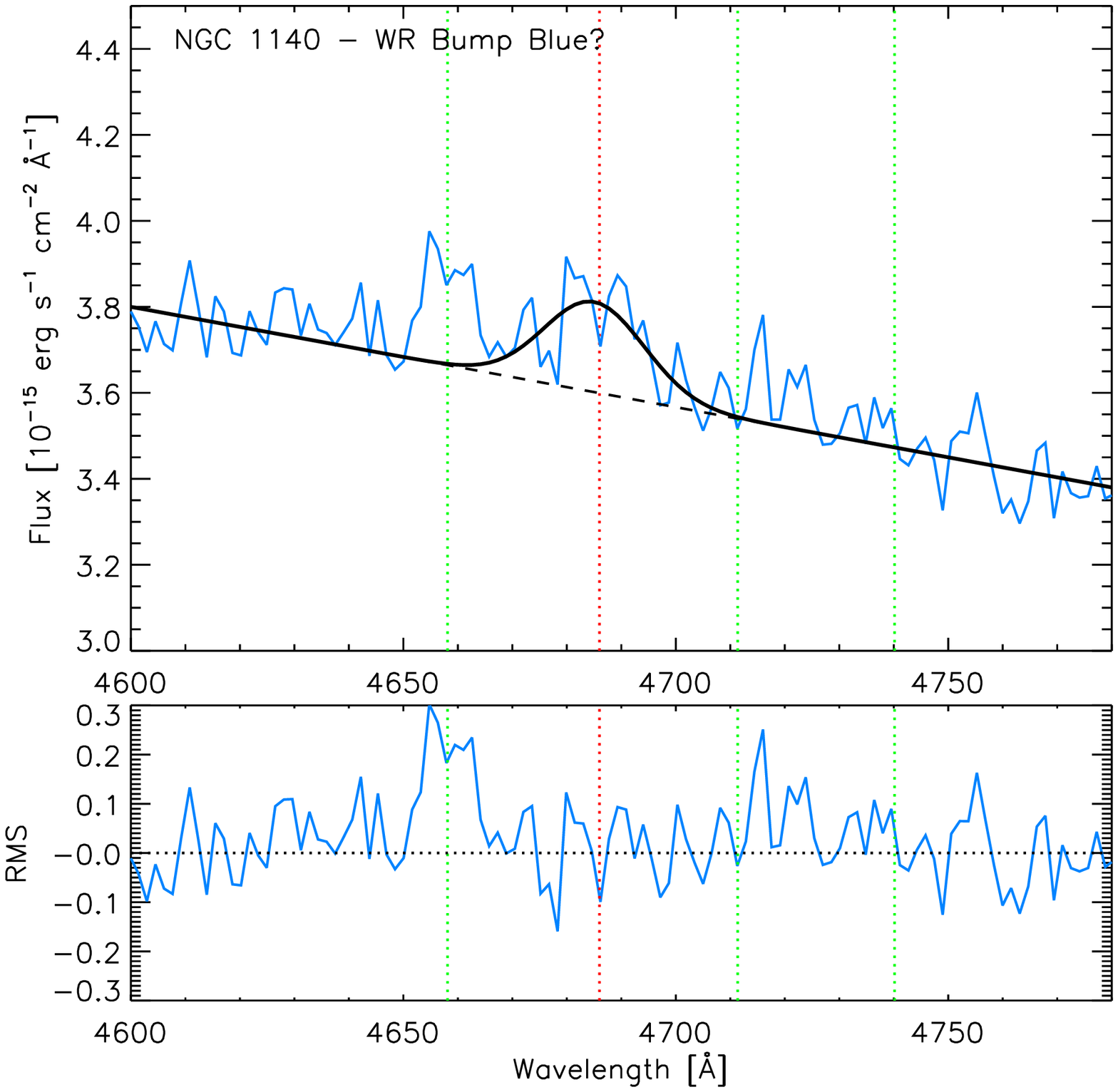}}
(b){\includegraphics[scale=0.44, angle=90]{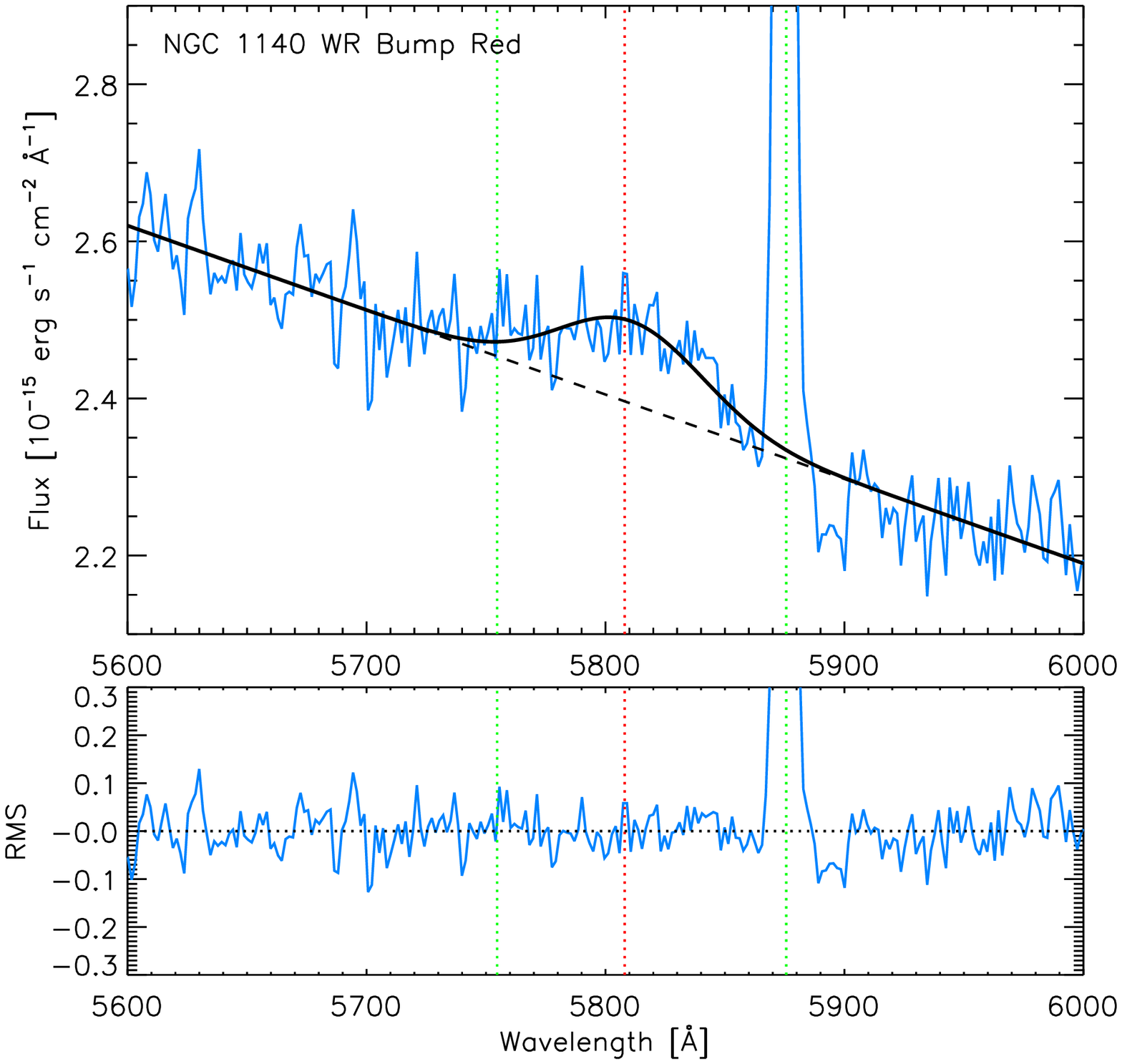}}
(c){\includegraphics[scale=0.44, angle=90]{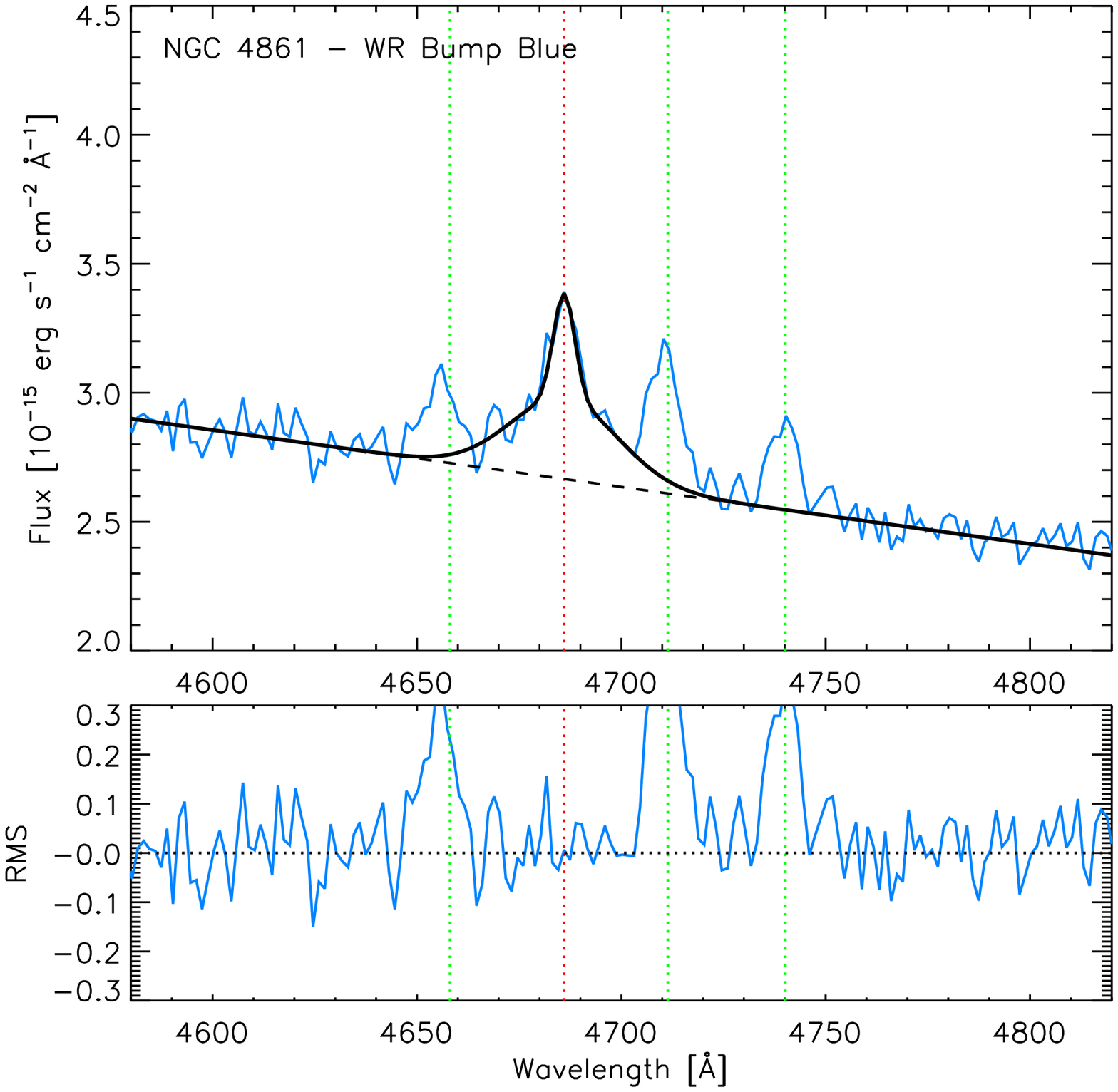}}
(d){\includegraphics[scale=0.44, angle=90]{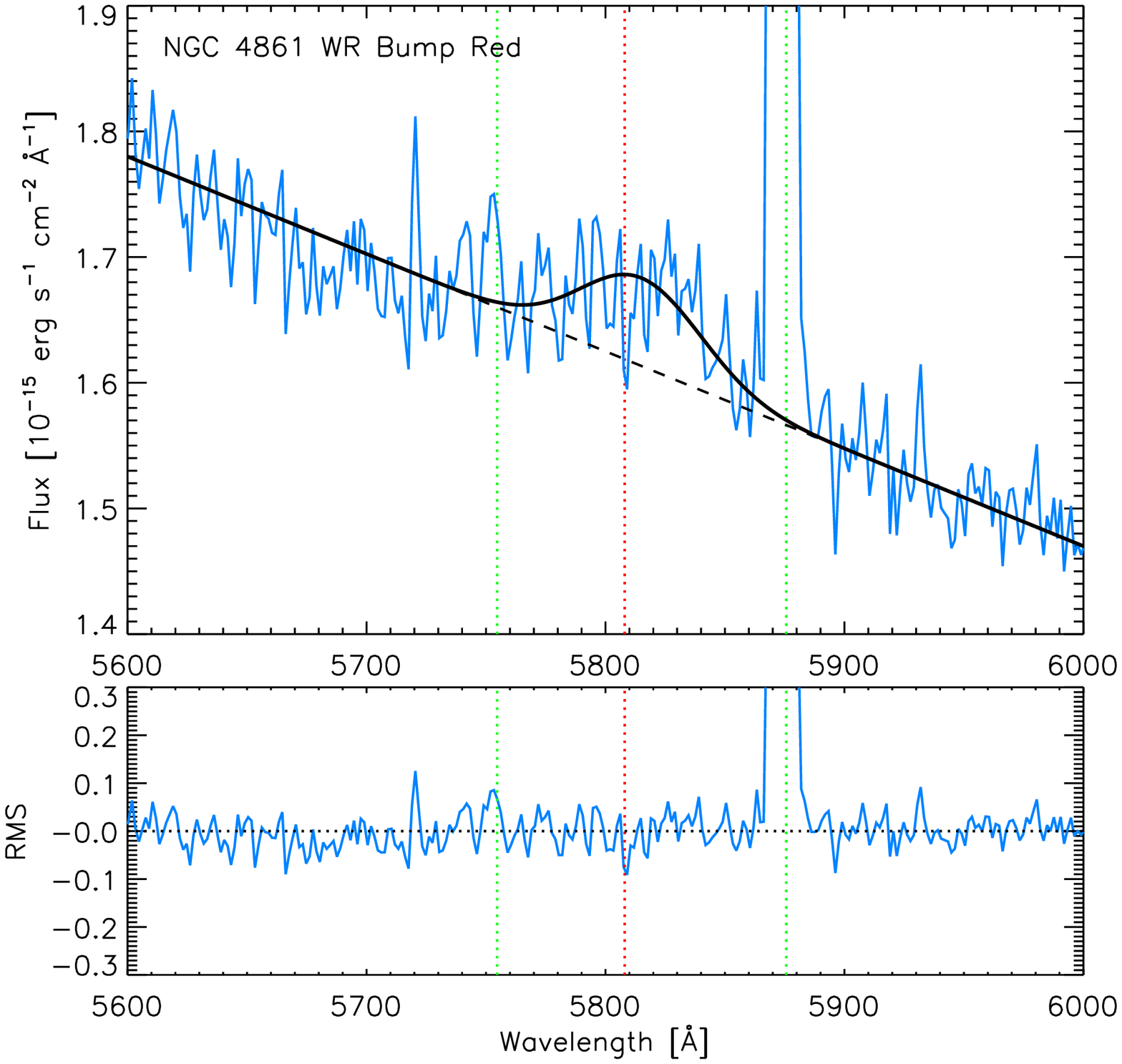}}
 
  \caption{Details of the optical spectra of NGC~1140 (top row) and NGC~4861 (bottom row) showing the broad emission lines of the Blue WR Bump (left) and the Red WR Bump (right). In the case of the low-metallicity galaxy NGC~4861, the nebular, narrow \ion{He}{ii}~$\lambda$4686 (dotted red line) is also clearly detected, as well as the [\ion{Fe}{iii}]~$\lambda$4658 and [\ion{Ar}{iv}]~$\lambda\lambda$4711,4740 emission lines (dotted green lines) in the area of the blue WR bump. The positions of the \ion{C}{iv}~$\lambda$5808 emission (dotted red line) and the [\ion{N}{ii}]~$\lambda$5755 and \ion{He}{i}~$\lambda$5875 emission lines have been also included in the panels showing the red WR bump.
In all cases the black dashed line represents the continuum fit. The best fit to the observed spectrum (blue continuous line) is shown with a black continuous line. The bottom panel below each diagram shows the residual spectrum after subtracting our fit model to the observed spectrum.
    \label{spectrawr} }
\end{figure*}

\subsection{Analysis of the WR features}

A magnified view of the spectrum of NGC 1140 around
4650\,\AA\ is presented in the top-left panel of Fig.~\ref{spectrawr}. The blue WR bump at around 4686~\AA,  which was 
previously  reported by \citet{GIT00}, 
is  identified in this spectrum, although with a low SNR.
However we clearly identify the red WR bump around \ion{C}{iv}~$\lambda$5808 in this object, as it is seen in the top-right panel of Fig.~\ref{spectrawr}.
We detect both WR features in the spectrum of the low-metallicity galaxy NGC~4861. In this case, both the broad and the narrow  \ion{He}{ii}~$\lambda$4686 emission lines are observed (see bottom row in Fig~\ref{spectrawr}).

In order to quantify the fluxes of the WR features detected in these two galaxies, we fitted a broad and a narrow Gaussian for the
stellar and nebular WR lines for each WR bump. In the case of the red WR bump we only considered a broad \ion{C}{iv}~$\lambda$5808 component as any narrow emission is observed. These fits are included in Fig.~\ref{spectrawr}, which also shows the residual spectrum after subtracting our fit model to the observed spectrum. The derived fluxes for the blue and red WR bumps in each case --which considers only the broad component-- have been compiled in Table~\ref{spectra}.

We followed the procedure described in \citet{GIT00}
to derive the number of WR stars and the WR/(WR+O) and WCE/WNE ratios in NGC~1140.
Using the values for the flux of the broad \ion{He}{ii}~$\lambda$4686 and \ion{C}{iv}~$\lambda$5808 emission lines 
derived from our fits (see Table~\ref{spectra})
and considering the metallicity-dependence of the WR luminosities --Eqs.~7 and 8 in \citet{LSE10a}, which considered both the broad-line WNL and WCE
luminosities given by \citet{CH06} for solar and $Z_{\odot}$/50 metallicites--,
we derive WNL = 92$\pm$36 and WCE = 59$\pm$15. From the total luminosity of the H$\beta$ line and considering their Eq.~10, we find that the total number of O stars in the burst is 1553$\pm$300. For this, the contribution of the WR
stars and other O stars subtypes to the ionizing flux must be considered.
This is done via the the $\eta_0(t)$=O7V/O parameter \citep{VC92,V94,SV98},
which depends on the age of the burst.
Considering 
the age of the most recent star-formation event ($\sim$5.0~Myr, see Table~\ref{ages})
and the models provided by \citet{SV98}, we assumed $\eta_0(t)\sim0.25$ for this object.
We then derive a WR/(WR+O) ratio of 0.089$\pm$0.034 and a WCE/WNL ratio of 0.64$\pm$0.30. 
Both values agree well with the ratios expected for an object with the oxygen abundance of NGC~1140. 
We note that our estimations of the WR/(WR+O) and WCE/WNL ratios are similar, within the errors, than those derived by \citet{Moll07} using HST data, WR/(WR+O) = 0.11 and WCE/WNL=0.36. However, their estimations of the total number of O and WR stars are around 4 times the values we derive here. The reason of this discrepancy is that we are probably observing a slightly different area within the galaxy and using a different aperture size.

We repeat this procedure for NGC~4861, a low-metallicity galaxy also showing both the  broad \ion{He}{ii}~$\lambda$4686 and \ion{C}{iv}~$\lambda$5808 emission lines. Note that for this object we subtracted the flux of the nebular, narrow \ion{He}{ii}~$\lambda$4686 emission line to get a proper estimation of the flux of the broad \ion{He}{ii}~$\lambda$4686 emission line.
In this case, we derive WNL = 225$\pm$35,  WCE = 67$\pm$30 and O = 4336$\pm$1200, assuming the appropriate oxygen abundance for the WR luminosities and $\eta_0(t)\sim0.25$ considering that the age of the most recent starburst is  $\sim$4.6~Myr, see Table~\ref{ages}. Hence, we estimate that the WR/(WR+O) and the WCE/WNL ratios for NGC~4861 are 0.062$\pm$0.021 and 0.30$\pm$0.12, respectively. Both the values are also in agreement with the WR ratios expected for a galaxy with its metallicity.    

Although the center of the galaxy NGC~6764 also shows some broad emission lines around the \ion{He}{ii}~$\lambda$4686 emission feature which may be attributed to WR stars, the fact that the ionization nature of this object is not photoionization (see Sect.~4.1) preclude us to derive any realistic estimation of the number of WR stars in this region.

\subsection{Analysis of individual galaxies}

\subsubsection{NGC\,1140}

 NGC~1140 is a SbPec  galaxy located at a distance of 17.9~Mpc.
 It is an object showing blue colors, \mbox{$B-V=0.01\pm0.02$~mag,} 
 and intermediate metallicity, 12+log(O/H)=8.38$\pm$0.10.
 \citet{Hunter94a} 
 used optical broad-band images, H$\alpha$ data, optical spectroscopy 
 and neutral hydrogen observations to identify the  
 central  giant star forming
 region and a  chain of other star forming  regions coinciding with the
 low light  level extension in the  south-west tail of  the galaxy (see Fig.~\ref{NGC1140image}). 
 Later, \citet{Hunter94b} used HST data to show that the central  
 region  consists  of a  supergiant  \ion{H}{ii}  region  and few  super-star
 clusters  with sizes  of  $<$ 10~pc.
 With  our ground-based images  it is not possible to  resolve the  individual
 clusters observed by the HST in  the central  giant  \ion{H}{ii} region  and  hence
   it has  been considered as a single star forming region (knot \#1).  
 The other star
 forming knots numbered  \#2, \#3 and \#4, which  are in the south-west
 part of the galaxy, also show  very blue colours. In fact knots \#3 and \#4 
 are bluer than the central knot. 
 The  bluer colour of the outer body  of the galaxy indicates that  extensive star formation
 is  going  on  throughout  this  galaxy.  
Indeed, using the optical colours we are not able to detect stars older than 25-50~Myr underlying the starburst.
The age of the most recent burst decreases from the center ($\sim$5.0~Myr) to the external regions ($\sim$3.2~Myr) of NGC~1140, suggesting that the arc-like plume where they are located has been originated very recently.

 WR features were previously detected in NGC~1140 by 
 \citet*{GIT00} and \citet{Moll07}. 
 We certainly detect both the blue and red WR bumps (see Fig.~\ref{spectrawr}), being this last one specially prominent. 
 As it was discussed before, 
 the derived WR properties within NGC~1140 agree well with those expected for a galaxy with its metallicity
 \citep*{GIT00,LSE10a}.

 The  star  formation  rates
 calculated using  the H$\alpha$ flux  for the four knots  reveal that
 the central region (knot~\#1)  is undergoing an intense starburst,
 while the  other three star forming  knots have a moderate rate of star
 formation.  
 Our estimation of the SFR within this galaxy,  0.65~$M_{\odot}$/yr,
 which agrees well with those determined by  \citet{Hunter94b} --$0.8~M_{\odot}$/yr--
 and  \citet{Moll07} --$0.7\pm0.3~M_{\odot}$/yr--.
FIR and 20-cm radio-continuum data are available for this galaxy. Using the flux densities for 60~$\mu$m and 100~$\mu$m
provided by IRAS \citep[{\it Infrared Astronomical Satellite},][]{Moshir90} 
and applying the relations provided by \citet{Condon92} and \citet{K98}, 
we derive SFR$_{60\, \mu m}$=0.25~$M_{\odot}$\,yr$^{-1}$ and SFR$_{\rm FIR}$=0.30~$M_{\odot}$\,yr$^{-1}$, respectively. 
On the other hand, considering 
the 1.4~GHz flux for this galaxy provided by \citet{Hunter94a} and the \citet{CCB02} 
calibration, we derive SFR$_{\rm 1.4\,GHz}$=0.20~$M_{\odot}$\,yr$^{-1}$. 
Hence, the SFR values derived using the FIR and radio data are 2--3 times lower than our H$\alpha$-based SFR.

Considering the enhancement of SF activity in its center and its peculiar optical and \ion{H}{i} morphology.
\citet{Hunter94a,Hunter94b}  concluded that NGC~1140 is undergoing recent violent disturbances which
may be explained assuming a merger of two low surface brightness galaxies. 
Our new data agrees with this scenario. Hence, NGC~1140 seems to be another example
of a WR galaxy in which starburst has been triggered by galaxy interactions, 
as it was found in the majority of the objects analyzed by \citet{LS10}.

 \subsubsection{IRAS\,07164+5301}

 \textbf{IRAS 07164+5301} is known as an extreme starburst source \citep{Allen91}. 
 The optical spectrum of the galaxy was previously studied by \citet{Huang96} 
 but the broad-band colours of IRAS~07164+5301 are analyzed here for the first time.
 We report blue colours in this galaxy, $U-B=-0.47\pm0.05$, $B-V=0.06\pm0.04$, 
 $V-R=0.06\pm0.05$,  and $V-I=0.30\pm0.02$.
 The comparison of the derived optical colours with the stellar population synthesis models 
 and  the analysis of the equivalent width of the H$\alpha$ emission line
 provide
 an age of $\sim$6.5~Myr for the most recent star-formation event.
 Interestingly, the optical colors do not show the presence of an important old stellar population underlying the starburst.
 The oldest age we derive for the dominant stellar populations in this galaxy is $\sim$25~Myr. 
 The youth of the starburst agrees with the detection of WR features 
 in the optical spectrum of the galaxy analyzed by \citet{Huang96}, who reported 
 the presence of broad lines around 4686~\AA, suggesting the presence of \ion{N}{iii}~$\lambda$4640, 
 \ion{C}{iii}~ $\lambda$4650 and \ion{He}{ii}~$\lambda$4686.
 These authors also indicated a tentative detection of \ion{O}{v}~$\lambda$5835 and
 a lack of \ion{C}{iv}$\lambda$5808. However, we do not detect any of these features, having our spectrum
 a similar or even better quality than the spectrum presented by  \citet{Huang96}.
  
We derive an oxygen abundance of 12+log(O/H)=8.50 for IRAS~07164+5301.
This value is much lower than the metallicity obtained by \citet{Huang96}, 
who found 12+log(O/H)=8.96 using the direct method. However, their estimation of the [\ion{O}{iii}]~$\lambda$4363 flux is very probably overestimated,
and even its detection is not clear because of both the relatively low SNR and spectral resolution (4.7~\AA/pix) of their spectrum.
  Taking into account that we have estimated the oxygen abundance of 
this object using several parameters ($N_2$, $N_2O_3$, $R_{23}$, $P$) and calibrations, all giving similar
values (except for the caveat of those methods based on photoionization models, as we discussed before)
we consider that our metallicity estimation of IRAS~07164+530 is more appropriate than that obtained by
\citet{Huang96}.

Using our data we compute a radial velocity of $V_r$=12,981~km\,s$^{-1}$ (redshift $z=0.0433$)
for IRAS~07164+5301. Hence this galaxy lies at a distance of 177~Mpc.
The high radial velocity this galaxy possesses is the reason we cannot detect its \Ha\ emission using the 
rest-frame \Ha\ filter. However, we used the flux of the \Ha\ line observed in our optical spectrum
to estimate a SFR of $\sim$7.6~$M_{\odot}$\,yr$^{-1}$.
Using the FIR and 20 cm radio-continuum data available for this galaxy,
we derive SFR$_{60\, \mu m}$=8.3~$M_{\odot}$\,yr$^{-1}$,
SFR$_{\rm FIR}$=8.1~$M_{\odot}$\,yr$^{-1}$, and
SFR$_{\rm 1.4\,GHz}$=5.7~$M_{\odot}$\,yr$^{-1}$. 
The 1.4\,GHz data were provided by \citet{Condon98}. 
All these values are in excellent agreement with the SFR we have estimated here using the H$\alpha$ emission.

 \subsubsection{NGC\,3738}

\textbf{NGC~3738} is classified as an irregular \citep{deVaucouleur91}, 
low-metallicity galaxy.  As it has a very low radial velocity, $V_r$=229~km\,s$^{-1}$, it is
not easy to estimate the distance to this galaxy., and hence 
The best estimations of its distance are ranging  from 4.0 to 5.4~Mpc 
\citep{Hunter82,HunterGR82,HunterHoffman99}.
The distance value listed in Table~\ref{Table1} for this galaxy, $d$=5.56~Mpc,  comes just from the value of the Hubble constant.
However, the most recent distance estimation for this galaxy using the luminosity of the tip of the red giant branch stars
is 4.90~Mpc \citep{Karachentsev2003}. 

The optical spectrum of NGC 3738 resembles those of \ion{H}{ii} regions 
with emission dominated by massive, hot stars. \citet{Martin97}  
points out the presence of a broad \ion{He}{ii}~$\lambda $4686 emission line in the integrated 
spectrum of this galaxy, and hence  NGC~3738 was also listed as WR galaxy by \citet*{SCP99}. 
However, our optical spectrum does not show any faint WR features.

We used the bright
emission lines to estimate the metallicity of the galaxy, 12+log(O/H)=8.31. This value was obtained
averaging the results provided by both the high- and low-branch metallicity calibrations and agrees within the uncertainties
with the result obtained by  \citet{Martin97}, which was 12+log(O/H)=8.23.

NGC 3738 has a bright star forming region in center, for which we derive a SFR of 0.035$\pm$0.003~$M_{\odot}$\,yr$^{-1}$. 
Our result agrees well with the SFR estimation provided using the H$\alpha$ flux provided by \citet{Kennicutt08}
for this galaxy, 0.030~$M_{\odot}$\,yr$^{-1}$.
Using the FIR fluxes available for this galaxy and the same equations cited in the previous subsection we derive SFR$_{60\, \mu m}$=0.015~$M_{\odot}$\,yr$^{-1}$ and SFR$_{\rm FIR}$=0.018~$M_{\odot}$\,yr$^{-1}$. 

These values are just slightly lower that the SFR derived using the H$\alpha$ images.

The age of the most recent star-formation event we estimate for this galaxy using both the $W(H\alpha)$ and the $U-B$ color is 6.0~Myr. 
However, from the $B-V$ and, specially, $V-R$ and $V-I$ colors it is clear that the galaxy possesses a very important old stellar population underlying the starburst, for which we estimate an age of $\sim600$~Myr.

  \subsubsection{UM\,311}
 
\textbf{UM~311} is an intriguing extragalactic \ion{H}{ii} region which shows a radial velocity of $V_r$=1675~km\,s$^{-1}$.
Therefore the distance to this galaxy is 18.7~Mpc.
An essential note in NED defines UM~311 as an ``\ion{H}{ii} region in the  GPair CGCG~0113.0-0107".
Our optical images (see Fig.~\ref{UM311image}) clearly show two galaxies in apparent interaction, 
a spiral galaxy (NGC~405) and a dwarf elliptical at its NE (UGC~807), which have
radial velocities of 1,761~km\,s$^{-1}$ and 11,431~km\,s$^{-1}$. Therefore, these
two galaxies are not physically associated. Because of the very similar radial velocity, 
UM~311 may indeed be an  \ion{H}{ii} region within NGC~405. However, 
its observed properties (blue colours, compactness and intensity of the star-formation activity)
does not discard the possibility that it is an independent Blue Compact Dwarf Galaxy (BCDG) which is interacting
with the spiral galaxy NGC~405. A similar example of a BCDG interacting with a spiral galaxy is the
stunning NGC 1512 / NGC 1510 system \citep{KoribalskiLS09}.
Interferometric \ion{H}{i} data are needed
to completely elucidate this issue.

The first study reporting the presence of WR stars in UM~311 was presented by \citet{Masegosa91},  
who detected the broad \ion{He}{ii}~$\lambda$4686 emission line. 
Later many authors have analyzed the WR content on this object. \citet{IT98} 
confirmed the presence of both the blue and red WR bumps. 
\citet*{GIT00} noticed that the blue WR bump was particularly strong. 
\citet*{Pindao99,Buckalew05,Zhang07} and \citet{BKD08} 
also included UM~311 in their studies of WR galaxies. However, the poor quality
of our optical spectrum does not allow to detect the faint broad features attributed to WR stars.
We find that the age of the most recent starburst is 3.2~Myr, however using the optical colours 
we are not able to observe the underlying old stellar population. This fact indicates the strength of the starburst in UM~311.

UM~311 has been also used in the analysis of the chemical abundances of low-metallicity extragalactic \ion{H}{ii} regions
\citep[e.g.][]{IT98,IT99,Izotov04,Izotov06,P01a}
and hence its oxygen abundance has been determined
with good precision, 12+log(O/H)=8.31$\pm$0.04 \citep{IT98}. Using several empirical calibrations 
and the data provided by our optical spectrum, we estimate that the oxygen abundance of UM~311 is
12+log(O/H)=8.27, in excellent agreement with the previous estimations.

Using our H$\alpha$ images, we derive a SFR of 0.065$\pm$0.005~$M_{\odot}$\,yr$^{-1}$. This value is very low
when it is compared with the SFR derived by \citet{Hopkins02} using  FIR and radio-continuum data, 
SFR$_{60\, \mu m}$=1.5~$M_{\odot}$\,yr$^{-1}$ and SFR$_{\rm 1.4\,GHz}$=1.1~$M_{\odot}$\,yr$^{-1}$.
However, we consider that our SFR estimation
is more appropriate, as both FIR and radio-continuum images do not enough spatial resolution to 
resolve UM~311 within the spiral galaxy NGC~405.

 \subsubsection{NGC\,6764}
 
\textbf{NGC\,6764} is barred spiral (SBb type) galaxy (see Fig~\ref{NGC6764image}) 
which is also classified as a classical LINER galaxy  \citep*{AlonsoHerrero00}. 
Indeed, using  optical emission line ratios
we confirm that the nucleus of this galaxy lies in the LINER region of the BPT diagram  (Fig.~\ref{diagnostic}). 
\citet{OsterbrockCohen82}  point out
the presence of broad \ion{N}{iii}~$\lambda$4640 and \ion{He}{ii}~$\lambda$4686 emission lines in the
spectrum of the nucleus of NGC~6764. These features were later reported by 
\citet*{Eckart96,GIT00} and \citet{FCCG04}. 
 \citet{OsterbrockCohen82} also reported
a signal excessive  widths of \ion{He}{i}~$\lambda$5876 and H$\alpha$ which they attributed
to emission from WR stars. Indeed, broad \ion{C}{iii}~$\lambda$5696 and \ion{C}{iv}~$\lambda$5808 lines 
from WC stars were discovered by \citet{FCCG04}. 
Our optical spectrum confirms the broad features in the blue WR bump, although the existence of the red WR bump is not 
clear (see Fig~\ref{spectrafig2}).
However, as the photoionization is not coming from massive stars, we cannot
compute the WR content in this object.

From our H$\alpha$ images of NGC\,6764 we identify 3 regions showing ionized gas emission: the center of the galaxy 
(knot~\#2) and the regions located at the end of the bar (knot~\#3 at the east; knot~\#1 at the west). 
The starburst activity is NGC~6764 was first observed by 
\citet{Eckart91,Eckart96}, 
who revealed a dense concentration of molecular gas and very recent (few tens of Myr) starburst at the nucleus of NGC~6764. 
Indeed, we estimate an age of  $\sim$15~Myr for the dominant stellar population in the center of the galaxy, although the $U-B$ color and the 
$W$(H$\alpha$) suggest that the age of the most recent starburst is  $\sim$6~Myr. 
This value agrees with the result obtained by \citet{Leon07}, 
who concluded that the nuclear starburst is 3-7~Myr old using their analysis of
the interplay between the central activity and
the molecular gas.
However, knots~\#1 and 3 show an important 
stellar population underlying the bursts, with has an age of $\sim$400~Myr.

We cannot compute the oxygen abundance of the nucleus of the galaxy because its LINER nature. However, we
derive a metallicity of 12+log(O/H)=8.65 for knot~\#3 using several independent empirical calibrations. So far,
this is the first gas-phase metallicity estimation reported in this galaxy.

 \subsubsection{NGC\,4861}

\textbf{NGC\,4861} 
is classified as a Magellanic irregular galaxy \citep{SandageTammann81} 
and a BCDG because of its blue colours and high UV continuum 
\citep{French80,ThuanMartin81}. 
This object is located at a distance of 12.9~Mpc.
NGC~4861 has a comet-like morphology (see Fig~\ref{NGC4861image}) with 
many star-forming regions located along its major axis.
The morphology of the galaxy was discussed by \citet{Dottori94} 
who concluded that NGC~4861 might have undergone a merger process.
From \ion{H}{i} maps, this galaxy is an edge-on and rotating disk system \citep{Conselice00}.
Some authors
have distinguished the bright star-forming region at its south (our knot~\#1)
as NGC~4861 and the dwarf irregular galaxy IC~3961 (the rest of the galaxy)
and recommended the definition of Mkn~59 to describe the full system.
In any case, NGC~4861 has been amply studied as an example 
of a compact blue galaxy
\citep[e.g.][]{KS98,Izotov97,IT99,Lee04,Esteban09}.

Since the discovery of the broad \ion{He}{ii}~$\lambda$4686 emission line in
the spectrum of this galaxy \citep{DinersteinShield86}, 
many studies have confirmed the presence of WR stars in NGC~4861
\citep*{SCP99}.
The broad \ion{C}{iv}~$\lambda$5808 emission  has been also detected in NGC~4861 \citep*{Izotov97,GIT00}.
Our optical spectrum confirms the presence of these broad WR features. 
We derive WR/(WR+O)=0.062 and WCE/WNL=0.30, in agreement with
previous estimations.

The most recent analysis of the ionized gas within this object
was presented by \citet{Esteban09}, who derived 
an oxygen abundance of 12+log(O/H)=8.05$\pm$0.04 
using many nebular and auroral emission lines
from optical spectra obtained with the 10m Keck~I telescope. 
Although slightly higher than the value we derive here using
the emission lines observed in our optical spectrum (including the auroral [\ion{O}{iii}]~$\lambda$4363 line), 
12+log(O/H)=7.95$\pm$0.05, both measurements agree well within the errors.

The analysis of our H$\alpha$ images allows us to quantify the star-formation
activity throughout NGC~4861. 
The net-H$\alpha$ image (Fig~\ref{NGC4861image})
reveals 12 well-defined star-forming regions.
Bright knot~\#1 hosts the majority of the
starburst activity and, indeed, has the highest SFR per area. For this region we derive a SFR of 0.47$\pm$0.03~$M_{\odot}$\,yr$^{-1}$
using the H$\alpha$ luminosity. 
The SFR derived for all the galaxy is 0.48$\pm$0.04~$M_{\odot}$\,yr$^{-1}$.
The SFR obtained from the available FIR and 1.4~GHz data 
are quite low (10 -- 18 times lower) when compared with the H$\alpha$-based value,
 SFR$_{60\, \mu m}$=0.086~$M_{\odot}$\,yr$^{-1}$ and SFR$_{\rm 1.4\,GHz}$=0.049~$M_{\odot}$\,yr$^{-1}$.
As both FIR and radio-continuum are tracing the star-formation activity in the last $\sim$100~Myr but the timescale of the H$\alpha$ emission is around
10~Myr, this may suggest that there has been an enhancement of the star-formation activity within this object in the last few million years.

The age of the most recent star-formation even in bright knot~\#1 is 4.6~Myr. Some knots seem to have even younger ages.
The analysis of the broad-band optical colors suggests that the majority of the star-forming regions host an important
underlying stellar population, with ages ranging from 30 to 100~Myr. 
This fact together with the presence of strong absorption features in the optical spectrum of
the galaxy indicate the presence of an intermediate-age stellar population
in this BCDG.

 \subsubsection{NGC\,3003}

\textbf{NGC 3003} is a  SBbc type galaxy located at 24~Mpc
which appears to be almost edge-on  (see Fig~\ref{iNGC3003}).
Its morphology shows two asymmetric spiral arms,
suggesting that it may be a disturbed galaxy. 
Its SW areas show a bright and compact region
which may be the remnant of a dwarf galaxy, also it may also be 
just an intense \ion{H}{ii} region within the spiral arms of NGC~3003.

\citet{Ho95} reported the presence of the broad blue WR bump 
in NGC~3003, and therefore the galaxy was included in the WR galaxy catalogue
created by \citet*{SCP99}. This feature has not been
observed again in this galaxy. Indeed it is not seen in our optical spectrum,
however it has a  low SNR. Using the brightest emission lines
and the typical empirical calibrations we derive an oxygen abundance of 
12+log(O/H)=8.57 for this galaxy. Besides NGC~3003 has been amply
studied in the past we have not found any previous estimation
of its gas-phase metallicity  in the literature.

Our net-H$\alpha$ image allows us to identify 25 independent star-forming
regions, including the bright nucleus (knot~\#13) and the intriguing 
bright object at the SW (knot~\#3). These two knots actually host the 
highest H$\alpha$ emission of the galaxy.
The total H$\alpha$ luminosity of NGC~3003 adding up the flux of all star-forming region
is 2.97$\times10^{40}$~erg\,s$^{-1}$, which can be translated
in a total SFR of 0.24~$M_{\odot}$\,yr$^{-1}$.
Our derived H$\alpha$ luminosity  is 4 times lower
than that derived by \citet{Hoopes99}
who reported $L_{\rm H\alpha}$= 1.18$\times10^{41}$~erg\,s$^{-1}$.
Some part of the missing H$\alpha$ flux may be consequence of not considering an adequate
value for the correction for extinction, which may be important in edge-on galaxies. 
However, we should also expect that there is some diffuse H$\alpha$ gas in the galaxy 
which does not belong to the analyzed regions and hence it was not
considered in our total flux.
The SFR derived from the FIR and radio-continuum data available for this galaxy are
SFR$_{60\, \mu m}$=0.39~$M_{\odot}$\,yr$^{-1}$, SFR$_{\rm FIR}$=0.15~$M_{\odot}$\,yr$^{-1}$, and
SFR$_{\rm 1.4\,GHz}$=0.58~$M_{\odot}$\,yr$^{-1}$. All these values suggest that the 
total H$\alpha$ luminosity computed by  \citet*{Hoopes99}, who translates in a SFR of 0.94~$M_{\odot}$\,yr$^{-1}$,
is probably slightly overestimated.

We estimated the age of the most recent star-formation event via the H$\alpha$ equivalent width and the $U-B$ colors, 
finding typical values between 3 and 6~Myr. 
The optical colours obtained for the star-forming regions, in particular the $V-R$ and the $V-I$ colors, clearly indicate 
the presence of an important old stellar population underlying the bursts, with ages in many cases older than 500~Myr. 
Indeed, this is the only galaxy for which we find ages higher than 1~Gyr in some knots.

\section{Summary and conclusions}

We presented a detailed photometric  and spectroscopic study  of sample of 7
Wolf-Rayet galaxies. The observed galaxies are NGC~1140, IRAS~07164+5301, NGC~3738, UM~311, NGC~6764, NGC~4861 and NGC~3003. 
Star-forming regions  within these galaxies have  been  identified  using
narrow-band H$\alpha$ images. Combining with the data obtained using our optical broad-band images,
we analyze the morphologies, colours, star-formation rates and stellar populations of these star-forming regions.

 We discussed the morphology of the galaxies using our broad-band images. In some cases (NGC~1140, UM~311, NGC~481 and NGC~3003) we find features which may indicate that the galaxy has experienced a recent interaction.

We used the H$\alpha$ images and optical broad-band colours in combination with Starburst99 models to derived the age of the most recent star-formation event.
We confirmed that almost all the analyzed regions show a very young (3 -- 6~Myr old) starburst. We also used the optical colours to estimate the internal reddening and to study the age of the dominant underlying stellar populations within all these regions.  Knots in NGC~3738, NGC~6764 and NGC~3003 generally show the presence of an important old (400 -- 1000~Myr) stellar population. However, the optical colours are not able to detect stars older than 20 -- 50~Myr in the knots of the other four galaxies. This fact suggests both the intensity of the starbursts and that the star-formation activity has been ongoing for at least some few tens of million years in these objects. Deep NIR data should be needed to detect the old stellar populations in these galaxies.

We  derived the SFR of each knot using the H$\alpha$ luminosity.
The H$\alpha$-based SFR derived for each galaxy usually agrees well with the SFR derived using FIR and radio-continuum data. 

The optical spectra were used to search for the faint WR features, 
to confirm that the ionization of the gas is consequence of the massive stars, and to quantify the chemical properties of each object. 
The high S/N optical spectrum of NGC~1140 and NGC~4861 allowed us to precise the oxygen abundance of the ionized gas using the direct method (i.e., via the detection of the faint [\ion{O}{iii}]~$\lambda$4363 emission line). We also derive the chemical abundances of N, S, Ne and Ar in these two galaxies. In NGC~4861, the N/O ratio is $\sim$0.25-0.35~dex higher than that expected from its oxygen abundance. This fact may be related to the presence of WR stars within this galaxy. Indeed, we clearly detected the features originated by WR stars in  NGC~1140 and NGC~4861 and
used them to derive the population of O, WNL and WCE stars. 
In both cases, the derived WR/(WR+O) and WCE/WNL ratios agree well with those expected for galaxies with similar oxygen abundances.
 
For the rest of the galaxies we provided an estimation of the oxygen abundance of the ionized gas using several and independent empirical calibrations. Here we presented the first oxygen abundances computed for NGC~6764 and NGC~3003, which are 12+log(O/H)=8.65 and 8.57, respectively. We also derived the oxygen abundance of IRAS~07164+5301, 12+log(O/H)=8.50, which is $\sim$34\% of the only available determination of the metallicity of this galaxy.

\label{lastpage}

\section{Acknowledgments}

We would like to thank the anonymous referee, whose comments helped to  improve the paper considerably.
Authors thanks S.Ramya(IIA), Ian Stevans (University of Brimingham) and  Claus
Leitherer (STSci)  for their comments on the initial draft of the paper. Author also thanks C. Eswaraiah (ARIES) for his support and encouragement. CK acknowledges the financial assistance through a sponsored project Grant No: SR/S2/HEP-10/2002  of Department of Science and Technology (DST), Government of India, New Delhi. He also acknowledges Prof. T. P. Prabhu, IIA for his constant encouragement throughout this work and support during his stay at CREST campus, IIA. This research has made use of the NASA/IPAC Extragalactic Database (NED) which is 
operated by the Jet Propulsion Laboratory, California Institute of Technology, 
under contract with the National Aeronautics and Space Administration. 
This research has made extensive use of the SAO/NASA Astrophysics Data System Bibliographic Services (ADS).


\begin{thebibliography}{}

\scriptsize{


\bibitem[\protect\citeauthoryear{Allen et al.}{Allen et al.}{1991}]{Allen91}
Allen, David A.; Norris, R. P.; Meadows, V. S. \& Roche, P. F. 	1991, MNRAS, 248, 528

\bibitem[\protect\citeauthoryear{Alonso-Herrero et al.}{Alonso-Herrero et al.}{2000}]{AlonsoHerrero00}
Alonso-Herrero, A., Rieke, M. J., Rieke, G. H. \& Shields, J. C. 2000, ApJ, 530, 688

\bibitem[\protect\citeauthoryear{Amor\'{i}n et al.}{Amor\'{i}n et al.}{2012}]{Amorin12}
Amor\'{i}n, R., P\'erez-Montero, E., V\'{\i}lchez, J.M. \& Papaderos, P. 2012, ApJ, in press (arXiv:1202.3419)


\bibitem[\protect\citeauthoryear{Asplund, Grevesse \& Sauval}{Asplund et al.}{2005}]{ASP05}
Asplund, M., Grevesse, N. \& Sauval, A. J. 2005, in ASP Conf. Ser. 335, \emph{Cosmic Abundances as Records of Stellar Evolution and Nucleosynthesis}, ed. F.N. Bash \& T.G. Barnes (San Francisco: ASP), 25

\bibitem[\protect\citeauthoryear{Baldwin et al.}{Baldwin, Phillips \& Terlevich}{1981}]{BPT81} 
Baldwin J., Phillips M., \& Terlevich R., 1981, PASP, 93, 5



\bibitem[\protect\citeauthoryear{Bresolin et al.}{Bresolin et al.}{2009}]{Bresolin09}
Bresolin, F., Gieren, W., Kudritzki, R-P., Pietrzy\'nski, G., Urbaneja, M.A. \& Carraro, G. 2009, ApJ, 700, 309


\bibitem[\protect\citeauthoryear{Brinchmann et al.}{Brinchmann, Kunth \& Durret}{2008}]{BKD08}
Brinchmann, J., Kunth, D., \& Durret, F. 2008, A\&A, 485, 657

\bibitem[\protect\citeauthoryear{Buckalew et al.}{Buckalew, Kobulnicky \& Dufour}{2005}]{Buckalew05}
Buckalew, B.A., Kobulnicky, H.A. \& Dufour, R.J. 2005, ApJS, 157, 30



\bibitem[\protect\citeauthoryear{Calzetti et al.}{Calzetti et al.}{2007}]{Calzetti07}
Calzetti, D. et al. 2007, ApJ, 666, 870

\bibitem[\protect\citeauthoryear{Cardelli et al.}{Cardelli, Clayton \& Mathis}{1989}]{Cardelli89}
Cardelli, J.~A., Clayton, G.~C. \& Mathis J.~S. 1989, ApJ, 345, 245

\bibitem[\protect\citeauthoryear{Condon}{Condon}{1992}]{Condon92}
Condon, J.J. 1992, ARA\&A 30, 575
\bibitem[\protect\citeauthoryear{Condon et al.}{Condon et al.}{1998}]{Condon98}
Condon, J.~J., Cotton, W.~D., Greisen, E.~W., Yin, Q.~F., Perley, R.~A., Taylor, G.~B. \& Broderick, J.~J.\ 1998, \aj, 115, 1693
\bibitem[\protect\citeauthoryear{Condon et al.}{Condon, Cotton \& Broderick}{2002}]{CCB02}
Condon, J.J., Cotton, W.D. \& Broderick, J.J. 2002, AJ, 124, 675
\bibitem[\protect\citeauthoryear{Conti}{Conti}{1976}]{Conti76} 
Conti, P.S., 1976,  MSRSL, 9, 193
\bibitem[\protect\citeauthoryear{Conti}{Conti}{1991}]{C91} 
Conti, P.S., 1991, \apj, 377, 115


\bibitem[\protect\citeauthoryear{Conselice et al.}{Conselice et al.}{2000}]{Conselice00}
Conselice, C.J., Gallagher, J.S., Calzetti, D., Homeier, N. \& Kinney, A. 2000, AJ, 119, 79



\bibitem[\protect\citeauthoryear{Crowther \& Hadfield}{Crowther \& Hadfield}{2006}]{CH06}
Crowther, P.A. \& Hadfield, L.J., 2006, A\&A, 449, 711



\bibitem[\protect\citeauthoryear{Dinerstein \& Shields}{Dinerstein \& Shields}{1986}]{DinersteinShield86}
Dinerstein, H.L. \& Shield, G.A. 1986, ApJ, 311, 45

\bibitem[\protect\citeauthoryear{Dopita et al.}{Dopita et al.}{2000}]{Do00}
Dopita, M.A., Kewley, L. J., Heisler, C.A. \& Sutherland, R.S. 2000, ApJ, 542, 224 
      
\bibitem[\protect\citeauthoryear{Dottori et al.}{Dottori et al.}{1994}]{Dottori94}      
H. Dottori, J. Cepa, J. V\'{\i}lchez \& C.S. Barth, 1994, A\&A, 283, 753

\bibitem[\protect\citeauthoryear{Eckart et al.}{Eckart et al.}{1991}]{Eckart91}
Eckart, A., Cameron, M., Jackson, J. M., Genzel, R., Harris, A. I., Wild, W. \& Zinnecker, H. 1991,	ApJ, 372, 67

\bibitem[\protect\citeauthoryear{Eckart et al.}{Eckart et al.}{1996}]{Eckart96}	
Eckart, A., Cameron, M., Boller, Th., Krabbe, A., Blietz, M., Nakai, N., Wagner, S. J. \& Sternberg, A. 1996, AJ, 472, 58

\bibitem[\protect\citeauthoryear{Esteban et al.}{Esteban et al.}{2004}]{Esteban04}
Esteban, C., Peimbert, M., Garc\'{\i}a-Rojas, J., Ruiz, M. T., Peimbert, A. \& Rodr\'{\i}guez, M., 2004, MNRAS, 355, 229
\bibitem[\protect\citeauthoryear{Esteban et al.}{Esteban et al.}{2009}]{Esteban09}
Esteban, C., Bresolin, F., Peimbert, M., Garc\'{\i}a-Rojas, J., Peimbert, A. \& Mesa-Delgado, A. 2009, ApJ, 700, 654



\bibitem[\protect\citeauthoryear{Fernandes et al.}{Fernandes et al.}{2004}]{FCCG04}
Fernandes, I.F., de Carvalho, R., Contini, T. \& Gal, R.R. 2004, MNRAS 355, 728

\bibitem[\protect\citeauthoryear{French}{French}{1980}]{French80}
French, H.~B. 1980, \apj, 240, 41

\bibitem[\protect\citeauthoryear{Garc\'{\i}a-Rojas et al.}{Garc\'{\i}a-Rojas et al.}{2005}]{GRE05}
Garc\'{\i}a-Rojas, J., Esteban, C., Peimbert, A., Peimbert, M., Rodr\'{\i}guez, \& M., Ruiz, 2005, MNRAS, 362, 301

\bibitem[\protect\citeauthoryear{Garnett}{Garnett}{1992}]{G92}
Garnett, D.R. 1992, AJ, 103, 1330
\bibitem[\protect\citeauthoryear{Garnett}{Garnett}{2004}]{G04}
Garnett, D.R. 2003, lectures on \emph{Cosmochemistry: The melting pot of the elements}. 
XIII Canary Islands Winter School of Astrophysics, Puerto de la  Cruz, Tenerife, Spain, 
November 19-30, 2001, edited by C. Esteban, R. J. Garc\'{\i}a L\'opez, A. Herrero, F. S\'anchez. 
Cambridge contemporary astrophysics. Cambridge, UK: Cambridge University Press, ISBN 0-521-82768-X, 2004, p. 171

\bibitem[\protect\citeauthoryear{Garnett et al.}{Garnett et al.}{1991}]{Garnett91}
Garnett, D.R., Kennicutt, R.~C.Jr., Chu, Y.-H. \& Skillman E.~D. 1991, ApJ, 373, 458 



\bibitem[\protect\citeauthoryear{Guseva et al.}{Guseva, Izotov \& Thuan}{2000}]{GIT00}
Guseva, N., Izotov, Y. I. \& Thuan, T.X. 2000, ApJ, 531, 776




\bibitem[\protect\citeauthoryear{Ho et al.}{Ho, Filippenko \& Sargent}{1995}]{Ho95}
Ho, L. C., Filippenko, A. V. \& Sargent, W. L. 1995, ApJS, 98, 477

\bibitem[\protect\citeauthoryear{Hoopes et al.}{Hoopes, Walterbos \& Rand}{1999}]{Hoopes99}
Hoopes, C.G., Walterbos, R. A. M. \& Rand, R.J. 1999, ApJ, 522, 669

\bibitem[\protect\citeauthoryear{Hopkins et al.}{Hopkins, Schulte-Ladbeck \& Drozdovsky}{2002}]{Hopkins02}
Hopkins, A.~M., Schulte-Ladbeck, R.~E. \& Drozdovsky, I.~O.\ 2002, AJ, 124, 862

\bibitem[\protect\citeauthoryear{Huang et al.}{Huang et al.}{1996}]{Huang96}
Huang, Jiehao; Gu, Qiusheng; Su, Hongjun; Shang, Zhaohui, 1996, Ap\&SS, 235, 109


\bibitem[\protect\citeauthoryear{Huang et al.}{Huang et al.}{1999}]{Huang99}
Huang, J.H., Gu, Q.S., Ji,L., Li, W.D., Wei, J.Y. \& Zheng, W. 1999, ApJ, 513, 215

\bibitem[\protect\citeauthoryear{Hunter}{Hunter}{1982}]{Hunter82}
Hunter D. A., 1982, ApJ, 260, 81
\bibitem[\protect\citeauthoryear{Hunter et al.}{Hunter, Gallagher \& Rautenkranz}{1982}]{HunterGR82}
Hunter D. A., Gallagher, J. \& Rautenkranz, D. 1982, ApJS, 49, 53


\bibitem[\protect\citeauthoryear{Hunter et al.}{Hunter et al.}{1994a}]{Hunter94a}
Hunter D. A., O Connell R. W., Gallagher III J. S., 1994a, AJ, 108, 84

\bibitem[\protect\citeauthoryear{Hunter et al.}{Hunter et al.}{1994b}]{Hunter94b}
Hunter D. A., van Woerden H., Gallagher III J. S., 1994b, ApJS, 91, 79


\bibitem[\protect\citeauthoryear{Hunter \& Hoffman}{Hunter \& Hoffman}{1999}]{HunterHoffman99}
Hunter D. A., \& Hoffman, L. 1999, AJ, 117, 2789

\bibitem[\protect\citeauthoryear{Izotov et al.}{Izotov, Thuan, \& Lipovetski}{1994}]{ITL94}
Izotov, Y.I., Thuan, T.X., \& Lipovetski, 1994, ApJ, 435, 647
\bibitem[\protect\citeauthoryear{Izotov et al.}{Izotov et al.}{1997}]{Izotov97}
Izotov, Y.~I., Foltz, C.~B., Green, R.~F., Guseva, N.~G. \& Thuan T.~X., 1997, ApJ, 487, L37 
\bibitem[\protect\citeauthoryear{Izotov \& Thuan}{Izotov \& Thuan}{1998}]{IT98}
Izotov, Y.I. \& Thuan, T.X. 1998, ApJ, 500, 188
\bibitem[\protect\citeauthoryear{Izotov \& Thuan}{Izotov \& Thuan}{1999}]{IT99}
Izotov, Y.I. \& Thuan, T.X. 1999, ApJ, 511, 639
\bibitem[\protect\citeauthoryear{Izotov et al.}{Izotov et al.}{2004}]{Izotov04}   
Izotov, Y.I., Papaderos, P., Guseva, N.G., Fricke, K.J. \& Thuan, T.X. 2004, A\&A 421, 539
\bibitem[\protect\citeauthoryear{Izotov et al}{Izotov et al}{2006}]{Izotov06}
Izotov, Y.I., Stasi\'nska, G., Meynet, G., Guseva, N.G. \& Thuan, T.X. 2006, A\&A, 448, 955

\bibitem[\protect\citeauthoryear{James, Tsamis \& Barlow}{James et al.}{2009}]{James09}
James, B.L., Tsamis, Y.G., Barlow, M.J.,  Westmoquette, M.S., Walsh, J.R., Cuisinier, F. \& Exter, K.M. 2009, MNRAS, 398, 2
\bibitem[\protect\citeauthoryear{Johnson et al.}{Johnson et al.}{1999}]{J99}
Johnson, K. E., Vacca, W.D., Leitherer, C., Conti, P. S. \& Lipscy, S. J., 1999,  AJ, 117, 1708
\bibitem[\protect\citeauthoryear{Johnson \& Conti}{Johnson \& Conti}{2000}]{JC00} 
Johnson, K. E. \& Conti, P.S. 2000, ApJ, 119, 2146

\bibitem[\protect\citeauthoryear{Karachentsev et al.}{Karachentsev et al.}{2003}]{Karachentsev2003}
Karachentsev, I.D.,  Sharina, M. E.,  Dolphin,  A. E.,  Grebel, E. K.,  Geisler, D., 
Guhathakurta, P., Hodge,  P. W.,  Karachentseva,  V. E.,  Sarajedini, A. \& 
Seitzer, P. 2003, A\&A, 398, 467

\bibitem[\protect\citeauthoryear{Kauffmann et al.}{Kauffmann et al.}{2003}]{Kauffmann03}
Kauffmann, G. et al. 2003, MNRAS, 346, 1055

\bibitem[\protect\citeauthoryear{Kennicutt}{Kennicutt}{1998}]{K98}
Kennicutt, R.C. Jr. 1998,ARAA, 36, 189 



\bibitem[\protect\citeauthoryear{Kennicutt, Bresolin \& Garnett}{Kennicutt et al.}{2003}]{Kennicutt03}
Kennicutt, R.C. Jr., Bresolin, F. \& Garnett, D.R. 2003, ApJ, 591, 801

\bibitem[\protect\citeauthoryear{Kennicutt et al.}{Kennicutt et al.}{2008}]{Kennicutt08}
Kennicutt, R. C., Jr., Lee, J.C., Funes, S.J., Jos\'e G., Sakai, S., \& Akiyama, S. 2008, ApJS, 178, 247

\bibitem[\protect\citeauthoryear{Kewley et al.}{Kewley et al.}{2001}]{KD01}
Kewley, L.J., Dopita, M.A., Sutherland, R.S., Heisler, C.A. \& Trevena, J. 2001, ApJS, 556, 121

\bibitem[\protect\citeauthoryear{Kewley \& Dopita}{Kewley \& Dopita}{2002}]{KD02}
Kewley, L.J. \& Dopita, M.A. 2002, ApJS, 142, 35


\bibitem[\protect\citeauthoryear{Kobulnicky et al.}{Kobulnicky et al.}{1997}]{Kobulnicky97}
Kobulnicky, H.A., Skillman, E.D., Roy, J.-R., Walsh, J.R. \& Rosa, M.R., 1997, ApJ, 277, 679

\bibitem[\protect\citeauthoryear{Kobulnicky \& Skillman}{Kobulnicky \& Skillman}{1998}]{KS98}
Kobulnicky, H.A, \& Skillman, E.D. 1998, ApJ, 497, 601

\bibitem[\protect\citeauthoryear{Kobulnicky \& Kewley}{Kobulnicky \& Kewley}{2004}]{KK04} 
Kobulnicky H.~A. \& Kewley L.~J. 2004, ApJ, 617, 240 


\bibitem[\protect\citeauthoryear{Koribalski \& L\'opez-S\'anchez}{Koribalski \& L\'opez-S\'anchez}{2009}]{KoribalskiLS09}
Koribalski, B.S. \& L\'opez-S\'anchez, \'A.R. 2009, MNRAS, 400, 1749



 

\bibitem[\protect\citeauthoryear{Landolt}{Landolt}{1992}]{L92}
Landolt A.U. 1992, AJ, 104, 340

\bibitem[\protect\citeauthoryear{Lee et al.}{Lee, Salzer \& Melbourne}{2004}]{Lee04}
Lee, J.C., Salzer J.J. \& Melbourne, J. 2004, ApJ, 616, 752L

\bibitem[\protect\citeauthoryear{Leitherer \& Heckman}{Leitherer \& Heckman}{1995}]{LH95}
Leitherer, C. \& Heckman, T.M. 1995, ApJS, 96, 9
\bibitem[\protect\citeauthoryear{Leitherer et al.}{Leitherer et al.}{1999}]{L99}
Leitherer, C., Schaerer, D., Goldader, J.D., Gonz\'alez-Delgado, R.M., Robert, C., Kune, D.F., de Mello, D.F., Devost, D. \& Heckman, T.M. 1999, ApJS, 123, 3


\bibitem[\protect\citeauthoryear{Leon et al.}{Leon et al.}{2007}]{Leon07}
Leon, S.; Eckart, A.; Laine, S.; Kotilainen, J. K.; Schinnerer, E.; Lee,
S.-W.; Krips, M.; Reunanen, J. \& Scharwachter, J. 2007, A\&A, 473, 747



\bibitem[\protect\citeauthoryear{L\'opez-S\'anchez}{L\'opez-S\'anchez}{2010}]{LS10}
L\'opez-S\'anchez, \'A.R. 2010, A\&A, 521, 63 

\bibitem[L\'opez-S\'anchez et al.(2004a)]{LSER04a}
L\'opez-S\'anchez, \'A.R., Esteban, C. \& Rodr\'{\i}guez, M. 2004a, ApJS, 153, 243
\bibitem[L\'opez-S\'anchez et al.(2004b)]{LSER04b}
L\'opez-S\'anchez, \'A.R., Esteban, C. \& Rodr\'{\i}guez, M. 2004b, A\&A 428,445
\bibitem[\protect\citeauthoryear{L\'opez-S\'anchez et al.}{L\'opez-S\'anchez, Esteban \& Garc\'{\i}a-Rojas}{2006}]{LSEGR06}
L\'opez-S\'anchez, \'A.R., Esteban, C. \& Garc\'{\i}a-Rojas, J. 2006, A\&A, 449, 997
\bibitem[\protect\citeauthoryear{L\'opez-S\'anchez et al.}{L\'opez-S\'anchez et al.}{2007}]{LSEGRPR07}
L\'opez-S\'anchez, \'A.R., Esteban, C., Garc\'{\i}a-Rojas, J., Peimbert, M. \& Rodr\'{\i}guez, M. 2007, ApJ, 656, 168
\bibitem[\protect\citeauthoryear{L\'opez-S\'anchez \& Esteban}{L\'opez-S\'anchez \& Esteban}{2008}]{LSE08} 
L\'opez-S\'anchez, \'A.R. \& Esteban, C. 2008, A\&A, 491, 131
\bibitem[\protect\citeauthoryear{L\'opez-S\'anchez \& Esteban}{L\'opez-S\'anchez \& Esteban}{2009}]{LSE09} 
L\'opez-S\'anchez, \'A.R. \& Esteban, C. 2009, A\&A, 508, 615
\bibitem[\protect\citeauthoryear{L\'opez-S\'anchez \& Esteban}{L\'opez-S\'anchez \& Esteban}{2010a}]{LSE10a} 
L\'opez-S\'anchez, \'A.R. \& Esteban, C. 2010a, A\&A, 516, 104
\bibitem[\protect\citeauthoryear{L\'opez-S\'anchez \& Esteban}{L\'opez-S\'anchez \& Esteban}{2010b}]{LSE10b} 
L\'opez-S\'anchez, \'A.R. \& Esteban, C. 2010b, A\&A, 517, 85
\bibitem[\protect\citeauthoryear{L\'opez-S\'anchez et al.}{L\'opez-S\'anchez et al.}{2011}]{LS+IC10+11} 
L\'opez-S\'anchez, \'A.R., Mesa-Delgado, A., L\'opez-Mart\`{\i}n, L \& Esteban, C. 2011, MNRAS, 411, 2076
\bibitem[\protect\citeauthoryear{L\'opez-S\'anchez et al.}{L\'opez-S\'anchez et al.}{2012}]{LSD12}
L\'opez-S\'anchez, \'A.R., Dopita, M.A., Kewley, L.J., Zahid, H.J., Nicholls, D.C. \&  Scharw\"achter, J. 2012, MNRAS, 426, 2630

                        


\bibitem[\protect\citeauthoryear{Martin}{Martin}{1997}]{Martin97}
Martin, C. L. 1997, ApJ, 491, 561

\bibitem[\protect\citeauthoryear{Masegosa et al.}{Masegosa, Moles \& del Olmo}{1991}]{Masegosa91}
Masegosa, M., Moles, M. \& del Olmo, A. 1991, A\&A, 244, 273
\bibitem[\protect\citeauthoryear{Mazzarrella \& Boronson}{Mazzarrella \& Boronson}{1993}]{MB93}
Mazzarrella, J.M. \& Boronson, T.A. 1993, ApJS, 85, 27


\bibitem[\protect\citeauthoryear{McGaugh}{McGaugh}{1991}]{McGaugh91}
McGaugh, S.S. 1991, ApJ, 380, 140

\bibitem[\protect\citeauthoryear{M\'endez \& Esteban}{M\'endez \& Esteban}{2000}]{ME00}
M\'endez, D.I. \& Esteban, C., 2000, A\&A, 359, 493

\bibitem[\protect\citeauthoryear{Meynet \& Maeder}{Meynet \& Maeder}{2005}]{MeynetMaeder05}
Meynet G. \& Maeder A. 2005. A\&A, 429, 581

\bibitem[\protect\citeauthoryear{Moll et al.}{Moll et al.}{2007}]{Moll07}
Moll, S.L., Mengel, S., de Grijs, R., Smith, L. J. \% Crowther, P.A. 2007, MNRAS, 382, 1877

\bibitem[\protect\citeauthoryear{Monreal-Ibero et al.}{Monreal-Ibero et al.}{2010}]{Monreal-Ibero+10}
Monreal-Ibero, A., V\'{\i}lchez, J.M.; Walsh, J.R. \& Mu\~noz-Tu\~n\'on, C. 2010, A\&A, 517, 27

\bibitem[\protect\citeauthoryear{Moshir et al.}{Moshir et al.}{1990}]{Moshir90}
Moshir M. et al. 1990, Infrared Astronomical Satellite Catalogs, The Faint Source Catalog, Version 2.0 

\bibitem[\protect\citeauthoryear{Moustakas et al.}{Moustakas et al.}{2010}]{Moustakas+10}
Moustakas, J., Kennicutt, R.C., Jr., Tremonti, C. A., Dale, D. A., Smith, J.-D. T. \& Calzetti, D. 2010, ApJS, 190, 233

\bibitem[\protect\citeauthoryear{Noeske et al.}{Noeske et al.}{2003}]{N03}
Noeske, K.G., Papaderos, P., Cair\'os, L.M. \& Fricke, K.J. 2003, A\&A, 410, 481
\bibitem[\protect\citeauthoryear{Noeske et al.}{Noeske et al.}{2005}]{Noeske05}
Noeske, K.G., Papaderos, P., Cair\'os, L.M \& Fricke, K.J. 2005, A\&A, 429, 115

\bibitem[\protect\citeauthoryear{Oke}{Oke}{1990}]{Oke90}
Oke, J. B. 1990, ApJ, 99, 1621

\bibitem[\protect\citeauthoryear{Osterbrock \& Cohen}{Osterbrock \& Cohen}{1982}]{OsterbrockCohen82}
Osterbrock, D.E. \& Cohen, R.D. 1982, ApJ, 261, 64

\bibitem[\protect\citeauthoryear{Peimbert \& Costero}{Peimbert \& Costero}{1969}]{PC69}
Peimbert, M. \& Costero, R. 1969, Bol. Obs. Ton. y Tac., 5, 3

\bibitem[\protect\citeauthoryear{Peimbert et al.}{Peimbert et al.}{2007}]{Peimbert07}
Peimbert, M., Peimbert, A.. Esteban, C.; Garc\'{\i}a-Rojas, J., Bresolin, F., Carigi, L., Ruiz, M.T. \& L\'opez-S\'anchez, \'A.R. 2007, RMxAC, 29, 72


\bibitem[\protect\citeauthoryear{P\'erez-Montero et al.}{P\'erez-Montero et al.}{2011}]{PerezMontero11}
P\'erez-Montero, E., V\'{\i}lchez, J. M., Cedr\'es, B., H\"agele, G. F.; Moll\'a, M. Kehrig, C., D\'{\i}az, A. I., 
Garc\'{\i}a-Benito, R. \& Mart\'{\i}n-Gord\'on, D. 2011, A\&A, 532, 141


\bibitem[\protect\citeauthoryear{Pettini \& Pagel}{Pettini \& Pagel}{2004}]{PP04}
Pettini, M. \& Pagel, B.E.J. 2004, \mnras, 348, 59


\bibitem[\protect\citeauthoryear{Pindao}{Pindao}{1999}]{Pindao99}
Pindao, M.\ 1999, in \emph{IAU Symp.~193: Wolf-Rayet Phenomena in Massive Stars and Starburst Galaxies}, 193, 614
\bibitem[\protect\citeauthoryear{Pindao et al.}{Pindao et al.}{2002}]{PSGD02}
Pindao, M., Schaerer, D., Gonz\'alez-Delgado, R.M. \& Stasi\'nska, G. 2002, A\&A 394, 443

\bibitem[\protect\citeauthoryear{Pilyugin}{Pilyugin}{2001a}]{P01a}
Pilyugin, L.S. 2001a, A\&A, 369, 594
\bibitem[\protect\citeauthoryear{Pilyugin}{Pilyugin}{2001b}]{P01b}
Pilyugin, L.S. 2001b, A\&A, 374, 412


\bibitem[\protect\citeauthoryear{Pilyugin \& Thuan}{Pilyugin \& Thuan}{2005}]{PT05}
Pilyugin, L.S. \& Thuan, T.X. 2005, ApJ, 631, 231
\bibitem[\protect\citeauthoryear{Pilyugin, V\'{\i}lchez \& Thuan}{Pilyugin et al.}{2010}]{PVT10}
Pilyugin, L.S., V\'{\i}chez, J.M. \& Thuan, T.X. 2010, ApJ, 720, 1738


\bibitem[\protect\citeauthoryear{Pustilnik et al.}{Pustilnik et al.}{2004}]{Pustilnik04}   
Pustilnik, S., Kniazev, A., Pramskij, A., Izotov, Y., Foltz, C., Brosch, N., Martin, J.-M. \& Ugryumov, A. 2004, A\&A 419, 469



\bibitem[\protect\citeauthoryear{Reines et al.}{Reines et al.}{2010}]{Reines10} 	
Reines, Amy E.; Nidever, David L.; Whelan, David G. \& Johnson, Kelsey E. 2010, ApJ, 708, 26





\bibitem[\protect\citeauthoryear{Rosales-Ortega et al.}{Rosales-Ortega et al.}{2011}]{RO11}
Rosales-Ortega, F.F., D\'{\i}az, A.I., Kennicutt, R.C. \& S\'anchez, S.F. 2011, MNRAS, 415, 2439


\bibitem[\protect\citeauthoryear{Salzer et al.}{Salzer, MacAlpine \& Boroson}{1989}]{Salzer89a}
Salzer, J.~J., MacAlpine, G.~M. \& Boroson, T.~A.\ 1989, ApJS, 70, 447


\bibitem[\protect\citeauthoryear{Sandage \& Tammann}{Sandage \& Tammann}{1981}]{SandageTammann81}
Sandage, A. \& Tammann, G. A.1981, {\it A Revised Shapley-Ames Catalog of Bright Galaxies} , Washington, DC; 
Carnegie Inst. Washington


\bibitem[\protect\citeauthoryear{Schaerer \& Vacca}{Schaerer \& Vacca}{1998}]{SV98}
Schaerer, D. \& Vacca, W.D. 1998, ApJ, 497, 618
\bibitem[\protect\citeauthoryear{Schaerer et al.}{Schaerer, Contini \& Pindao}{1999}]{SCP99}
Schaerer, D., Contini, T. \& Pindao, M. 1999, A\&AS 136, 35


\bibitem[\protect\citeauthoryear{Schlegel et al.}{Schlegel, Finkbeiner \& Davis}{1998}]{SFD98}
Schlegel, D.J., Finkbeiner, D.P. \& Davis, M. 1998, ApJ, 500, 525

\bibitem[\protect\citeauthoryear{Shaw \& Dufour}{Shaw \& Dufour}{1995}]{SD95}
Shaw, R.A. \& Dufour, R.J. 1995, PASP, 107, 896


\bibitem[\protect\citeauthoryear{Stasi\`nska}{Stasi\'nska}{1978}]{S78}
Stasi\'nska, G. 1978, A\&A, 66, 257

\bibitem[\protect\citeauthoryear{Storey \& Hummer}{Storey \& Hummer}{1995}]{SH95}
Storey, P.J. \& Hummer, D.G. 1995, MNRAS 272, 41

\bibitem[\protect\citeauthoryear{Thuan \& Martin}{Thuan \& Martin}{1981}]{ThuanMartin81}
Thuan, T.X. \& Martin, G.E. 1981, ApJ, 247, 823


\bibitem[\protect\citeauthoryear{Vacca \& Conti}{Vacca \& Conti}{1992}]{VC92} 
Vacca, W.D. \& Conti, P.S., 1992, ApJ, 401, 543
\bibitem[\protect\citeauthoryear{Vacca}{Vacca}{1994}]{V94} 
Vacca, W.D. 1994, ApJ, 421, 140

\bibitem[\protect\citeauthoryear{Vanzi et al.}{Vanzi et al.}{2000}]{Vanzi00}
Vanzi, L., Hunt, L.K., Thuan, T.X. \& Izotov, Y.I. 2000, A\&A, 363, 493
\bibitem[\protect\citeauthoryear{Vanzi et al.}{Vanzi et al.}{2002}]{Vanzi02}
Vanzi, L., Hunt, L.K. \& Thuan, T.X. 2002, A\&A, 390, 481

\bibitem[\protect\citeauthoryear{Veilleux \& Osterbrock}{Veilleux \& Osterbrock}{1987}]{VO87}
Veilleux, S. \& Osterbrock, D.E. 1987, ApJS, 63 295

\bibitem[\protect\citeauthoryear{de Vaucouleur et al.}{de Vaucouleur et al.}{1991}]{deVaucouleur91}
de Vaucouleurs G., de Vaucouleurs A., Corwin Jr. H.G., Buta
R.J., Paturel G., Fouqu\'e P. 1991, ÒThird Reference Catalogue
of Bright GalaxiesÓ (New York: Springer Verlag),

\bibitem[\protect\citeauthoryear{Waller}{Waller}{1990}]{Waller90}
Waller, W.H. 1990, PASP, 102, 1217

\bibitem[\protect\citeauthoryear{York et al.}{York et al.}{2000}]{York00}
York, D.G. et al. 2000, AJ, 120, 1579

\bibitem[\protect\citeauthoryear{Zhang et al.}{Zhang et al.}{2007}]{Zhang07}
Zhang, W., Kong, X., Li, C., Zhou, H.-Y., Cheng, F.-Z. 2007, ApJ, 655, 851





}

\end{thebibliography}
\end{document}